\documentclass[useAMS,usenatbib]{mn2e}
\usepackage{graphicx}
\usepackage{txfonts}
\usepackage{amssymb}

\def\apj{ApJ}

\def\beq#1{\begin{equation}\label{#1}}
\def\eeq{\end{equation}}
\def\beqa#1{\begin{eqnarray}\label{#1}}
\def\eeqa{\end{eqnarray}}

\def\comment#1{\relax}

\begin{document}

\title[Variable neutron star free precession in Her X-1]{Variable neutron star free precession in Hercules X-1 from evolution of RXTE X-ray pulse profiles 
with phase of the 35-day cycle}
\author[Postnov et al.]{K. Postnov$^{1}$\thanks{E-mail:
kpostnov@gmail.com (KAP)},
N. Shakura$^{1}$,
R. Staubert$^{2}$, 
A. Kochetkova$^{1}$, 
D. Klochkov$^{2}$, 
J. Wilms$^{3}$\\
%\and L. Rodina
$^1$Moscow M.V. Lomonosov State University, Sternberg Astronomical Institute, Universitetskij pr., 13, 119992, Moscow, Russia\\	
$^2$
Institute of Astronomy and Astrophysics, Eberhard-Karls-Universit\"at, T\"ubingen, 
Sand 1, D-7276, Germany\\
$^3$
Dr. Remeis Observatory, Bamberg, Erlangen University 
}
\date{Received ... Accepted ...}

\pagerange{\pageref{firstpage}--\pageref{lastpage}} \pubyear{2012}

\maketitle

\label{firstpage}

\begin{abstract}

Accretion of matter onto the surface of a freely precessing neutron star with a complex 
non-dipole magnetic field can explain the change of X-ray pulse profiles of Her X-1 observed by 
RXTE with the phase of the 35-day cycle. 
We demonstrate this using all
available measurements of X-ray pulse profiles in the 9-13 keV energy
range obtained with the RXTE/PCA. The measured profiles guided the
elaboration of a geometrical model and the definition of locations
of emitting poles, arcs, and spots on the NS surface which
satisfactorily reproduce the observed pulse profiles and their
dependence on free precession phase.
We have found that the observed trend of the times of the 35-day turn-ons 
on the O-C diagram, which can be 
approximated by a collection of consecutive linear segments around the mean value, can be 
described by our model by assuming a variable free precession period, with a fractional period 
change of about a few percent. Under this assumption and using our model, we have found that the 
times of phase zero of the neutron star 
free precession (which we identify with the maximum separation 
of the brightest spot on the neutron star surface with the NS spin axis) occur 
about 1.6 days after 
the the mean turn-on times inside each 'stable' epoch, producing a linear trend on the O-C diagram with 
the same slope as the observed times of turn-ons. 
We propose that the 2.5\% changes in the free precession period that occur on time
scales of several to tens of 35-day cycles can be related to wandering
of the principal inertia axis of the NS body due to variations in the
patterns of accretion onto the NS surface.
The closeness of periods of the disk 
precession and the NS free precession can be explained by the presence of a 
synchronization mechanism in the system, which modulates the dynamical interaction of the gas 
streams and the accretion disk with the NS free precession period.  
\end{abstract}
\begin{keywords}
accretion - pulsars:general - X-rays:binaries
\end{keywords}

\section{Introduction}

The X-ray pulsar Her X-1 is an accreting neutron star (NS) with a spin period of $P\approx 1.24$~s in 
a 1.7-day orbit around 1.8-2 $M_\odot$ star \citep{Tananbaum_ea72}. Since its discovery in 1972 by the 
\textit{UHURU} satellite, the detailed nature of the source has been a subject of intensive studies, primarily 
because of its very complex time behavior. 
%observed. 

The 
%form of the 
long-term X-ray light curve of Her X-1 is modulated with a $\sim 35$-day period. 
The cycle includes the 
main-on state with an average duration of $\approx 7$ days and the secondary (short-on) state of smaller 
intensity which lasts $\approx 5$~days. Most 35-d cycles last for about 20, 20.5, or 21 orbital cycles; see 
\citep{Staubert_ea83, Shakura_ea98}. These two on-states are separated by time intervals with a duration 
of about four orbital periods during which the X-ray flux almost vanishes. These changes, as well as spectral measurements from the RXTE satellite \citep{Kuster_ea05}, clearly indicate the eclipse 
of the X-ray source by a precessing accretion disk.

The 35-day cycle in Her X-1 is generally interpreted to be a manifestation of the precession motion of the 
accretion disk around the NS in the retrograde orbital direction \citep{Gerend&Boynton76, Shakura_ea99}. 
Soon after the discovery of this X-ray pulsar it was proposed that free precession of the NS can also 
be responsible for the observed long-term modulation of the X-ray flux \citep{Brecher72}. 
%(which, however,is now believed to be caused by the precession of the disk). 
Later, \citet{Truemper_ea86} had interpreted the evolution of the X-ray pulse profile shape with the 
phase of the 35-day cycle as observed by EXOSAT as due to the free precession of the NS in Her X-1.

One of the observed properties  of the 35-d cycle in Her X-1 is its turn-ons occurring predominantly around orbital 
phases $\phi_{orb}\approx 0.2$ or 0.7 (see, however, the recent debate in \citet{Leahy_10} based on the analysis 
of the RXTE ASM data). This feature can be explained by tidal nutation motion of outer parts of the accretion disk 
at about double orbital frequency \citep{Katz_ea82, Levine&Jernigan82, Boynton87}, as at these orbital phases the disk inclination 
angle to the line of sight changes most rapidly (see also \citep{Klochkov_ea06}). 

Immediately after the beginning of active studies of Her X-1 it became clear that the tilted accretion disk in this
binary system must have a twisted form. During the retrograde precession, the outer parts of such a disk 
open the view to the central X-ray source (the turn-on of the source), 
and the inner parts of the disk cover it 
at the end of the main-on
\citep{Boynton87}. The ingress and egress to and from the two 35-day low states 
are asymmetric due to scattering of X-ray emission by 
the hot rarefied corona above the disk. Indeed, the opening of the X-ray source at the beginning of the main-on 
is observed to be accompanied by significant changes in the soft X-ray band suggesting strong absorption \citep{Becker_ea77, DavisonFabian77, Parmar_ea80, Kuster_ea05}. 

The change of X-ray pulse profiles with the 35-d phase is a well-established property of Her X-1 and has been observed by various satellites
(see e.g. \citet{Truemper_ea86, Deeter_ea98, Scott_ea00, Staubert_ea13}). In the present paper we show that the changes
in X-ray pulse profiles observed by \textsl{RXTE}/PCA in the 9-13 keV range with 35-d phase can be 
explained by a model which assumes a freely precessing neutron star with complex surface 
magnetic field structure. In addition to the canonical magnetic poles (of a dipole field), 
bright 
arc-like or quasi-circular structures must be present around the poles. As was first considered in 
\citet{Shakura_ea91, Panchenko&Postnov94}, to form such structures around the poles, the magnetic 
field of NS must have substantially non-dipole structure. 
The appearance of complex emitting regions on the surface of
an accreting NS with non-dipole magnetic field 
was confirmed by detailed numerical 3D MHD simulations (\citet{Long_ea08, Long_ea12}).

Due to the free precession, the emitting regions 
on the NS surface change location with respect to the line of sight leading to gradual change of X-ray 
profiles with the 35-d cycle phase. More rapid changes in pulse profiles (which are observed at the 
end of the main-on state) can be additionally caused by the eclipse of the NS surface by the inner parts 
of the disk. At low states (when the X-ray source is hidden behind the accretion disk), only 
a sine-line 
component (mostly pronounced in soft X-rays) due to reflection of X-ray emission 
of the NS by the inner warped parts of the twisted accretion disk remains.  

We also show that in our model the times of phase zero of the NS precession 
(defined by the maximum angle between the brightest spot on the NS surface and the NS angular 
momentum vector), as derived from modeling of pulse profiles in individual 35-day cycles, make straight line segments on the O-C diagram, which closely follow the shape of the 
turn-ons of the source, suggesting the presence of a stable clock over periods up to several tens of 35-day cycles. 

The change in slope of the linear segments in the O-C diagram implies a variable free precession period 
with a fractional period change $\Delta P_{pr}/P_{pr}\sim 2-3\%$. 
The possible reason for the free precession period variations is the corresponding $\sim 2-3\%$ change in angle between 
the principal axis of inertia of the neutron star body and the total angular momentum during the accretion of
matter onto the NS with complex surface magnetic field.

\section{Physical motivation}

\subsection{Previous works}

The apparent change of X-ray pulse profiles with 35-day cycle has been
previously addressed in several papers. Tr\"umper et al. (1986) discussed
change of pulse profiles in the main-on and short-on states observed by
\textit{EXOSAT} in terms of the neutron star free precession. Later, however, the
free precession model was questioned. For example,
 Bisnovatyi-Kogan et al. (1989) argued that the model of a freely precessing
neutron star approximated by a Maclaurin spheroid of rotation, with the
geometrical parameters of (symmetric) dipole magnetic field as derived from
EXOSAT observations, is unable to explain the observed X-ray pulse profile
evolution. They proposed the model of a progradely precessing accretion disk
with the inner radius at $\sim 10$ NS radii around a NS with rather low
surface
magnetic field of $\sim 10^{10}$-$10^{11}$~G. In this model, the pulse
profile
is shaped by the inner disk eclipsing a broad emission region on the NS surface. To explain 
difference in pulse profiles between the main-on and
short-on states of the source
observed by \textit{EXOSAT},  \cite{Petterson_ea91} proposed an alternative qualitative model. In their model, the inner radius of the retrogradely precessing accretion disk around non-precessing neutron star was assumed to circulate synchronously with the neutron star rotation (1.24 s) and to be partially responsible for the observed changes in the pulse profiles. The complex pulse shape in the main-on state was, however, attributed to the complex radiation transfer in the accretion column. The main concern about the free precession model advocated by \cite{Truemper_ea86} was  
the secular
alignment of the spin axis of the accreting NS with the orbital angular momentum.   
Later, Scott et al. (2000) analyzed the \textit{Ginga} observations
of Her X-1, and thereby provided a very detailed look at the
evolution of the X-ray pulse profiles across the entire 35-day cycle.
They discarded the NS free precession model 
(again, in the simplest version  with symmetric magnetic field and under 
model assumptions on the X-ray pattern) and  proposed an alternative model, in 
which the pulse profile is shaped by the combination of X-ray pattern
effects (pencil beam from the magnetic poles mostly seen in hard X-ray band
plus softer fan beam formed at a distance of
1.5 NS radii above the NS surface), light bending effects in the NS strong
gravitational field distorting the fan beam and forming the soft subpulses
flanking the main pulse, and eclipses by the inner edge of the
retrogradely precessing accretion
disk located at $30-40$ NS radii. The model was shown to be capable of qualitatively 
reproducing the observed changes in X-ray pulse profiles with the 35-day phase.
The need for a central off-set of the magnetic dipole to explain the 
observed changes in the pulse shape asymmetries was also noted.   

Clearly, all the models proposed to explain the observed change of X-ray  pulse
profiles in Her X-1 are based on a number of assumptions, whose physical
justification is not always obvious. Therefore, it seemed to us that a return to the free precession model of the neutron star
as the underlying clock mechanism for the 35-day cycle and the cause of the
visible change of X-ray pulse profiles could be useful to understand
the nature of this complex source.
  
 The wealth of excellent observational data obtained by dedicated RXTE PCA
observations of Her X-1 at different phases of the 35-day period make it
possible to check a distinctive  feature of the NS free precession
model. If the regular pulse shape changes are due to stable 
NS free precession and turn-ons
of the X-ray source are caused by the (less stable) precession of the tilted twisted accretion disk
around the NS, then assuming the stable NS free precession period one would
expect to observe the same pulse shapes at different phases relative to the
turn-on times. In other terms, on the O-C diagram for the turn-ons,  
the pulses of the same shape would not follow the turn-on trend and,
instead, would stay along a strict line. The slope of this line would
correspond to the 'true' period of the NS free precession. This program was
pursued in the recent study by \citet{Staubert_eaDublin, Staubert_ea13}, and gave a perplexing result: contrary
to expectations, phases of the 'template' profiles constructed from the
RXTE database during main-on states of Her X-1 closely trailed the O-C trend
for turn-ons, suggesting only one 35-day clock in the system. This finding
implies that either the precessing accretion disk is solely responsible for
both the 35-day variability and the corresponding X-ray pulse changes (e.g.
like in \cite{Petterson_ea91} and \cite{Scott_ea00} models), or the NS free precession is still there,
but the precession period in not stable, but varies by $2-3\%$, and 
there exists a strong synchronization between the NS free precession
period and the precessional motion of the accretion disk. The Occam's razor
principle would incline us in favor of the first possibility, but then we
have to understand why the 35-day disk precession persists over many years
at around the same value? What is the reason for the accretion disk tilt? 

The possible answer could be the corresponding behavior of the optical component HZ Her
itself (e.g. \cite{Petterson75}). 
However, we think that the NS free precession as the clock mechanism of the observed
35-day periodicity remains an attractive alternative. If present, it allows
synchronization of the precessional motion of the disk via dynamical
effect of gas streams acting on the disk (see \cite{Shakura_ea99}) and
keeps the phase of the disk's precessional motion after anomaloue low states (ALS) 
of the source (those states where the X-ray emission from the NS is blocked by the accretion disk all the 
times \citealt{Coburn_ea00, Vrtilek_ea01}). The aim of this
paper is to show that accretion onto a freely precessing prolate NS with complex non-dipolar
magnetic field can successfully reproduce the observed X-ray pulse profiles
and their changes with the 35-day NS free precession period.

\subsection{The present model}

The present model has several features. 

\textbf{A) NS free precession}.

It is well known that free precession of a non-spherical solid body changes the angle between a given point 
on the body's surface and the total angular momentum vector (if none of the inertia axes coincides with the angular momentum vector), therefore the location of the magnetic poles, 
where most of the X-ray emission is presumably generated, will change 
with respect to the line of sight to produce periodic modulation of the observable 
X-ray flux. 

The NS 
ellipticity $\epsilon\sim 10^{-6}$, which is required to explain the $\sim 35$-day free precession period in Her X-1, can be a natural consequence of the presence of a strong internal toroidal magnetic field of $\sim 10^{14}$~G in the NS \citep{Braithwaite09}. Such a field leads to a prolate form of the NS body and causes its prograde free precession. In an isolated NS (e.g. a single pulsar), electromagnetic torques would secularly damp the free precession \citep{Goldreich70}. But this can not be the case for an accreting NS subjected to torques with different signs
from the disk-magnetospheric interaction.

\textbf{B) Disk-magnetosphere interaction}.

The disk-magnetosphere interaction is an extremely complicated process which is not fully understood as yet and can lead to different quasi-periodic phenomena (see, for example, recent 3D MHD simulations by \cite{Romanova_ea12}).
When a strongly magnetized NS is surrounded by a thin diamagnetic disk, the NS magnetic field will exert
a torque on the disk (see \cite{Lai99} and references therein).
If such a NS freely precesses, the torque changes with the phase of precession. A torque of 
the same magnitude but with the opposite sign is exerted on the NS from the disk 
via the magnetic field. The 
value and sign of this torque averaged over the NS spin period (1.24 s in Her X-1) will depend on the angle between 
the magnetic pole and the NS angular momentum vector $\theta$. For a dipole magnetic field, there is a critical 
value of this angle $\theta_{cr}\approx 54^\circ 16'$ 
at which the spin-averaged torque on the NS vanishes
\citep{Lipunov&Shakura80}, and at which the inner parts of the disk are
effectively not subjected to any forces from the NS magnetic field.
If $\theta>\theta_{cr}$, the inner twisted parts of the disk take a spiral-like shape with a certain 
direction of twisting. The inner edge of the disk then tends to co-align with the NS rotational equator. If 
$\theta<\theta_{cr}$, the inner disk is also twisted and tends to co-align with the NS rotational equator, but 
with the opposite direction of twisting. 

Generally, the NS spin axis can be misaligned with the orbital angular momentum. 
The initial misalignment of the NS angular 
momentum and the orbital angular momentum can be due to, for example, the natal kick received by the NS 
during its formation in a core-collapse supernova. The accreting matter brings the orbital momentum to the 
NS, so the spin axis of a non-precessing NS (i.e., with some fixed value of the angle $\theta$) should tend 
to eventually co-align with the orbital angular momentum of the binary (this argument, we remind, was used by \cite{Petterson_ea91}
as evidence against the NS free precession in Her X-1). However, if the NS precesses, the 
disk-magnetosphere interaction could substantially slow down (or stop altogether) the secular NS--orbit alignment. 
Indeed, the interaction of the NS with the disk tends to co-align its spin axis with the orbital momentum 
if $\theta<\theta_{cr}$, and counter-align for $\theta>\theta_{cr}$. The net NS angular momentum misalignment 
will be determined by the sum of differentials $\Delta \theta$ due to the magnetospheric torque and the 
angular momentum of matter supplied to NS. During one free precession period
$\Delta \theta$ changes sign, and the net angular momentum misalignment 
can be strongly reduced in the case of accretion onto a precessing NS in comparison with 
a non-precessing NS. 
Therefore, it is very likely that due to the disk-magnetosphere interaction the spin axis of a freely precessing 
NS in a binary system can remain tilted to the orbital angular momentum. This
misalignment is important to produce a tilt of the outer parts of the accretion disk 
via gas streams which are non-coplanar with the orbital plane due to asymmetric 
illumination of the optical star atmosphere (\cite{Shakura_ea99}, but see also
\cite{Katz73}, \cite{Gerend&Boynton76}
for other discussions of the causes of the outer disk
tilt). 

\textbf{C) Complex structure of the NS surface magnetic field}. 

In the simplest case where most of the accreting energy is released at the magnetic poles (e.g. for a pure dipole magnetic field), the 
X-ray pulse profile would be formed by one or two components depending on the visibility conditions (as in radio pulsars). However, in Her X-1 the pulse profile has a very complex 
multi-component shape (e.g. \cite{Staubert_ea13} and below), which changes with the 35-d phase first 
smoothly and slowly and then more rapidly and strongly. 

%Rapid changes observed at the late phases
%of the main-on state may be due to obscuring of the NS 
%surface by the inner edge of precessing accretion disk.  

Observations show \citep{Deeter_ea98, Kuster_ea05} that the form of the pulse profiles at one precession 
phase is strongly dependent on energy.  In the main-on state, the hard central peak flanked by two softer 
details is clearly seen. The sub-pulse is much less pronounced and is better seen at lower photon energies. 
The ratio of the height of the sub-pulse to that of the
central peak also changes with the 35-d phase. 
To explain the observed smooth evolution of profiles at early phases of the 35-d cycle (during the main-on state) we introduce 
arc-like (or horseshoe-like) structures around the magnetic poles which may be due to non-dipole magnetic 
field structures near the NS surface \citep{Shakura_ea91}. This complex field structure near the surface 
can be formed either initially \citep{Ruderman01, Braithwaite09} or emerge when accreting matter 
is solidified with the NS crust. 
Indeed, during accretion the crust thickness near the magnetic poles increases, and the ductile spreading 
(readjustment) of the crust from poles toward the equator can begin. Assuming the strong freezing of the 
magnetic field into the conducting crust, 
the deformation of the initial dipole field into quadrupole and more 
complex configurations will naturally occur\footnote{A non-dipolar character of 
surface magnetic field of a freely precessing NS has recently been used to explain 
the anomalous braking index observed in radio pulsars \citep{BarsukovTsygan10}.}. As a result, at a distance of several NS radii above the NS surface, 
regions with close to zero field strength are formed in the magnetosphere of the accreting NS, where the 
accreting flux can split into different channels. Therefore, it is natural to expect that the X-ray emission can be 
generated both in the usual magnetic poles (which can be shifted due to the dipole off-set in the NS body) 
and in arclets, horseshoes, or ring-like structures surrounding the magnetic poles. This natural process during accretion onto 
a magnetized NS can split the initial large-scale magnetic field into smaller substructures, which could dissipate 
more rapidly due to magnetic field reconnection. 

\section{Parameters of the present model}

To explain the observed evolution of X-ray pulse profiles with 35-day phase, the following model is proposed. 
The model includes the geometrical parameters of the NS (shape, viewing angle) and properties of the 
emitting regions (their location on the NS surface and emission diagrams). 

\begin{figure*}
\centering
\includegraphics[width=0.7\textwidth]{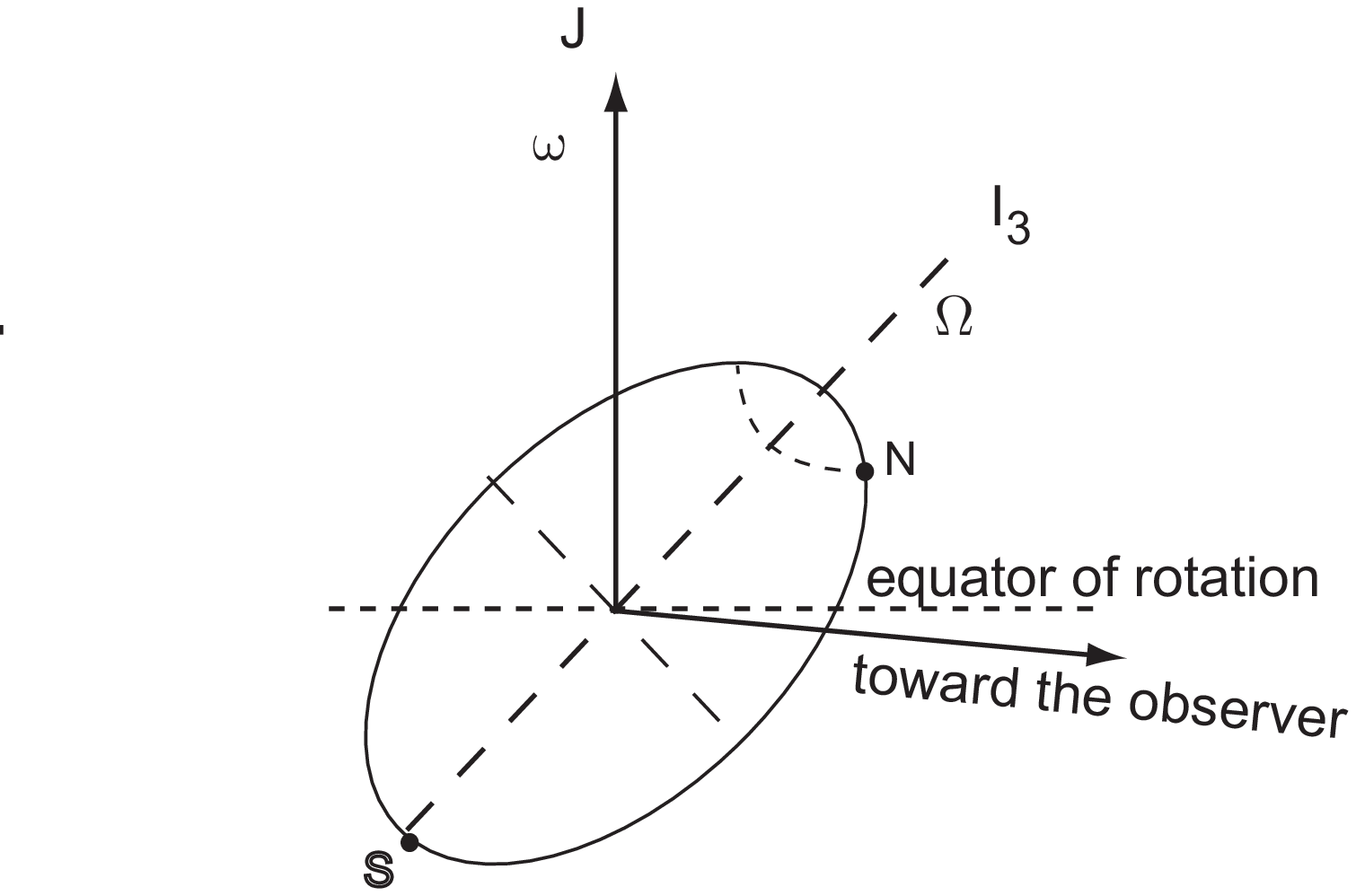}
\caption{Schematic view of the slowly precessing axially symmetric 
neutron star in Her X-1. 
The total motion of the NS can be presented as a sum of (rapid) rotation 
with angular frequency $\omega$ around the total angular momentum, which in
the case of slow free precession almost coincides with 
the NS angular momentum $J$, and slow rotation with precessional
angular frequency $\Omega\ll \omega$ around the principal inertia axis
$I_3<I_2=I_1$ of the NS, which coincides with its symmetry axis.}
\label{f:NSbody}
\end{figure*}

\textbf{A) NS body}. 
The NS represented by a solid body with two but not three equal
principal moments of inertia, $I_1 = I_2$ (Fig. \ref{f:NSbody}). The axis of the
larger principal moment of inertia (axis $I_3$ in the figure) is inclined relative to the 
angular momentum vector $\omega$ by an angle $\theta=50^o$. If there are no external torques, such 
a body should freely precess around its principal axis $I_3$ in the sense coincident with the spin  
(prograde precession) or opposite to it (retrograde precession), depending on whether the precession 
occurs around the minimum ($I_3<I_2=I_1$, prolate body) or maximum ($I_3>I_2=I_1$, oblate body) moment of inertia, respectively. 
To describe features on the NS surface, we choose a reference frame fixed inside the
 NS body with the 
polar axis directed along the inertia axis $I_3$. Below, all coordinates on the NS surface will be given 
in terms of the latitude $b$ (with zero at the inertia equator) and longitude $l$ counted counter-clockwise 
in the northern hemisphere, with the zero meridian passing through the axis of inertia and the north 
magnetic pole.

\textbf{B) Viewing angle of the NS spin axis.} 
The Her X-1 binary system is observed virtually edge-on, 
at a binary inclination angle of $i\simeq 81-88^o$ \citet{Cheng_ea95, Reynolds_ea97}. 
The observed evolution of pulses is best reproduced for the angle between
the direction towrd the observer and the rotational equator of the neutron star $-3^o$ (i.e. 
when this direction makes an angle of $93^o$ with the NS spin axis, see Fig. 1). 
This means that during one rotation of the NS 
the direction toward the observer draws on the NS surface 
a small circle lying $-3^o$ below the NS rotational equator. This circle is shown 
in Fig. 2 by the solid (red) line.
For the nearly edge-on binary system, 
this geometry is natural if the NS spin axis is coaligned with the 
orbital angular momentum. However, as discussed above, the NS  initial 
spin-orbit misalignment can be maintained for a long time due to disk-magnetosphere interaction around
the freely precessing NS; in that case the NS spin axis lying almost 
in the plane normal to the line of sight is by chance.

\textbf{C) Emission regions}. As discussed above, complex shape of the X-ray profiles in Her X-1 requires 
several emission components. We associate them with different emission regions on the NS surface. 
The two principal emission spots include the magnetic poles where accreting matter is channeled by the 
NS magnetosphere. However, if there is a more complex structure of the NS surface magnetic field, there 
could be additional emission regions. From the analysis of X-ray profiles in Her X-1 it was suggested earlier 
\citep{Shakura_ea91, Panchenko&Postnov94} that the NS surface magnetic field can have a quadrupole 
structure. In the simplest case of an additional quadrupole component aligned with the magnetic dipole, matter can accrete onto a ring around the magnetic pole. Such a structure of 
emitting regions around magnetic poles was confirmed by 3D numerical MHD-simulations by \citet{Long_ea08}. 
It was found in the latter paper that for an inclined dipole plus quadrupole the matter accretes onto ring-like 
structures (arcs) around the magnetic pole. The form of these arcs can be more complex, 
especially if the center of symmetry of the magnetic dipole
is offset from the center of the NS. 

\textbf{C) Magnetic poles}. The north (N) magnetic pole has coordinates $N(b=60^o,l=180^o)$, that is the angle 
between the N-pole and the inertia axis is $30^o$. The south (S) pole has coordinates $S(b=-85^o, l=0^o)$, 
i.e. lies very close to the inertia axis; it is observed always by about the same angles during the NS precession period. 

\textbf{D) Arcs}. 
It is convenient to describe the arcs by the coordinates of their centers, the radius (in degrees), 
and the coordinates of the arc beginning and end in degrees counted along the circle with the specified radius 
from the zero meridian. Several emission arcs were found to be needed to describe the X-ray pulse profiles 
in Her X-1. Their positions are specified in Table \ref{t:emregions}. 

\textbf{E) Transient features}. In addition to permanently emitting regions on the NS surface (poles, arcs),  
transient emitting features appear in several cycles. They represent some spots and arcs which 
ordinarily do not contribute to the total emission, but sometimes appear quite strong and are responsible 
for the emergence of distinct features in the pulse profile. The
appearance of these features sometimes repeated in different cycles, 
so we believe them to be real transient emitting regions on the NS surface. Their location is also specified in 
Table \ref{t:emregions} (P1 --  transient 'pole', A1, A2, A3 -- transient 'arcs'). 

\begin{table*}
\caption{Emission regions on the NS surface
\label{t:emregions}}
\begin{tabular}{lcccccc}
\hline
\hline
\multicolumn{7}{c}{Poles}\\
\hline
   & b  & l &&&& $\Delta\vartheta$ (deg)\\
\hline
N  & 60  & 180 &&&& 16\\
S  & -85 & 0   &&&& 16\\
P1 & -17 & 80  &&&& 14\\
%P2 & 8   & 280 &&&& 8 \\
\hline
\hline
\multicolumn{7}{c}{Arcs}\\
\hline
No &\multicolumn{2}{c}{center} & radius & beg. & end & $\Delta\vartheta$ \\
   & b              &  l       & (deg)  & (deg)& (deg)& (deg)\\ 
\hline
N1 cyan        & 70 & 160 & 50 & 200-210 & 250 & 14\\
N2 dark green  & 75 & 180 & 43 & 250 & 300 & 13\\
N3 light green & 65 & 180 & 43 & 302 & 350 & 12\\
N4 violet      & 65 & 170 & 53 & 355 & 400 & 11\\
N5 grey        & 60 & 160 & 55 & 45  & 60  & 11\\
N6  yellow     & 60 & 150 & 50 & 65  & 95  & 11\\
N7  dark rose  & 58 & 160 & 55 & 95  & 120 & 11\\
N8  red        & 70 & 140 & 60 & 120 & 160 & 13\\
\hline
S1 dark rose   & -97 & 180 &  44  & 330 & 350 & 11\\
S2 light rose  & -97 & 180 &  44  & 20  & 50  & 11\\
S3 light green & -97 & 180 &  44  & 50  & 100 & 11\\
S4 light blue  & -95 & 200 &  42  & 110 & 170 & 8\\
\hline
A1 green       & -100& 200 &  65  & 205 & 260 & 11\\
A2 blue        &  60 & 180 &  80  & 245-255 & 275-285 & 11 \\
A3 dark green  &  75 & 180 & 105  & 90 & 120 & 11 \\
\hline 
\end{tabular}
\end{table*}

\begin{figure*}
\begin{minipage}[b]{\textwidth}
%\centering
\includegraphics[width=0.53\textwidth]{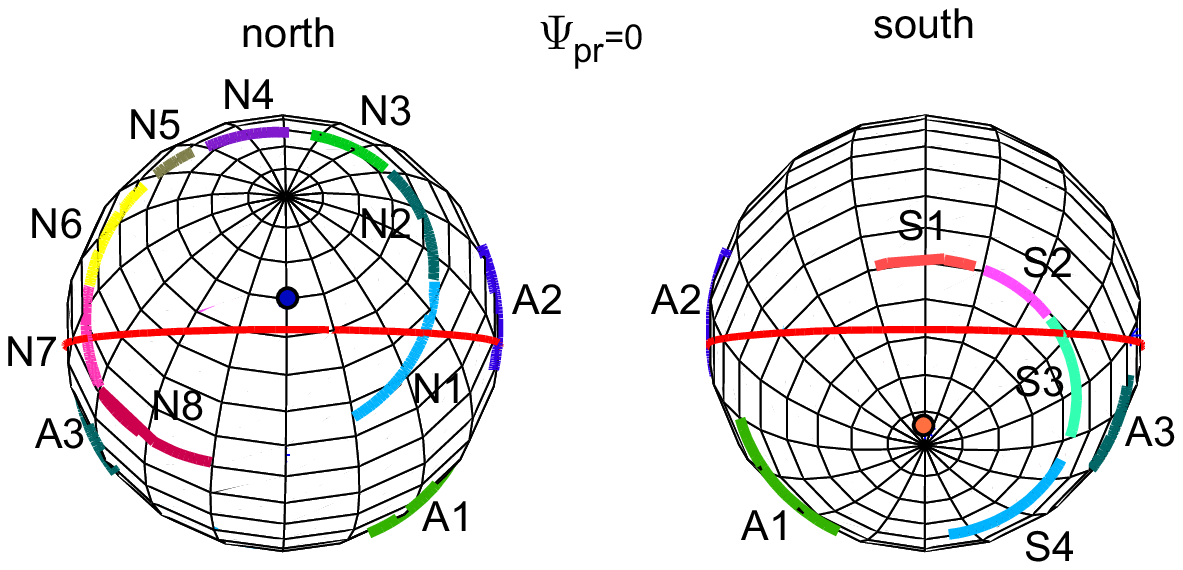}
\includegraphics[width=0.47\textwidth]{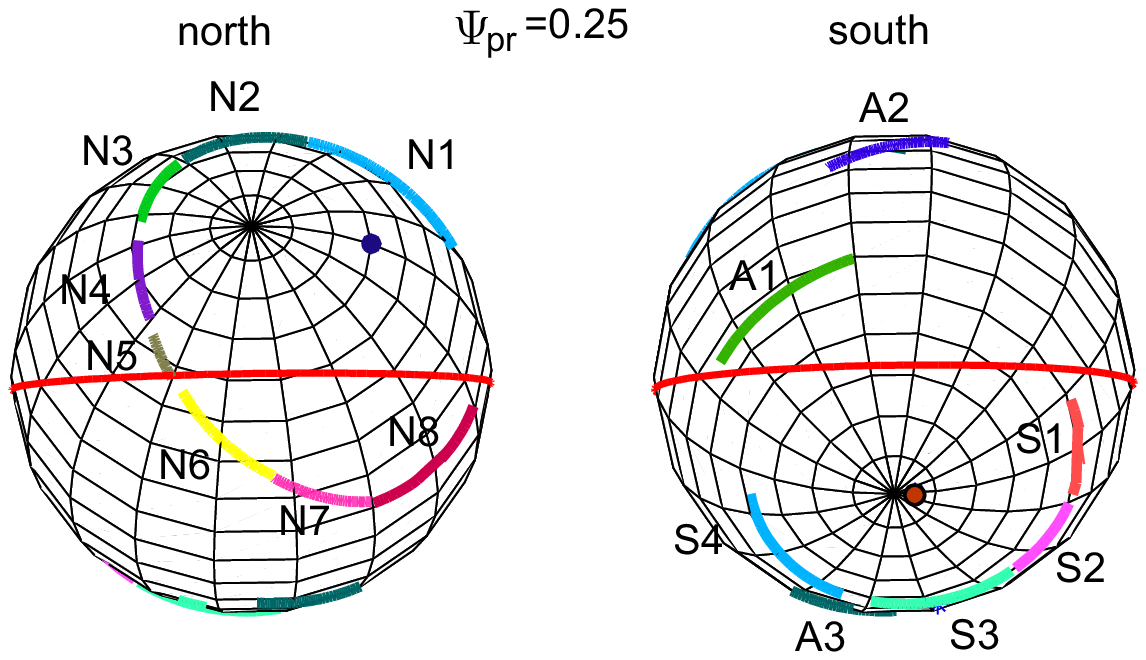}
\end{minipage}
%\hspace{0.5cm}
\begin{minipage}[b]{\textwidth}
%\centering
\includegraphics[width=0.47\textwidth]{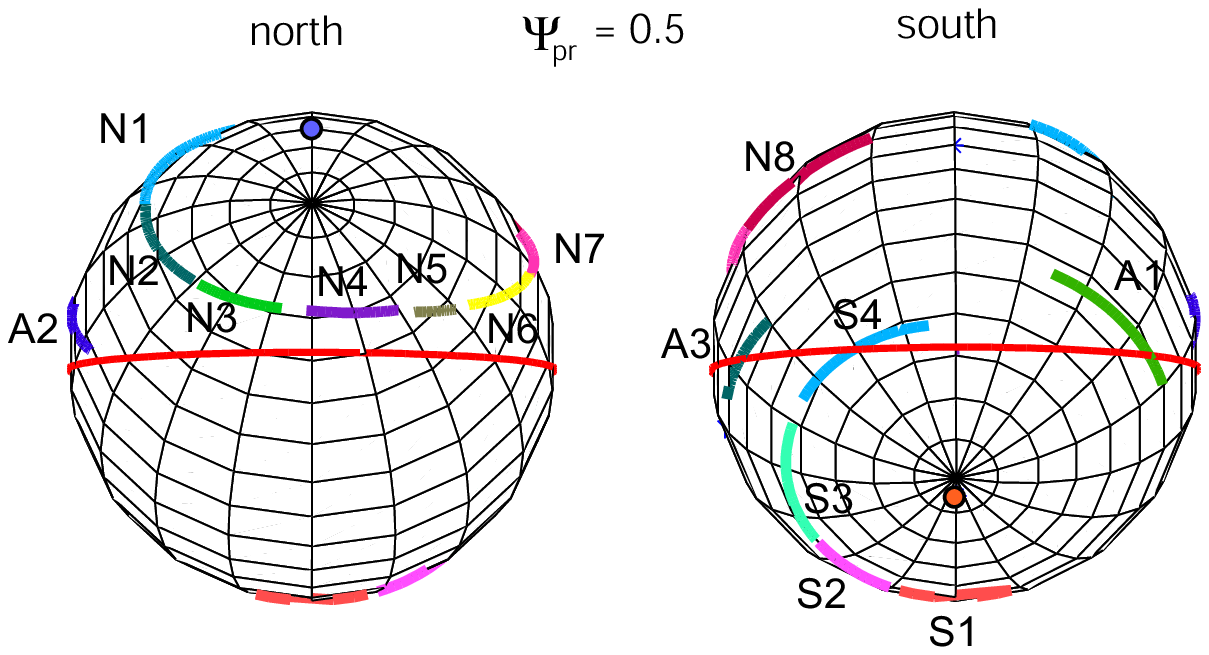}
\includegraphics[width=0.47\textwidth]{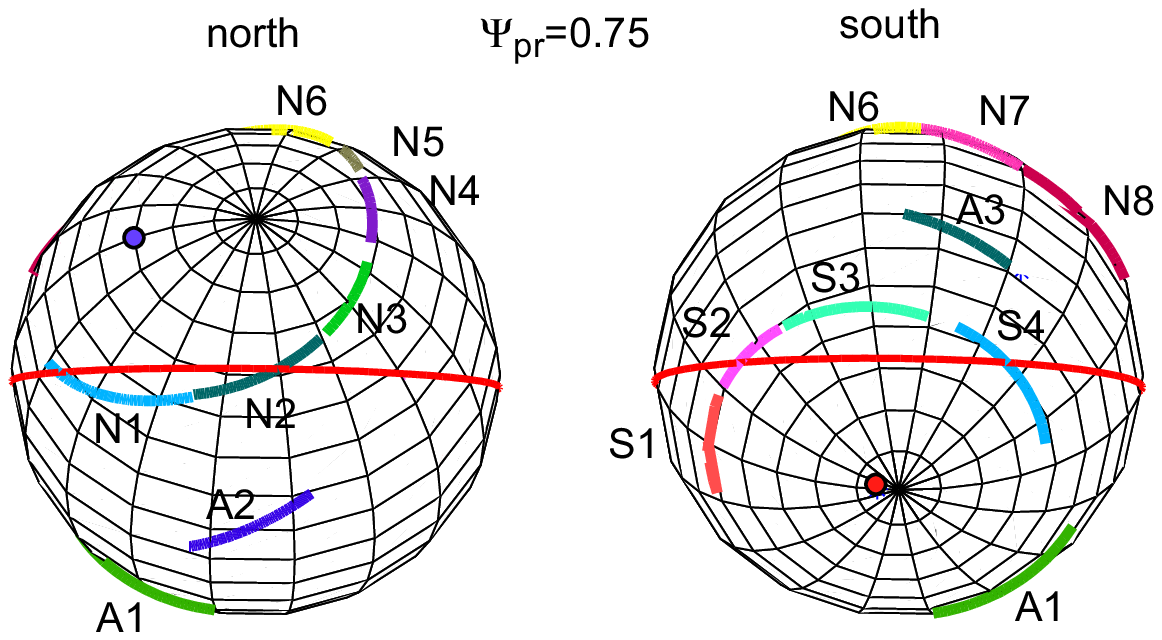}
\caption{Projection of the NS surface on the sky plane at different precession phases $\Psi_{35}=0,0.25,05,075$.
Left: north hemisphere (N), right: south hemisphere (S). The solid (red in the on-line version) line
shows small circle lying $3^o$ below the NS rotational equator, 
along which the line of sight crosses the NS surface in one rotation. 
The NS spin axis makes an angle of $93^o$ with the line  of sight and is directed upward (north). 
The NS sense of rotation 
seen from the top is counter-clockwise. The NS free precessional motion is prograde, i.e. counter-clockwise as seen 
along the principal axis of inertia from the north hemisphere (the north pole of the coordinate grid in the plots).
Emitting regions are shown in color (in the electronic version) and
marked by their notations as in Table \ref{t:emregions}.}
\label{f:precview}
\end{minipage}
\end{figure*}

The X-ray emission is assumed to form a Gaussian-like beam with
the characteristic  width
$\Delta\vartheta$: $I(\vartheta(\Psi,\phi))=I_0 e^{-(\vartheta/\Delta\vartheta)^2}$, 
where $\vartheta(\Psi,\phi)$ is the angle between the local normal and the diection toward
the observer
at the precession 
phase $\Psi=mod[(2\pi/P_{pr})t,2\pi]$, and the pulse phase $\varphi=mod[(2\pi/P)t,2\pi]$;
the angle $\Delta\vartheta$ characterizes the width of the X-ray beam.
(One should distinguish 
the NS precession phase $\Psi$ and the phase $\Psi_{35}$ of the 35-day period; the latter is zero by
definition at the moment of the turn-on, while the former can be non-zero 
at the turn-on (see section \ref{psi0} below).

When calculating the model pulse 
profiles, the emission poles were treated as point-like, while the emission arcs are represented as the sum of 
one-degree-long segments. At a given precessional phase $\Psi$, 
the total emission flux received by a distant observer as a function of
the pulse phase $\varphi$ 
was calculated by summing over all visible regions (marked by variable $i$) at given 
phases $\Psi$, $\varphi$ plus a sine-like component:
$$
F(\Psi,\varphi)=e^{-\tau_{eff}(\Psi)}\sum\limits_{visible\; i} I_{0,i}e^{-\left(\frac{\vartheta_i(\Psi,\varphi)}{\Delta\vartheta_i}\right)^2}\cos\vartheta_i(\Psi,\varphi)+B(\Psi)\sin(\varphi+\varphi_0)\,.
$$ 
Here $\vartheta_i$ is the angle between the local normal of 
the $i$-th emitting area (or its elementary 1-degree part, if the area is extended)  
on the NS surface and the direction toward the observer,  $\tau_{eff}(\Psi)$ is the effective optical depth which takes into account the 
fading of the X-ray emission observed through the accretion disk (we remind that 
the binary inclination angle in Her X-1 is close to 90$^\circ$, so the line of sight
slides closely above the accretion disk surface). 
The sine-like component is produced by the reflection of the X-ray emission generated near the NS surface from the warped twisted disk. The sine-like component is most clearly visible 
when the NS is screened by the accretion disk during 'off' states between the 'main-on' and
the 'short-on' states and 
dominates during the 'short-on' state of the source, but 
its presence is also required to describe pulse profiles in the 'main-on' state (see model pulse profiles below).        
The parameters ($\tau_{eff}(\Psi)$, $B(\Psi)$ and $\varphi_0$) have clear physical 
meaning and should be taken into account 
in addition to purely geometrical change in the aspect ratio of the
emitting regions on the neutron star surface and are necessary to describe the 
observed pulse profile changes with the 35-day phase. 
We also stress that within a given 35-day cycle we fixed relative intensities of all emitting regions and poles on the NS surface. However, some of them were 
found to vary from cycle to cycle (see Table \ref{t:relint}).

No light ray bending in the gravitational field of NS was taken into account, since the width of emission beams $\Delta\vartheta$ 
is typically 8-14 degrees and most of the radiation escapes almost normally to the NS surface.

The projection of the NS surface onto the sky plane (perpendicular to the line of sight) at different precession 
phases $\Psi$ at the spin phase corresponding to the minimum angle between the line of sight and the precession 
axis $I_3$ is shown in Fig. \ref{f:precview}. The coordinate grid on the NS surface is connected to the axis of 
precession. The solid (red) line shows the small circle lying 
$3^o$ below the NS rotational equator, which is
the trace of the line of sight that intersects the center of
the NS on the NS surface
over one spin period. The NS 
rotates from left to right and precesses counterclockwise (i.e., prograde) around the axis of inertia $I_3$ 
(the pole of the grid). We emphasize that the prograde direction of NS free precession is in line with the expectation of the prolate NS body form as being due to strong internal toroidal magnetic field \citep{Braithwaite09}.

\section{Model X-ray pulse profiles}

\begin{figure*}
%\begin{minipage}[b]{\textwidth}
%\centering
\includegraphics[width=0.45\textwidth]{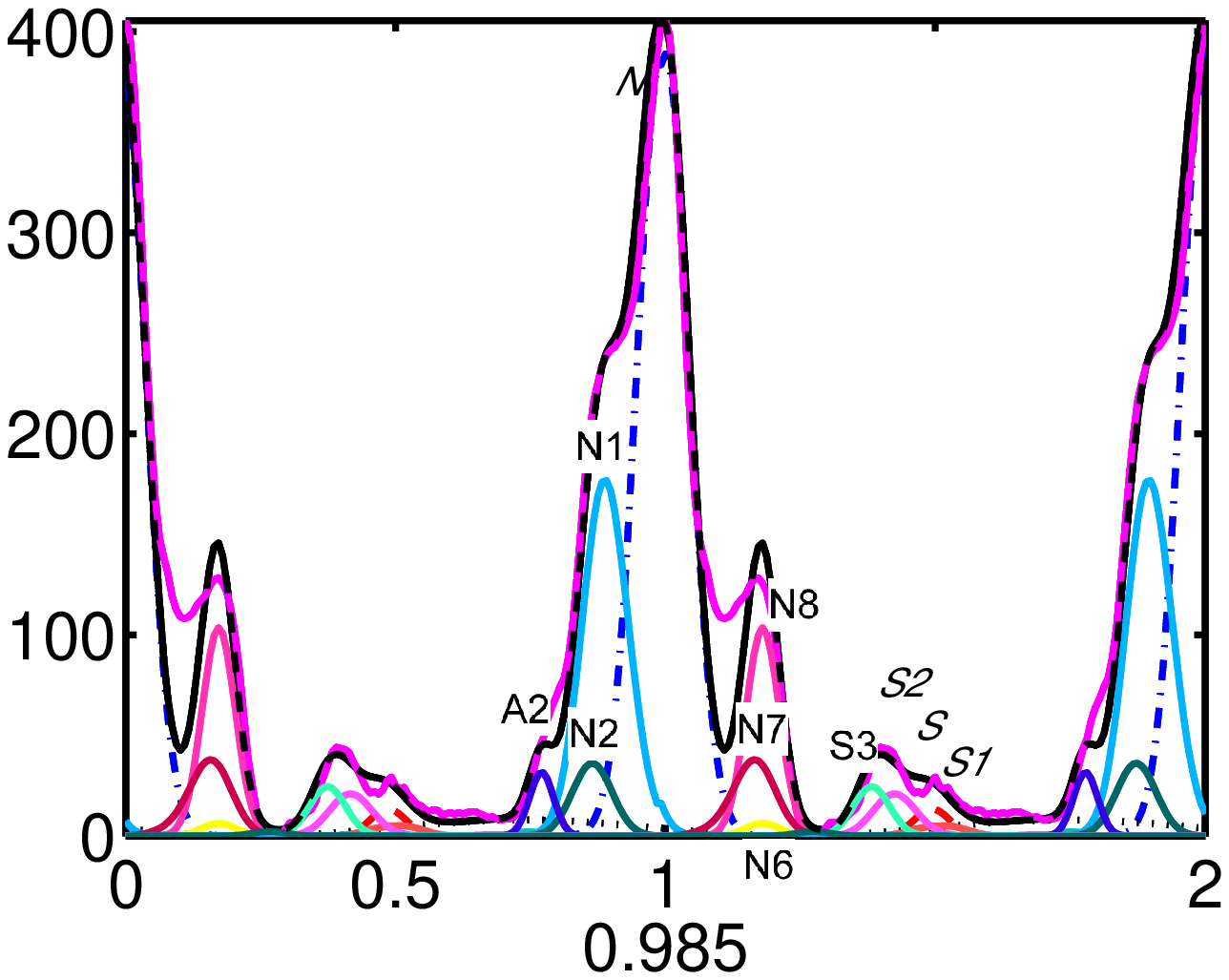}
\hfill
\includegraphics[width=0.45\textwidth]{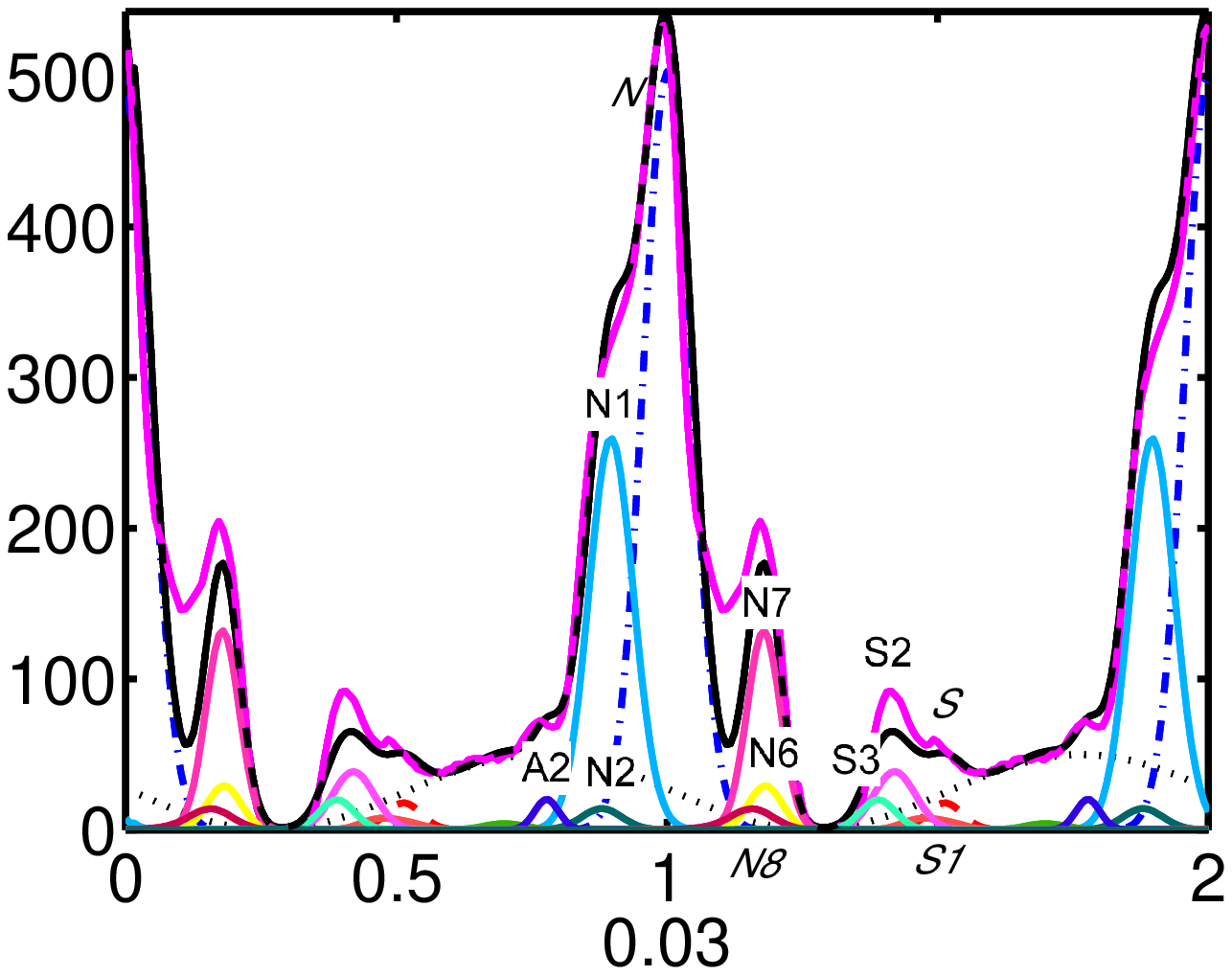}
%\end{minipage}
%\hspace{0.5cm}
%\begin{minipage}[b]{\textwidth}
%\centering
\includegraphics[width=0.45\textwidth]{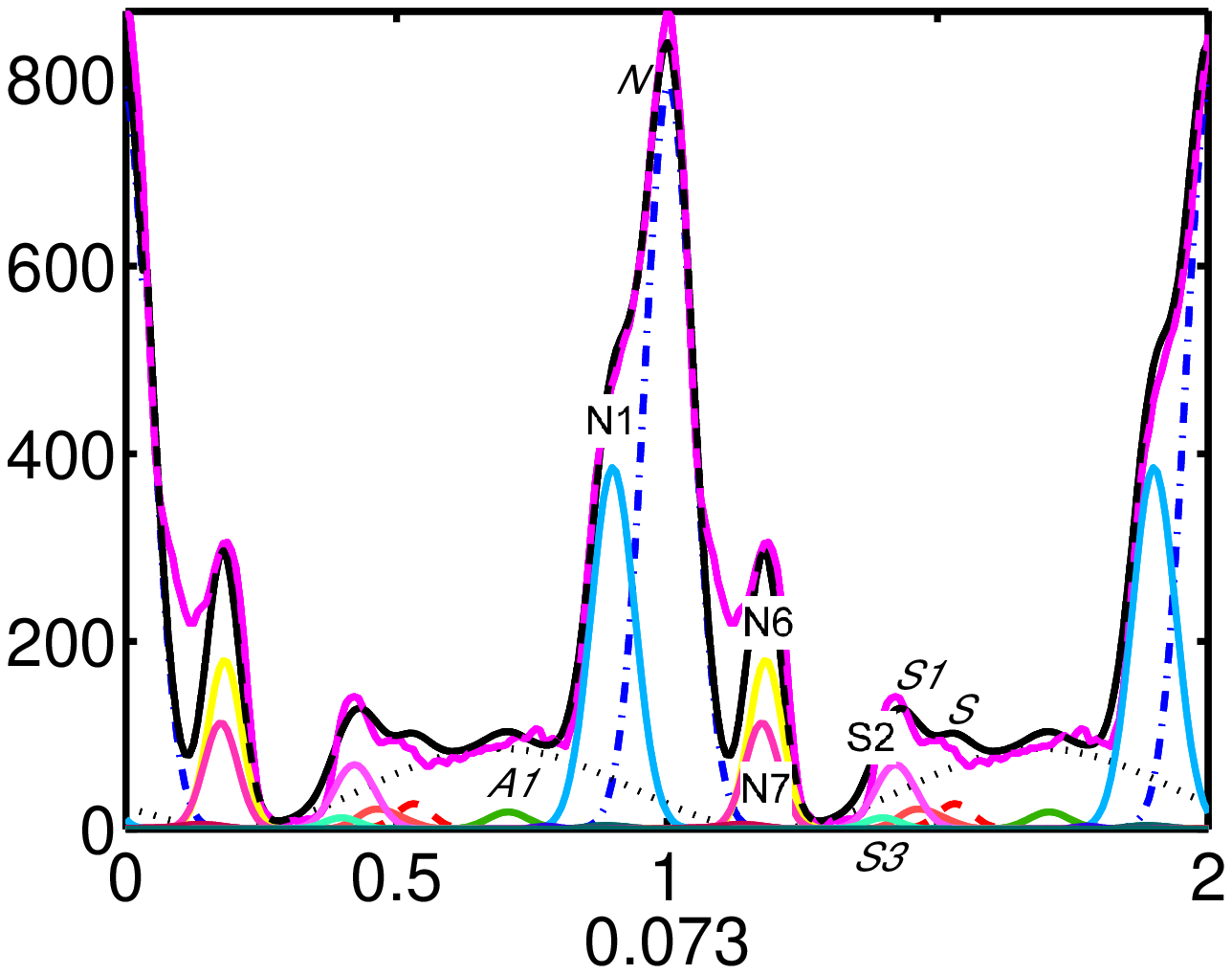}
\hfill
\includegraphics[width=0.45\textwidth]{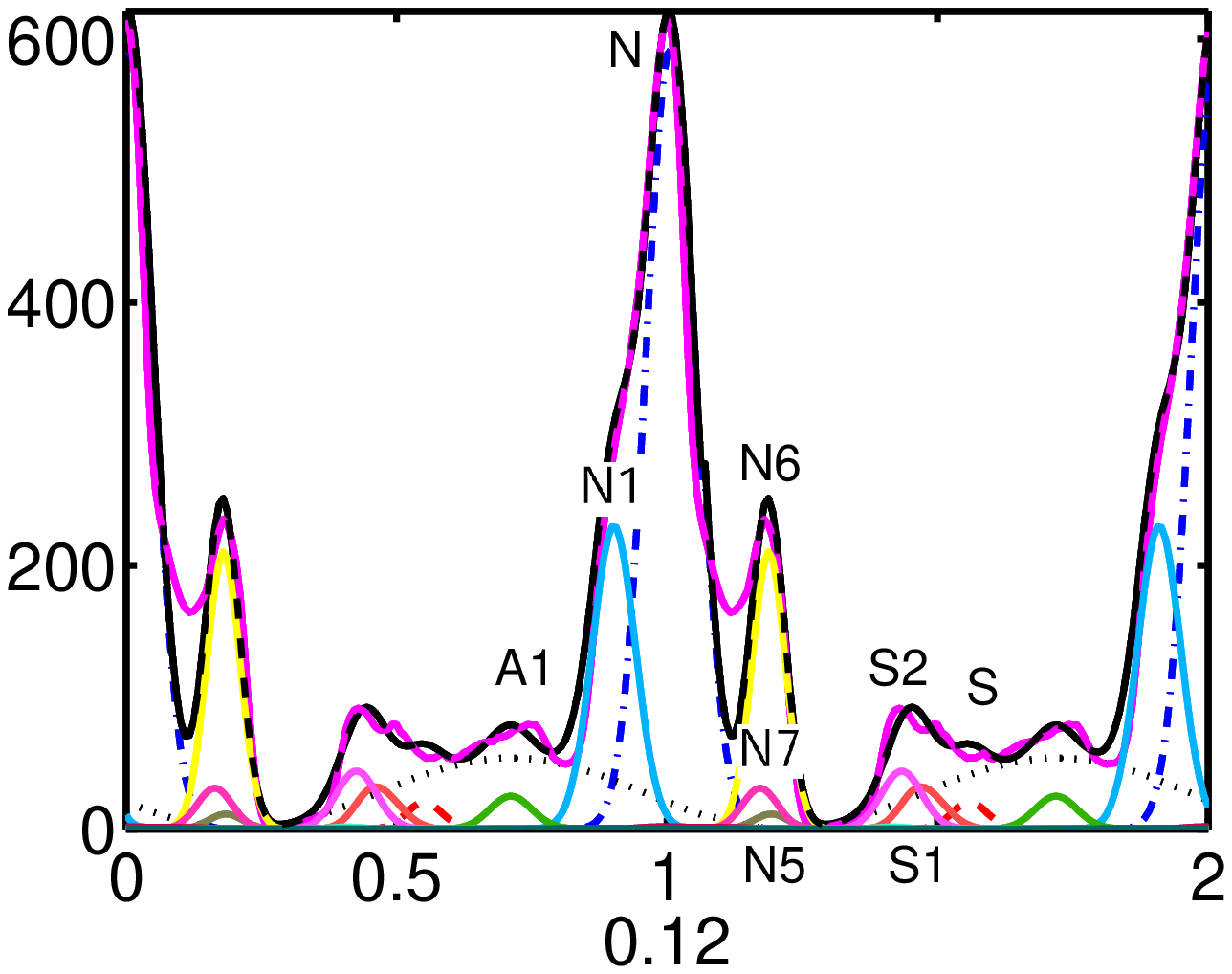}
%\begin{minipage}[b]{\textwidth}
%\centering
\includegraphics[width=0.45\textwidth]{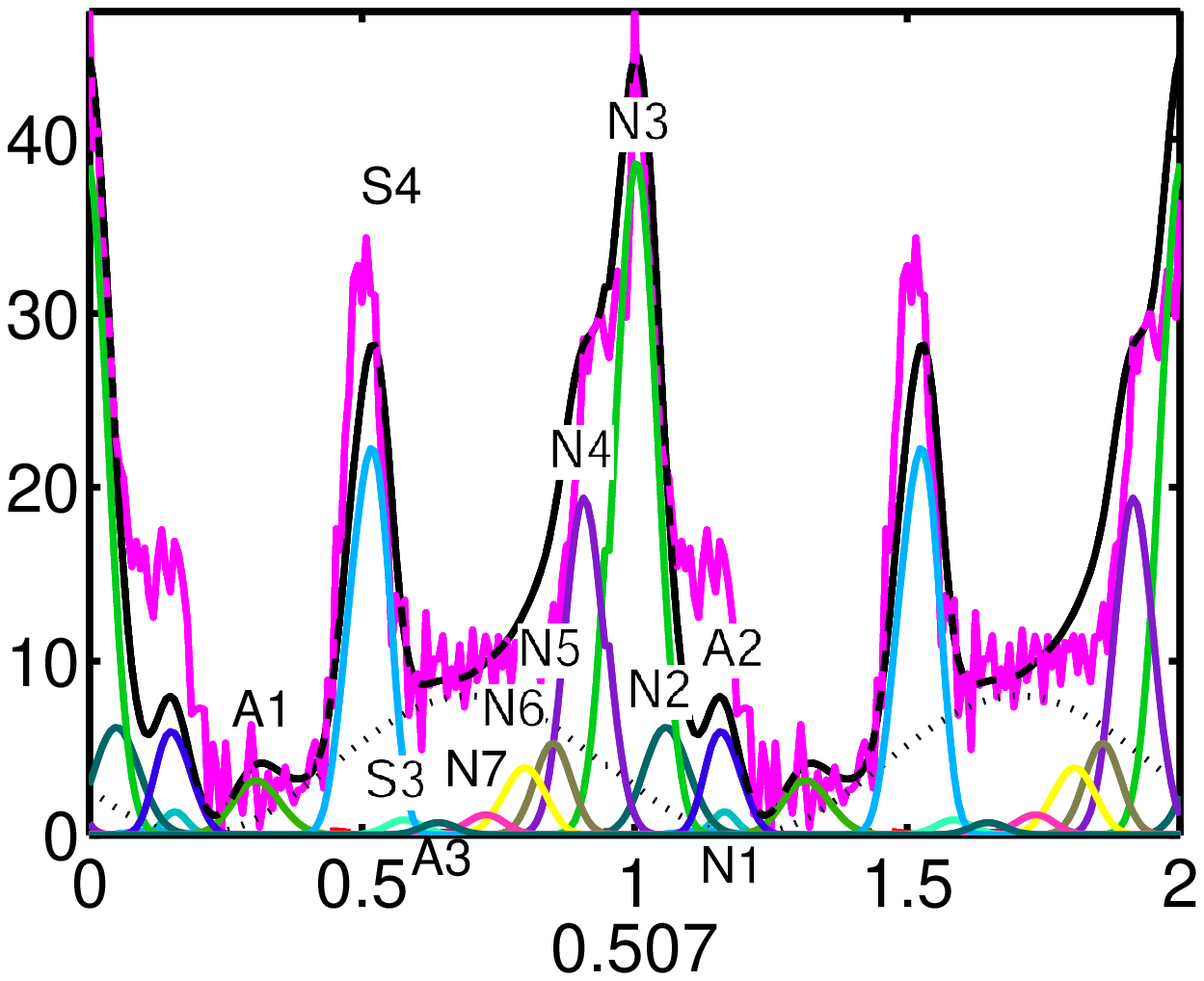}
\hfill
\includegraphics[width=0.45\textwidth]{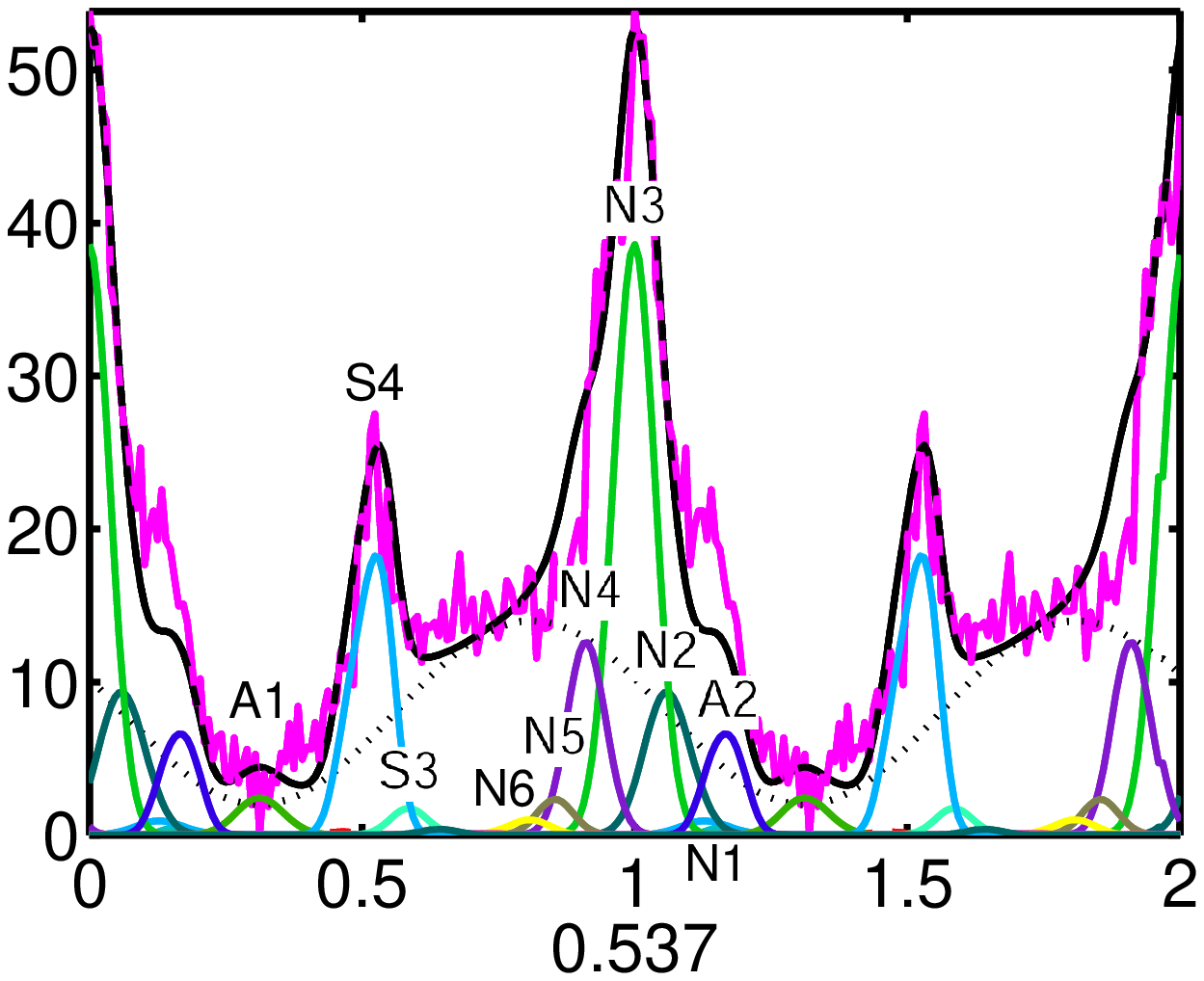}
%\end{minipage}
%\hspace{0.5cm}
\caption{Model RXTE profiles (9-13 keV) for cycle \#308 with identified emission features
(notations as in Table 1). A constant baseline component has been subtracted from the observed profiles. 
The model total flux is shown in solid black line. The observed profiles with a 
constant baseline component subtracted are shown in gray solid line (magenta in the on-line version).
Model NS free precession phase is indicated.}
\label{f:308nn}
%\end{minipage}
\end{figure*}

Individual RXTE observations used in our analysis 
are listed in the Appendix A. The RXTE PCA light curves were extracted for five energy bands: 2.0 -- 4.5 keV, 4.5 -- 6.5 keV, 6.5 -- 9 keV, 9 -- 13keV, and 13  -- 19 keV.
For our analysis, 
the requirement on the energy band was not to be close to CRSF 
energy $E_{cyc}\approx 38$ keV, but still to be 
hard enough to avoid soft X-ray absorption.  
In the 14-19 keV band the pulse proﬁles and their evolution with 35-d phase 
are similar to those in the 9-13 keV band,
so the use of 9-13 keV was chosen (somewhat arbitrarily).
The model X-ray pulse profiles for RXTE cycle \#308 (the turn-on
time at MJD 52067.8), with contributions from individual regions
identified, are presented in Fig. \ref{f:308nn} as a function of the 
NS free precession phase $\Psi$.  
The model X-ray pulse profiles for \textit{Ginga} cycle \#181 and all 
analyzed 
RXTE/PCA 35-day cycles specified in Table \ref{t:data}
(see Appendix A)
are presented in Fig. \ref{f181}-\ref{f324}
in Appendix B. In each observation, a constant baseline component
has been subtracted. The observed puls profiles are shown by solid gray lines
(magenta in the on-line version). For each individual 35-day cycle, the emission intensities $I_0$
from each emission region specified in Table
\ref{t:emregions} were adjusted to best fit the observed pulse shape. 
Relative intensities of the emission regions for different cycles are
listed in Table \ref{t:relint}.
Within the given 35-day cycle, only the effective optical depth 
$\tau_{eff}(\Psi)$, the amplitude $B(\Psi)$ and phase $\varphi_0$ of the reflected sine-line component were varied. As an example, the values of these parameters for the profiles observed in 
RXTE cycle \#308 shown in  Fig. \ref{f:308nn} are listed in Table \ref{t:308}.

It follows from Table \ref{t:relint} that there is a strong asymmetry 
in accretion rate onto the North and South hemispheres of the NS. Averaged over the NS rotation 
and free precession, this
asymmetry in the X-ray emission should lead to the appearance of a secular 
radiative drag acting on the binary system (see e.g. \cite{Pal'shinTsygan98}).
The acceleration of the binary system due to this force 
naturally explains  
the high galactic latitude of Her X-1/HZ Her, corresponding to $\sim 3$~kpc above 
the Galactic plane.

\begin{table*}
\footnotesize
\caption{Model parameters which do not change with NS precession phase in each 35-day cycle:
Relative intensities of emission regions on the NS surface
normalized to the intensity of the brightest N
pole, in per cents. 
\label{t:relint}}
\begin{tabular}{lccc ccc ccc ccc ccc ccc l}
\hline
\hline
 Cycle No & 181 & 252 & 257 & 258 & 259 & 268 & 269 & 271 & 277 & 303 & 304
& 307 & 308 & 313 & 319 & 320 & 323 & 324 & mean\\
\hline
N         & 100 & 100 & 100 & 100 & 100 & 100 & 100 & 100 & 100 & 100 & 100
& 100 & 100 & 100 & 100 & 100 & 100 & 100 & 100\\
S  	  & 27  & 31  &  27 &  45 &  32 &  25 &  27 &  33 &  30&  27 &   31
&  40 &  43 &  24 &  25 & 27  &  44  & 48 & 32.6\\
P1        & 2.5 &  0  &   0 &   1 &  1.5& 0.2 &  2.7&  0.9 & 2.7& 0.8 &  3.3
& 0.6 &   0&  0  &  0  &  0  &  0.7  & 0  & 0.94\\
\hline
N1 cyan   & 50 & 110   & 45 &   45 &  50 & 50 &  60  & 46 &  47  & 53 &  50
&   50 & 54 &  59 & 50  & 55  & 56   &  48 & 54.3\\
N2 d. green &110& 72   & 45 &   64 &  50 & 70 &  47  & 50 &  47  & 53 &  33
&   50 & 32 &  35 & 50  & 55  & 56   &  36 & 53.1 \\
N3 l. green &50& 50   &  50 &   45 &  50 & 35 &  46  & 50  &  50  & 50 &  50
&   50 & 54 &  76 & 56  & 61  & 50   &  54 & 51.5 \\
N4 violet & 18 & 20   &  20 &   12 &  14 & 12 &  20  & 25 &  20 & 13  &   27
&  10 & 21 & 31  &  25 & 20  & 20  & 20  & 19.3 \\
N5 grey   &  2&   2 &     2 &   2 &    2 &  2 &   5  &  8 &  1.5 & 8 &  7
&    6 & 6 & 5 & 6 & 6 & 5 & 5  & 4.5 \\
N6  yellow&  18 &  15 &  14&   18 & 31 & 10 & 18   & 33 & 17   & 27 & 21
&20 &  22 &  22 & 19  & 21 & 13 & 24 & 20.2\\
N7  d. rose& 27 & 36 & 18 & 18 & 63 & 20 & 23 & 38 & 23 & 33& 29   & 25& 22
& 22 & 22 & 24       & 22 & 21 & 25.8 \\
N8  red   & 27 &  52 & 27 & 27 &  13 & 30 & 33 & 42 & 33 & 33&    8  & 25&
27  & 29 & 31 & 27   & 28 & 30 & 29\\
\hline
S1 d. rose& 3.6 & 5.2 & 4.5 & 3.6 & 3.8 & 4 & 3.3& 8.3 & 0 & 5.3 & 0    & 4
& 4.3& 4.7 & 6.3  &    7  & 4.4 & 5 & 4.1\\
S2 l. rose&  6.3 &  7.2& 4.5 &  8 & 19 & 9   &  6 &    4.1& 6.7& 10.6 & 9
&6.5 & 6.5 & 8.2 & 6.3 &   7  & 5 & 6 & 7.6 \\
S3 l. green& 7.2&  21 & 27 & 6.3 & 7.5& 10 & 15 & 10  & 3.4 & 6.7& 4.2 & 10
& 6.5 & 6   & 9 &    9 & 10  & 14 & 10.2\\
S4 l. blue & 11 &  21 & 18 & 9   & 10  & 27 & 6.7 & 13 & 3.4 & 33 &  4.2 &
25& 19 & 18 & 22 &   22 & 20 & 23 & 17\\
\hline
A1 green   & 1 & 3 & 1   &  9    & 10 & 1 &   0 & 4      &   3.4& 12 & 12 &
1 & 3.2 & 1  & 1     &  1   & 1 &  1 & 3.6 \\
A2 blue    & 9 & 21& 6.4 & 7   &   6.3& 8 & 17 & 8.3  &    13 &  8  & 2.5& 9
& 6.5 & 12 & 14  & 14 & 17 & 12 & 11\\
A3 d. green& 1 & 1 & 1 &    9  &  1.3 & 0 & 6.7 & 1   &   3.4 &   0  & 1  &
0 & 1  &     0 & 6.3 & 6.3 & 6.3& 6.3 & 2.7\\
\hline 
\end{tabular}
\end{table*}

\begin{table}
\caption{Parameters changing with NS precession phase $\Psi$ for RXTE cycle
\#308. Zero optical depth corresponds to the observation with maximum count rate  
\label{t:308}}
\begin{tabular}{lcccccc}
\hline
\hline
$\Psi$  & 0.985 & 0.03 & 0.073 &0.12 & 0.507 & 0.537 \\
\hline
$\tau_{eff}(\Psi)$      &0.8 & 0.55 & 0 & 0.1 & 2.2 & 2.4  \\
$B (\Psi)$                 & 4 & 25 & 43 & 27 & 4 & 6            \\
$\varphi_0(\Psi)$           & 0.5 & 0.5 & 0.6& 0.6 & 0.5 & 0.4   \\
\hline
\end{tabular}
\end{table}

It is seen from Fig. \ref{f:308nn} and Figs. \ref{f181}-\ref{f324} that 
the agreement between the observed pulse profiles and the model is not perfect.
For example, in the framework of our model it is impossible to reproduce the 
pulse shape between the 'right shoulder' (formed by the 'N6-N8' arcs)  and the 'main peak' 
(formed by the 'N' pole in the main-on state). The additional light there cannot 
be explained by some intermediate Gaussian from an emission region located between N pole and N8 
arc, and the asymmetric form of the right shoulder rather suggests more complicated 
wings of the X-ray emission diagram, which cannot be calculated at present. Therefore, 
the formal comparison of the model with observations yields poor reduced $\chi^2$ value. 
For example, in the case of RXTE cycle 308 shown in Fig. \ref{f:308nn} we find 
$\chi^2\approx 1815$ for the number of degrees of freedom $6\times 129-(6\times 4+19)=731$.
When calculating the last number, we used 129 points per each of six observed profiles
in the cycle minus 
the number of adjustable parameters: 19 (intensities of 18 emitting regions and the 
'phase zero' of NS free precession $\Psi_0$, which were
not changed during the cycle) plus 4 variables per each profile (the phase of NS free precession 
$\Psi_i$, the effective optical depth $\tau_{eff}(\Psi_i)$, the amplitude $B(\Psi_i)$ and the phase $\varphi_0(\Psi_i)$ of the sine-like component. Therefore, formally we get $\chi^2/
dof\approx 2.5$, mostly due to failure to fit the 'main-on' profile between the 'main peak' and the 'right shoulder' as discussed above. However, 
the general behavior of the pulse profile evolution in our purely geometrical model 
with the NS precession phase is well reproduced. Similar comments are pertinent for 
other cycles.

We would like to stress that here the important parameter is the 
identification of the 'phase zero' of the NS free precession $\Psi=0=\Psi_0$. We remind that this phase corresponds to 
the minimum angle between the line of sight 
and the direction to the brightest (North) magnetic pole (see Fig. \ref{f:precview}, left top figure), which mostly contributes to the X-ray pulse amplitude during the main-on state. Once determined in a given 35-d cycle, the precession phase zero allowed us to 
assign precession phases for all other profiles observed in this cycle, since
the time intervals between individual RXTE pointings are known (see Table \ref{t:data}). Noteworthy here are cycles for which both 'main-on' and 'short-on' profiles were observed
(e.g., 181, 308, 313) and consecutive cycles (\#268-269, 
\#303-304, \#307-308). These cycles were used to estimate the 
error in the determination of the phase zero, 
which was found to be $\Delta \Psi_0\sim \pm 0.05$. The identification of the 
zero precession phase $\Psi_0$ allowed us to 
explore the properties of the assumed NS free precession 
as a function of time by plotting the moments of $\Psi_0$ on the O-C diagram (see the next Section). 

\section{Evidence for variable NS precession period from the O-C diagram}

\subsection{O-C diagram for $\Psi_0$}
\label{psi0}

Still \& Boyd (2004) pointed to the fact the O-C diagram for the turn-ons of the 35-day 
cycle of Her X-1 can be described by several  straight segments, which means that the local
 value of the 
35-day period during these intervals is different from the mean value used to obtain the 'calculated' points
(see also \citet{Staubert_ea10, Staubert_eaDublin, Staubert_eaIoffe, Staubert_ea13}).
% www.ioffe.ru/astro/NS2011, and Staubert et al. 2010, in Proc. INTEGRAL Conf., 
%Dublin, PoS (INTEGRAL 2010) 048).
The slope appears to change abruptly from
one segment to the next, often close to anomalous low states (ALS), 
when the source is blocked by the accretion disk all the times.
%of the source 

\begin{figure*}
\centering
\includegraphics[width=0.7\textwidth]{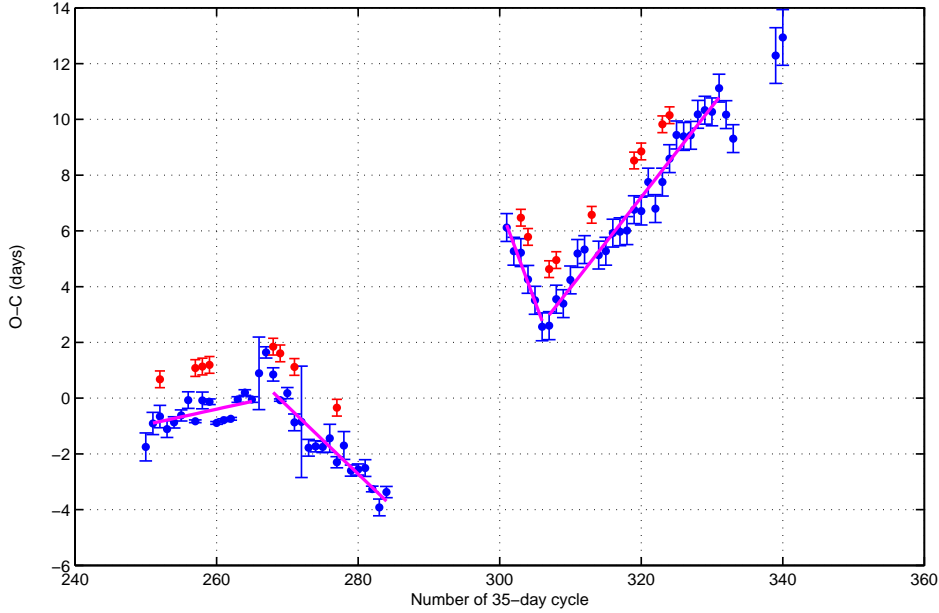}
\caption{The position of times of 35-day turn-ons  on the O-C diagram (black (blue) 
filled circles with error bars) and the location of the times of 
zero phases of the NS free precession $\Psi_0$  
for the modeled \textsl{RXTE} cycles (gray (red) filled circles with error bars).
Gray (magenta) solid lines show linear best-fit to the observed times of 35-day turn-ons
within 'stable' epochs from Table \ref{t:vp}.}
\label{f:omc}
\end{figure*}

If NS free precession with constant period were
responsible for
the long-term stability of the 35-day cycle, 
the expected location of NS precession zero phase $\Psi_0$ on the O-C diagram would make, 
within errors, a straight line with the slope 
determining the value of the true precession period $P_{pr}$. To check this hypothesis we plotted
MJD($\Psi_0$) for each 35-day cycle with known RXTE profiles from Table \ref{t:data} on the 
O-C diagram (Fig. \ref{f:omc}) by open diamonds. 
To obtain each O-C value in this plot, we took
each estimated time of free-precession phase zero $MJD(\Psi_0)$, and
then subtracted the time of the turn-on of the given cycle calculated
using the ephemeris of \citep{Staubert_ea83}:
$$
TO_C(MJD)=42409.84+20.5\times P_{orb}\times (N_{35}-31)
$$
where $N_{35}$ is the number of the 35-day cycle, and $P_{orb}=1.7001678$~d is the binary orbital period. 

In Fig. \ref{f:omc} we see that the times of the free-precession zero phases actually follow straight lines within the 'stable' 
epochs, where the MJD($\Psi_0$) make straight segments,  
with a time delay of $\approx 1.65$~d with respect to the mean turn-ons calculated using the linear 
approximation. This time delay is found to be the same for all four 'stable' epochs identified 
in the studied data set (Table \ref{t:vp}). 

\begin{table*}
\caption[]{Variable precession period in four 'stable' epochs}
\label{t:vp}
\begin{tabular}{ll|cccc}
\hline
Epoch N &Cycles & $TO_{obs}$ (MJD) & $TO_C$ (MJD) & MJD $\Psi_0$& $P_{35}$ (d)\\
\hline
I&252 & 50111.8 & 50112.46& 50113.14 &\textbf{34.90}\\
 &257 & 50285.9 & 50286.73& 50287.81\\
 &$258^*$ & 50321.5 & 50321.58& 50322.71\\ 
 &259 & 50356.3   & 50356.43& 50357.62\\
\hline
II&$268^*$ & 50670.96 &50670.11 &50671.96 &\textbf{34.61}\\
  &269 & 50704.9    &50704.97 &50706.57\\
  &271 & 50773.8    &50774.67 &50775.79\\
  &$277^*$ & 50981.5    &50983.80 &50983.46\\
\hline
III&303 & 51895.2 &51889.98 &51896.45 &\textbf{34.18}\\ 
   &304 & 51929.1 &51924.84 &51930.62\\
\hline
IV&307 &  52032.0 &52029.40 &52034.03 &\textbf{35.18}\\ 
  &$308^\dag$ & 52067.8  &52064.25 &52069.20\\ 
  &$313^\dag$ & 52243.7  &52238.52 &52245.10\\ 
  &319 & 52454.4  &52447.64 &52456.17\\
  &320 & 52489.2  &52482.49 &52491.34\\
  &323 & 52594.8  &52587.05 &52596.89\\  
  &324 & 52630.5  &52621.91 &52632.06\\
\hline
\end{tabular}
\newline
$^*$ -- cycles with pulse profiles available only during the short-on states
\newline
$^\dag$ -- cycles where both main-on and short-on states were observed
\end{table*}

This result rules out the hypothesis of a stable free 
precession period of NS as a clock mechanism 
for the 35-day cycle. A similar conclusion was
reached by \cite{Staubert_ea13} using a 
completely different analysis, in which \textit{no assumption} was 
made on the mechanism of pulse profile change, 
and the observed pulse profile template was used
for the timing analysis. The apparent tight connection 
between the $\Psi_0$ and the turn-ons of the source 
due to accretion disc precession suggests that 
either there is only one clock in the system, i.e.
the precessing accretion disk around a non-precessing NS, which also produces
the change in the observed X-ray pulse profiles, or
the free precession of NS is there, but with changing period. 
Here we will discuss the second possibility, i.e. variable free precession with 
a fractional change of the free precession period of up to $0.03$.

\subsection{Variable free precession}

The free precession period
of an axially symmetric solid body  is (Landau \& Lifshits, Mechanics)
\begin{equation}
\label{prec}
P_{pr}=P\frac{I_3}{(I_1-I_3)\cos\vartheta}\,.
\end{equation}
Here $P=2\pi/\omega$ is the NS spin period, $I_1$, $I_2$, $I_3$ stand for the principal moments of inertia 
of the body, $I_1=I_2$ for an axially symmetric body. The precession takes place around the axis $I_3$,
and the sign of the difference $I_1-I_3$ determines the sense of the precessional motion (prograde or 
retrograde for positive and negative sign, respectively). The angle $\vartheta$ is the angle between the axis $I_3$ and 
the angular momentum of the body. In our picture, the axis of inertia $I_3$ is pointed far away from the NS angular 
momentum $J$. 
The large value of the angle $\vartheta\simeq 50^o$ can be due to a complex pattern of 
accretion onto the NS surface with a non-dipole magnetic field, as described above.

Generally, the free precession 
period of the NS can vary either because of the change in the relative difference of moments 
of inertia $\Delta I/I$, or due to the change in the angle $\vartheta$. In our case the $10^{-6}$ non-sphericity 
of the NS is caused by internal properties of the NS crust and appears to be independent 
of the external accretion. 
In addition, possible accretion-induced jumps of the non-sphericity parameter 
$\Delta I/I$ by a few percent are expected to be all positive or negative. 
Oppositely, the angle $\vartheta$ can vary near some mean value $\vartheta_0$ as a result of accretion of matter onto
different spots on the NS surface, and its changes can be of any sign.   

What happens if the angle $\vartheta$ changes abruptly by an amount $\Delta\vartheta$ due to, for example, 
the increase of stresses in the NS crust above some critical value in the course of accretion? 
Clearly, the total angular momentum $J$ of the NS must be conserved for an instantaneous change in the 
angle $\vartheta$. The total angular momentum can be expressed through the NS spin angular velocity 
$\omega=2\pi/P$ as $J=I_1\omega$ and is independent of the angle $\vartheta$. This means that the change 
in the angle $\vartheta$ should not affect the NS spin, only the precessional motion will be changed. Then from 
(\ref{prec}) we find that the fractional change of the free precession period is 
\begin{equation}
\frac{\Delta P_{pr}}{P_{pr}}=\frac{\sin\vartheta}{\cos^2\vartheta}\Delta \vartheta
\end{equation}
that is to within a numerical factor of order one $\Delta P_{pr}/P_{pr}\simeq \Delta\vartheta$. This means that 
the fractional change in the free precession period can be both positive or negative, and can reach 
a few percent level for $\Delta\vartheta \sim 0.03$~rad.

Unlike the total angular momentum and the total energy, the kinetic energy of NS rotation will change.
It can be shown that the fractional change of the kinetic rotational energy is 
$$
\left|\frac{\Delta E_k}{E_k}\right|\simeq \left|\frac{\Delta \omega_{pr}}{\omega}\right|=
\left|\frac{\Delta P_{pr}}{P_{pr}}\right|\frac{P}{P_{pr}}\,.
$$
Clearly, this fractional change is too small to have any  observational consequences.  

In our model, the mean free 
precession period during the 'stable' epochs can be calculated from 
the O-C diagram for the times of the 35-day turn-ons (see Fig. \ref{f:omc}, straight lines, 
and Table \ref{t:vp}). 
The precession of the disk itself is subjected to action of tidal and dynamical torques which can fluctuate (see the Discussion below), and therefore is expected to be less stable than the NS precession. The moments of the disk opening of the X-ray source (the turn-on of the source,
i.e. the start of the main-on state, mostly derived from the  \textsl{RXTE}/ASM light curves, the column TO(MJD) in 
Table \ref{t:data}), should follow the free precession period only \textit{on average}.

\section{Discussion}

\textbf{A) Features of the model}.

The proposed model describes the observed X-ray pulse
profile of Her X-1 in terms of emission from several regions including
poles and arcs on the NS surface (see Table 1). The locations of the
emission show the complex pattern of accretion that occurs in the
presence of a non-dipole surface magnetic field. The narrow X-ray
emission beam is a crucial assumption 
for our model, and we think it is worth to give a more detailed 
justification to this statement. Generally, the width of the X-ray beam, which
may be characterized by $\theta_0$, depends on the optical depth for
scattering in the region where most of the X-ray emission is generated
according to $\theta_0 \sim 1 \sqrt{\tau^*}$  
\citep{Basko76, BaskoSunyaev76, Dolginov_ea79}.

For example, in the case of Coulomb braking $\tau^*\approx 20$
 \citep{ZeldovichShakura69, Nelson_ea93}. Therefore, if the accretion braking is due to Coulomb interactions (as is 
suggested by the observed positive correlation of the cyclotron line energy
with X-ray flux, \cite{Staubert_ea07, Klochkov_ea11}), $\theta_0\sim 1/5\sim 12^o$.   The fact that the observed X-ray pulse profile can be decomposed into several
narrow components was also discussed by \cite{BlumKraus00}. When the braking of the accretion flow is due to radiation pressure 
\citep{Davidson73, BaskoSunyaev76, WangFrank81}, most of the radiation flux
escapes from an optically thick accretion column which produces a wider X-ray emission diagram. The transition between the two regimes
of the accretion matter braking occurs at some critical X-ray luminosity, which depends on the total area on the NS surface $A=\pi r_0^2$ onto which the accretion proceeds. It lies between the local Eddington luminosity 
$L_{Edd}^*\approx L_{Edd}(A/4\pi R_{NS}^2)$ (here $L_{Edd}\simeq 2\times 10^{38}$~erg/s is the canonical Eddington luminosity for a NS mass of  1.5 $M_\odot$), and the X-ray luminosity $L^*\approx L_{Edd}^*(R_{NS}/r_0)$, above which the accretion column height is proportional to the X-ray luminosity \citep{BaskoSunyaev76}. Apparently, in Her X-1 the effective area is large enough (because of the complex magnetic field structure near the NS surface as discussed above) that even with an X-ray luminosity of $\sim 10^{37}$~erg/s the source is in the Coulomb braking regime.   

Note that a sharp hardening of the X-ray continuum
in the main-on pulse is observed 
when the left shoulder (arc 'N1') and the main peak ('North pole') 
cross the line of sight (see Fig. 6 in \cite{Vasco_ea13}),
which suggests a narrow X-ray beam. No such hardening is observed  
%emission diagram from the emitting regions
in the intermediate pulse 
('South pole') -- indeed, in our model the latter
never crosses the line of sight as it is located close to the axis of inertia of the NS. 
The intensities of accretion onto the 'North' and 'South' parts of the NS 
are found to be strongly different, suggesting a notable radiative drag force acting on the binary
system. The secular acceleration of the system due to this force may explain the 
observed high galactic latitude of Her X-1.    

We have found that to reproduce the observed change of X-ray pulse profiles with 35-day cycle, 
the NS should freely precess around one of its principal moments of inertia (assuming
two-axial NS body), which is assumed to be  
strongly inclined to the angular momentum (by about 50 degrees). 
For the assumed geometry of emission regions on the NS surface, the direction of free precession
is found to be prograde, which corresponds to the prolate shape of the NS (Fig. 1), 
as expected if the NS ellipticity is
produced by a strong internal toroidal magnetic field.

\textbf{B) Features of the free precession period changes}. 
As we noted above, if moments of inertia of the NS body do not change, variations in the free precession period 
due to jumps of the precession axis with conserved total angular momentum do not affect the pulsar spin period 
($J=\omega I_1$). Therefore, in principle, precise timing of the X-ray pulsar allows one to differentiate between the 
jumps in the precession axis and the free precession change due to jumps in the moments of inertia. However, 
disk-magnetosphere interaction appear to have a stronger effect on the pulse timing than precession effects. 

\textbf{C) Uncertainties of the model}. 
Clearly, a purely geometrical
model for generation of variable pulse profiles as considered here can not perfectly fit the observed ones. 
The reasons for this include time variations of the mass accretion rate onto individual zones 
on the NS surface which can change the emission diagram from the spots and arcs \cite{WangFrank81}, temporal appearance of 
additional 'hot spots' (we considered only the most stable features that form the pulse profile which are 
seen in most cycles), etc. For example, in the present model the prominent peak right of the main peak, which is seen almost at all main-on 
phases on the model pulse profile (e.g., Fig. \ref{f181}, \ref{f303}, \ref{f304}, etc.), is produced by
photons originating in the arc-like regions  most distant from the North pole (N7, N8 in Table 1). 
Magnetic field lines in these regions can be inclined to the NS surface under
some angle different from $\pi/2$.  In that case, due to relativistic light bending, the effective X-ray diagram 
from these emission spots will be more inclined toward the magnetic pole, and hence the part of the pulse profile that is 
formed when the line of sight crosses these structures, will appear more extended towards the main peak.  
In addition, X-ray photons 
scattered in the accreting flow can heat the NS surface around the poles and arcs 
thus producing more extended beams with non-Gaussian wings. Therefore, we do not expect to 
perfectly fit the observed pulse profiles by a set of Gaussians in the frame of our geometric model. 
Variable accretion rate onto different emitting spots on the NS surface in different cycles 
(see Table \ref{t:relint}), as well as changing effective optical depth, may also cause the 
pulse profile to look quite different at the same NS free precession phases $\Psi$ 
in different 35-day cycles (see especially unusual pulse profiles in RXTE cycles 259 and 303). 
However, we stress that it is the general behavior of the pulse profiles at the right precession phases in each 
35-day cycle considered that is important.

Clearly, the larger number of individual profiles and the longer the time span measured during one
35-day cycle, the more 
precise is the determination of the time of the NS free precession zero phase $\Psi_0$. 
Here especially important are cycles 
$\#308$ and $\#313$ (see Table \ref{t:data}), for which both the main-on and short-on states were observed by 
\textsl{RXTE}. It is very difficult to formally calculate the 
uncertainty in the determination of the free precession phase we ascribe to the pulse profiles in a 
particular 35-day cycle, but our experience with $\#308$ and $\#313$ shows that that an error equivalent 
to a phase shift $\pm 0.05$ 
is admissible. This minimal error is adopted as a measure of our accuracy in the determination of the 
time of the free 
precession  phase zero $\Psi_0$ in a specific cycle. 
Of course, in a 'poor' cycle where only a few measurements of pulse profiles are available, the phase 
error would be larger. But we note that for consecutive series of cycles within the epochs with a 'stable' 
precession period ($\#257-259, \#268-269-271, \#303-304, \#307-308, \#319-320, \#323-324$, see Table \ref{t:vp}) 
the phase shift applied to the first measurement in a series translates to all other measurements in a given series, 
and in all cases the admissible phase shift was found to be less than $\pm 0.1$. So in Fig. \ref{f:omc} we have 
plotted the derived times of zero precession 
phases in the studied cycles (open diamonds) with the error $\pm 0.05$.

\section{Conclusions}

To explain the changing X-ray pulse profiles of Her X-1 as observed by PCA
\textit{RXTE}, we consider a model of accretion of matter onto the surface of freely precessing neutron star with a 
complex non-dipole magnetic field. The required ellipticity of the NS body is provided by the internal toroidal magnetic field, leading to the prolate shape of the NS and prograde free precession, 
and the free precession is not damped out because of strong disk-magnetospheric interaction.
We have 
constructed a geometrical model that includes a number of poles, arcs and spots on the 
NS surface due to a complex character of the surface magnetic field of the neutron star, which 
emit in narrow pencil beams normally to the surface. 
The model satisfactorily reproduces the observed pulse
profile changes with 35-day phase, which are related 
to the phase of the NS free precession. 
We have found that the observed trend 
of the 35-day turn-ons on the O-C diagram, 
which can be approximated by a collection of consecutive linear 
segments around the mean value, can be described by our model by assuming a variable free precession period, 
with a fractional period change of about a few percent. Under this assumption and using our model we have found that the
times of the zero phase of the free precession (which we identify with the maximum separation of the brightest 
pole on the NS surface with the NS spin axis) occur about 1.6\,d after the 
mean turn-on times inside each 
'stable' epoch, producing a linear trend on the O-C diagram with the same slope as the observed 
turn-on times. 
We argue that the few-percent change in the free precession period occurring every few years, often within a short time interval of only a few 
35-day cycles, may be related to the wandering of the 
inertia axis around the mean value inside the NS body due to a complex pattern of accretion onto its surface.  
The striking closeness of periods of the disk 
precession and the NS  free precession can be 
explained by the presence of a synchronization mechanism in the system, which involves the dynamical interaction 
of the gas streams with the accretion disk modulated by the NS free precession period.  

\section*{Acknowledgements}

The authors thank the anonymous referee for careful reading of the manuscript 
and useful notes. 
This work was
supported by DFG-Schwerpunktprogramme (SPP 1177), grants of Russian
Foundation of Basic Research 10-02-00599a and 12-02-00186a. 
The authors acknowledge L. Rodina for help in RXTE pulse profile analysis.

\appendix

\section{RXTE data}
\label{AppendixA}

Table \ref{t:data} lists  
MJD of the turn-ons and of the center of \textit{Ginga} (cycle 181) 
and \textit{RXTE/PCA} (cycles 252-324) observations of Her X-1 used for pulse
profile modeling in the present paper.
The corresponding model phase of the NS precession $\Psi$ for each date is presented.  
Turn-on times of 35-day cycles are taken from \cite{Staubert_ea13}.

%\begin{multicols}{2}
%\begin{supertabular}{llrr}
\begin{table}
%\begin{minipage}[b]{0.5\linewidth}
%\centering
\caption{Data used for the pulse profile modeling}
\begin{tabular}{llrr}
\hline\hline
Cyc. N & Turn-on (MJD) & $<MJD>$ & Model $\Psi$\\
\hline
%\endhead
%\hline
\hline
\textbf{181}$^\dag$ & 47642.2&\multicolumn{2}{l|}{$^\dag$ \textit{Ginga} data \cite{Deeter_ea98}}\\
\hline
\multicolumn{2}{l|}{main-on} & 47643.9   & 0.0 \\
&& 47647.8 & 0.112\\
&& 47649.8 & 0.166\\
&& 47650.7 & 0.193\\
&& 47650.8 & 0.197\\
\hline
\multicolumn{2}{l|}{short-on} & 47662.8 & 0.540\\
&& 47663.0 & 0.545\\
&& 47663.6  & 0.563\\
&& 47664.7  & 0.591\\
\hline
\hline
\textbf{252} & 50111.8\\
\hline
\multicolumn{2}{l|}{main-on} &50114.18 & 0.03\\
\hline\hline
\textbf{257} & 50285.9\\
\hline
%&&50290.51 & 0.03\\
\multicolumn{2}{l|}{main-on} &50291.24 & 0.1\\
\hline\hline
\textbf{258} & 50321.5\\ 
\hline
\multicolumn{2}{l|}{short-on} &50340.90 & 0.52\\
&&50341.75 & 0.54\\
&&50342.70 & 0.57\\
% &&50343 & -\\
% &&50344 & - \\
% &&50345 & -\\
\hline\hline
\textbf{259} & 50356.3\\
\hline
\multicolumn{2}{l|}{main-on} &50356.15 & 0.97\\
&&50357.09 & 0\\
&&50358.55 & 0.04\\
&&50359.54 & 0.07\\
&&50360.49 & 0.09\\
&&50361.21 & 0.12\\
&&50362.71& 0.16\\
&&50363.85 & 0.19\\
% &&50364 & -\\
% &&50365 & -\\
% &&50380 & -\\
\hline\hline
\textbf{268} & 50670.96\\
\hline
\multicolumn{2}{l|}{short-on} &50690.50 & 0.54\\
&&50691.30 & 0.56\\
&&50694.20 & 0.64\\
\end{tabular}
%\end{minipage}
\end{table}

\begin{table}
\contcaption{}
\begin{tabular}{llrr}
\hline\hline
Cyc. N & Turn-on (MJD) & $<MJD>$ & Model $\Psi$\\
\hline\hline
\textbf{269} & 50704.9\\
\hline
\multicolumn{2}{l|}{main-on} &50704.67 & 0.94 \\
&&50705.53 & 0.97\\
&&50706.10 & 0.99\\
&&50707.33 & 0.03\\
&&50709.06 & 0.07\\
&&50710.47 & 0.11\\
&&50711.27 & 0.14\\
&&50713.20 & 0.19\\
%&&50714.10 & 0.21??\\
%&&50724.65 & 0.5??\\
% &&50725 & -\\
% &&50726 & -\\
% &&50727 & - \\
% &&50728 & -\\
\hline\hline
\textbf{271} & 50773.8\\
\hline
\multicolumn{2}{l|}{main-on} &50773.92 & 0.95\\
% &&50774.09 & 0.94\\
\hline\hline
\textbf{277} & 50981.5\\
\hline
%&&51000 & - \\
\multicolumn{2}{l|}{short-on} &51003.85 & 0.59\\
&&51004.20 & 0.60\\
%&&51005 & - \\
&&51006.70 & 0.67\\
%&&51007 & - \\
%\hline
%288 & 51366.8 & ALS \\
%\hline
%&&51370\\
%&&51371\\
%&&51372\\
%&&51373\\
%&&51374\\
%&&51375\\
%\hline
%296 & 51610.8& ALS\\ 
%\hline
%&&51621\\
%&&51622\\
%&&51623\\
%\hline
% 302 & 51860.4 & \\
% \hline
% &&51869.1 & 0.91\\
%\hline\hline
%\end{tabular}
%\end{minipage}
%\hspace{0.5cm}
%\begin{minipage}[b]{0.5\linewidth}
%\centering
%\begin{tabular}{llrr}
%\hline\hline
%Cyc. N & TO (MJD) & $<MJD>$ & Model $\Psi_{35}$\\
%\hline
%\endhead
\hline
\hline
\textbf{303} & 51895.2\\ 
\hline
\multicolumn{2}{l|}{main-on} &51894.23 & 0.94\\
&&51895.27 & 0.96\\
&&51896.67 & 0.01\\
% &&51897.02 & 0.03\\
&&51898.26& 0.05\\
&&51899.75 & 0.097\\
&&51900.03 & 0.105\\
&&51901.48 & 0.15\\
&&51902.07 & 0.16\\
\hline\hline
\textbf{304} & 51929.1\\
\hline
\multicolumn{2}{l|}{main-on} &51933.66 & 0.09\\
\hline\hline
\textbf{307} &  52032.0\\ 
\hline
\multicolumn{2}{l|}{main-on} &52032.99 & 0.97\\
%&&52033.03 & 0.97\\
&&52034.68 & 0.01\\
&&52036.27 & 0.06\\
&&52037.98 & 0.105\\
% &&52038.10 & 0.12?\\
&&52039.60 & 0.15\\
\end{tabular}
\end{table}

\begin{table}
\contcaption{}
\begin{tabular}{llrr}
\hline\hline
Cyc. N & Turn-on (MJD) & $<MJD>$ & Model $\Psi$\\
\hline\hline
\textbf{308} & 52067.8\\ 
\hline
% &&52067 & 0.95\\
\multicolumn{2}{l|}{main-on} &52068.69 & 0.985\\
&&52070.34 & 0.03\\
&&52071.95 & 0.07\\
% &&52072.11 & 0.1\\
&&52073.64 & 0.12\\
%&&52086 & -\\
\hline
\multicolumn{2}{l|}{short-on} &52087.25 & 0.51\\
&&52088.30 & 0.54\\
%&&52095 & -\\
%&&52097 & -\\
%&&52098 & -\\
%&&52099 & - \\
\hline\hline
\textbf{313} & 52243.7\\ 
\hline
\multicolumn{2}{l|}{main-on} &52243.55 & 0.96 \\
&&52244.55 & 0.98 \\
&&52245.64 & 0.02\\
&&52246.42 & 0.04\\
%&&52261 & - \\
%&&52262 & -\\
\hline
\multicolumn{2}{l|}{short-on} &52263.50 & 0.52\\
&&52264.20 & 0.54\\
&&52265.25 & 0.57\\
&&52266.40 & 0.61\\
%&&52267.25 & 0.63\\
%&&52268.4 & 0.66\\
%&&52269.3 & 0.69\\
\hline\hline
\textbf{319} & 52454.4\\
\hline
\multicolumn{2}{l|}{main-on} &52454.31 & 0.95\\
&&52460.89 & 0.13\\
\hline\hline
\textbf{320} & 52489.2\\
\hline
% &&52492.91 & 0.02? remove?\\
\multicolumn{2}{l|}{main-on} &52493.09 & 0.05\\
\hline\hline
\textbf{323} & 52594.8\\  
\hline
% &&52597.95 & 0.0 remove?\\
\multicolumn{2}{l|}{main-on} &52598.51 & 0.05\\
&&52599.49 & 0.07\\
&&52600.55 & 0.10\\
&&52601.53 & 0.13\\
&&52602.36 & 0.16\\
&&52603.51 & 0.19\\
&&52604.23 & 0.21\\
&&52605.20 & 0.24\\
%&&52606 & -\\
\hline\hline
\textbf{324} & 52630.5\\
\hline
% &&52633.90 & 0.0 0.03 remove?\\
\multicolumn{2}{l|}{main-on} &52634.10 & 0.04\\
\hline\hline
% 326 & 52701.0\\ 
% \hline
% &&52734 &- \\
% \hline
% 331 & 52877.0\\
% \hline
% &&52908 & - \\
%\hline
%333 & 52944.9\\
%\hline
%&&52950 & 0.94:\\
%\hline
%340 & 53193.5\\
%\hline
%&&53199.07 & 0.10\\
% &&53202 & 0.19\\
%\hline
%343 & 53296.8\\
%\hline
%&&53300.89 & 0.06\\
% &&53301.08 & 0.08? remove?\\
%\hline
\end{tabular}
%\end{minipage}
\label{t:data}
%\end{supertabular}
\end{table}

\section{Model pulse profiles}

The observed 9-13 keV 
pulse profiles and those predicted from our modeling are shown for each RXTE cycle 
at the
found NS precession phases $\Psi$ corresponding
to the mean MJD time of observations. In the figures (electronic version) the observed 
9-13 keV profiles are shown in magenta, 
the model profile is in black; the 
contributions from each emitting feature on the NS surface are shown in colors 
as in Table \ref{t:emregions}. The contributions from the north and south magnetic  
poles are shown by the dash-dotted blue and dash red lines, respectively. The sine-like component 
reflected from the disk is shown by the black dotted line.  

%\newpage
\onecolumn

\pagebreak
\begin{figure}
%\centering
\includegraphics[width=0.24\textwidth]{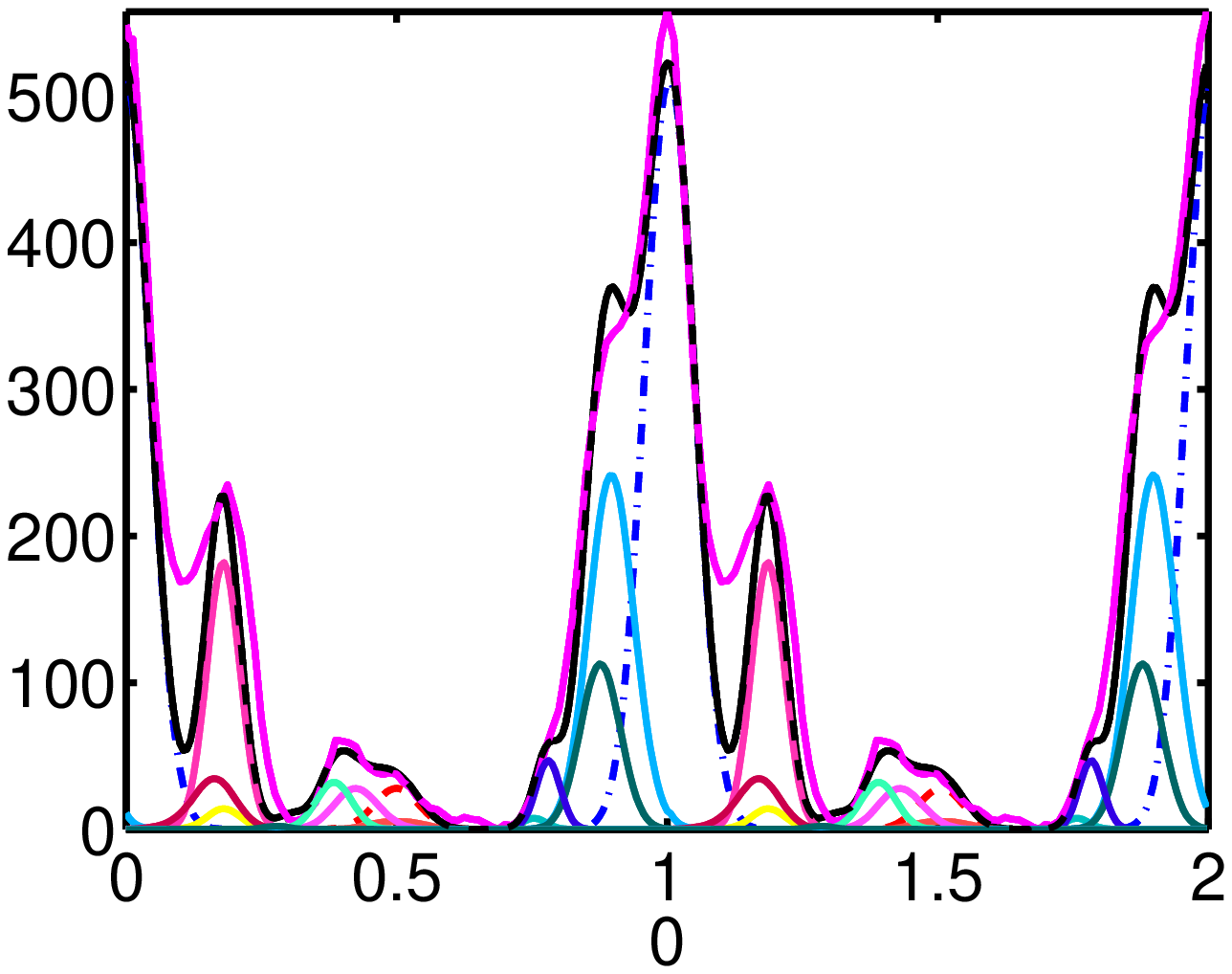}
\includegraphics[width=0.24\textwidth]{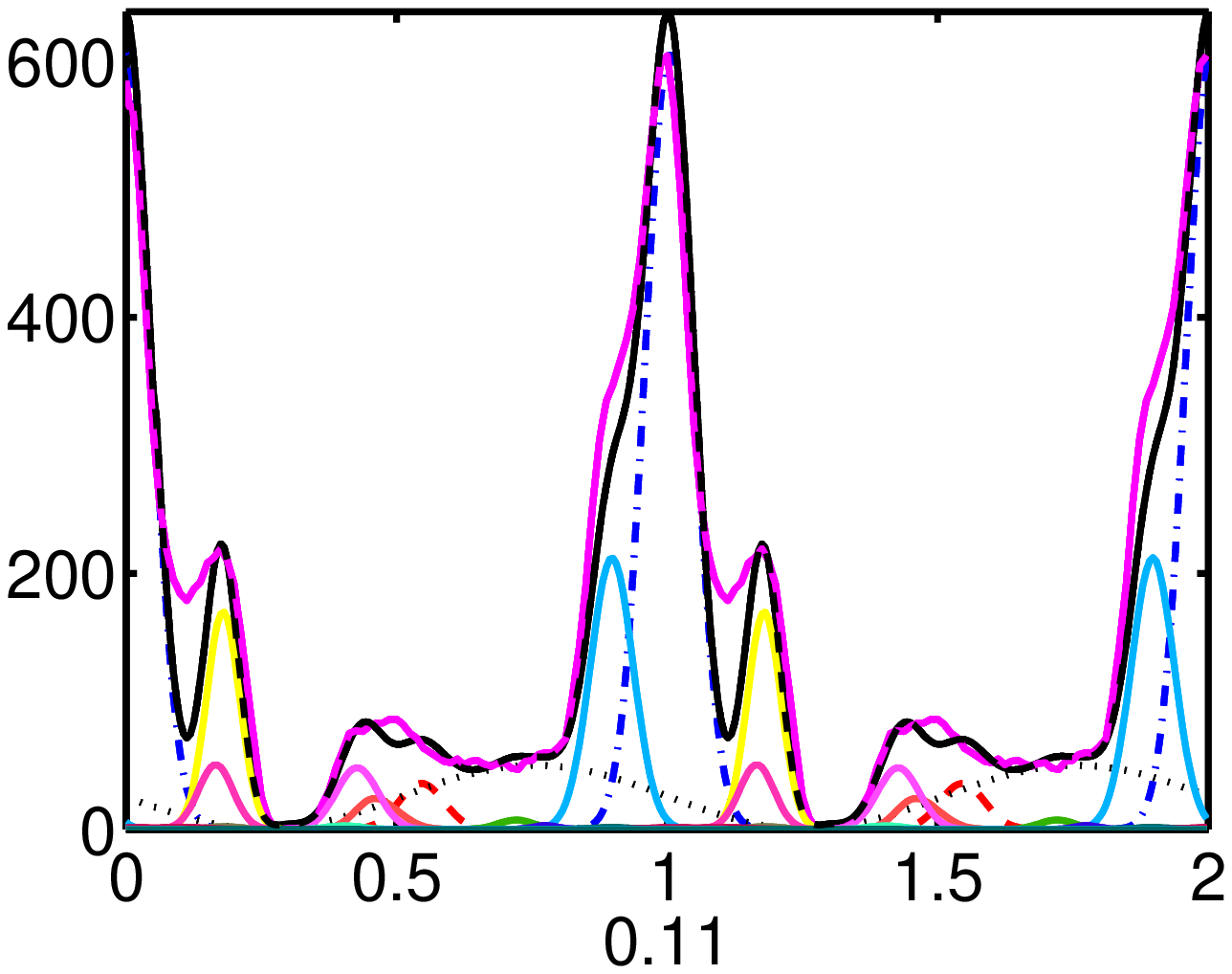}
\includegraphics[width=0.24\textwidth]{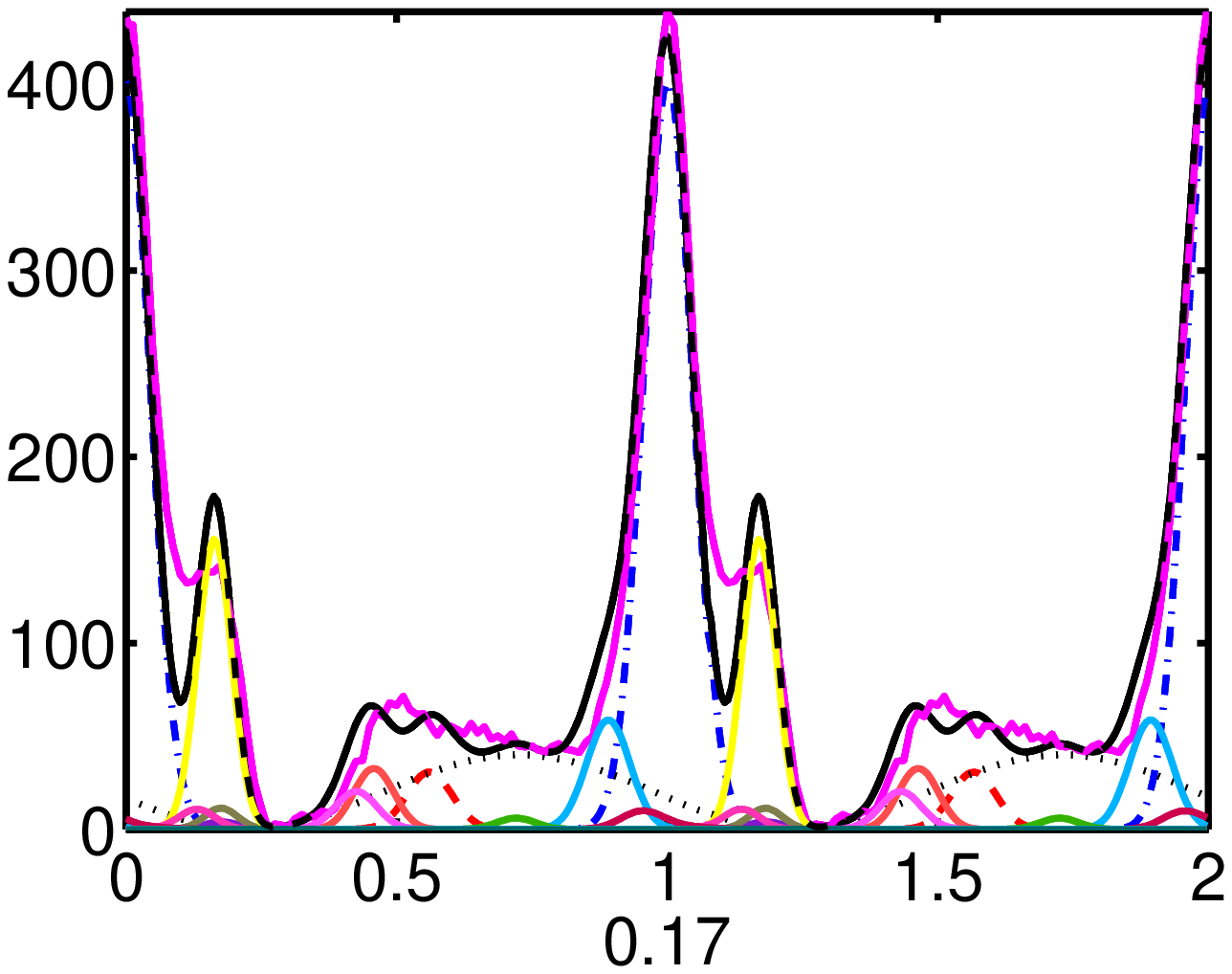}
\includegraphics[width=0.24\textwidth]{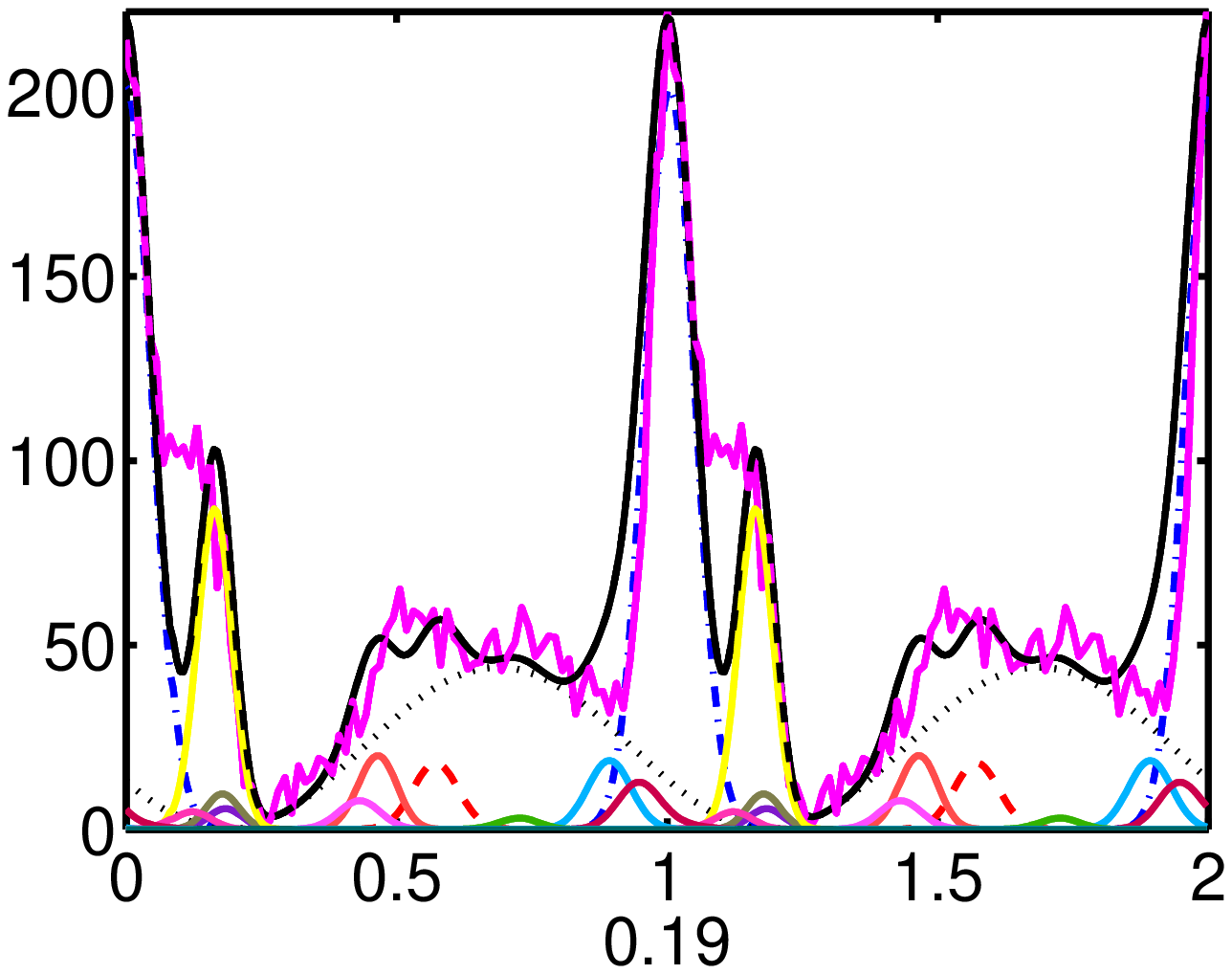}
\includegraphics[width=0.24\textwidth]{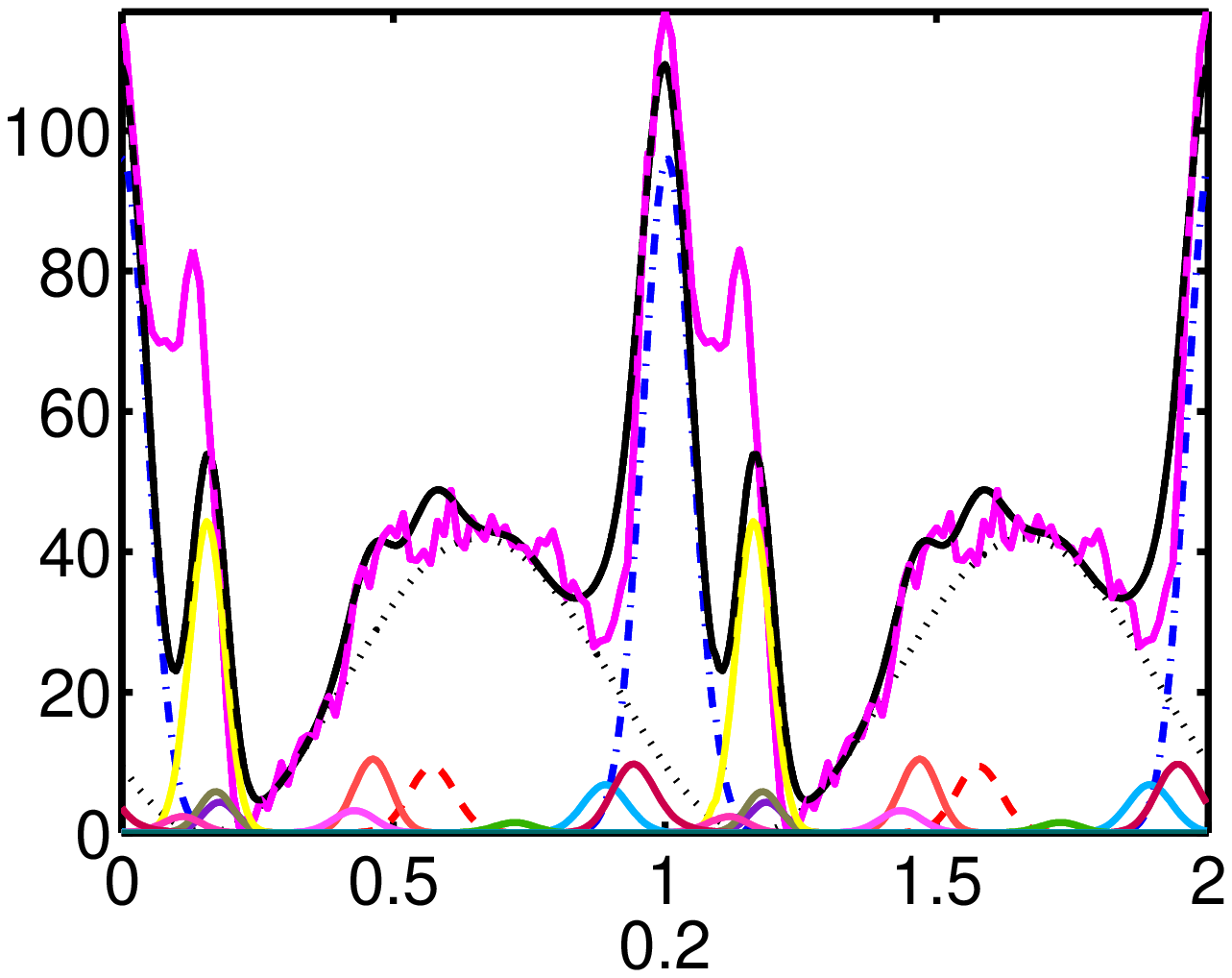}
\includegraphics[width=0.24\textwidth]{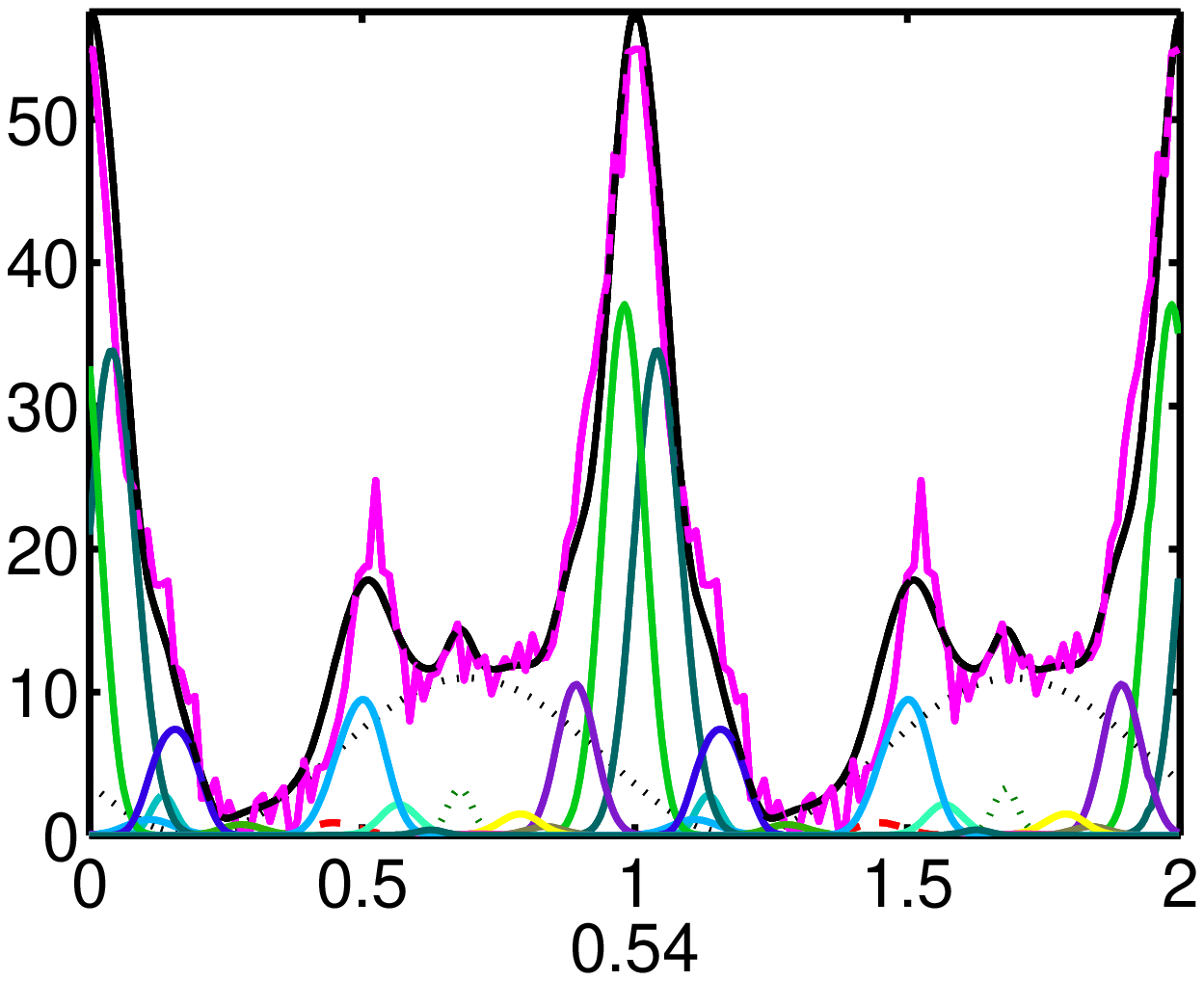}
\includegraphics[width=0.24\textwidth]{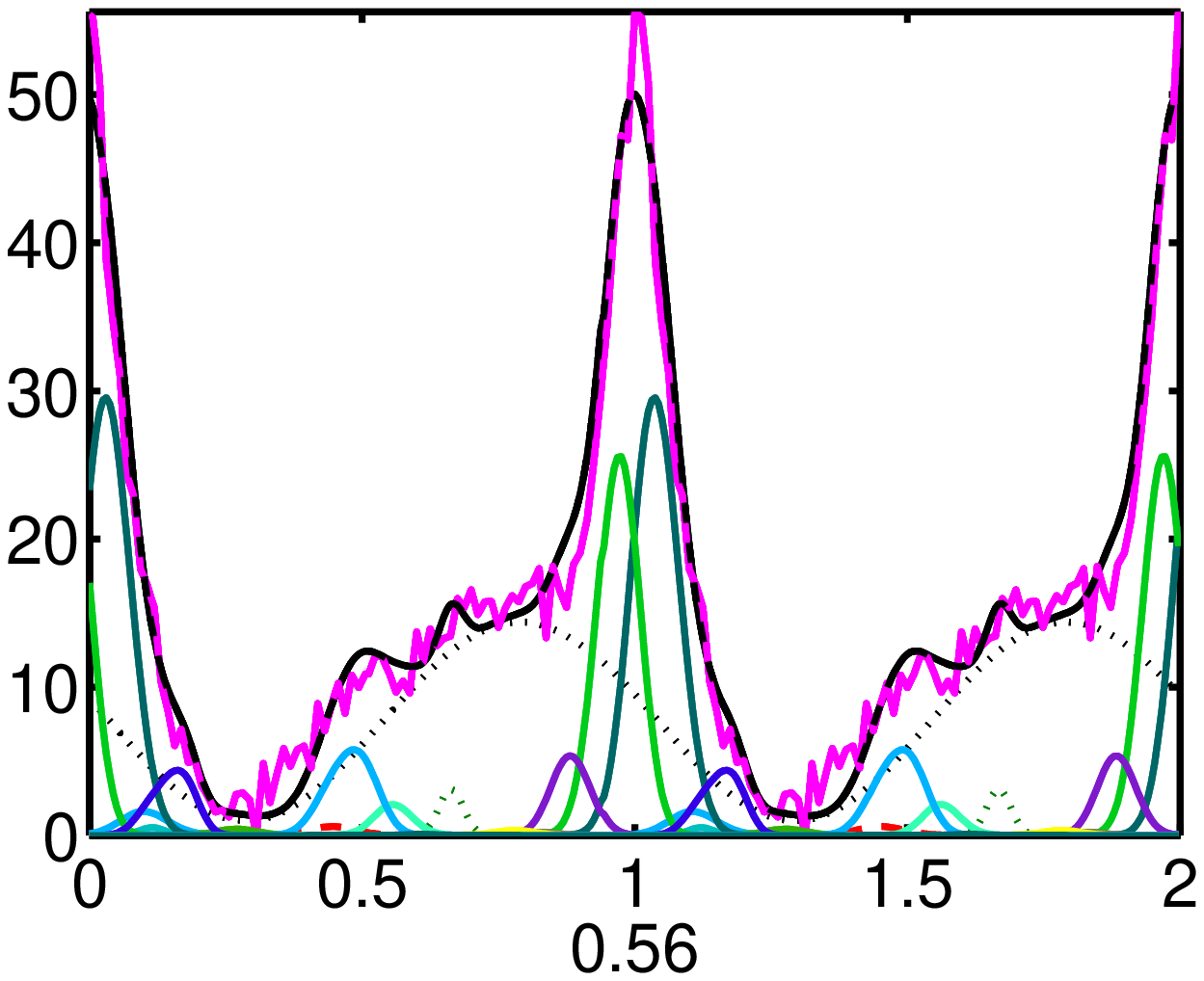}
\includegraphics[width=0.24\textwidth]{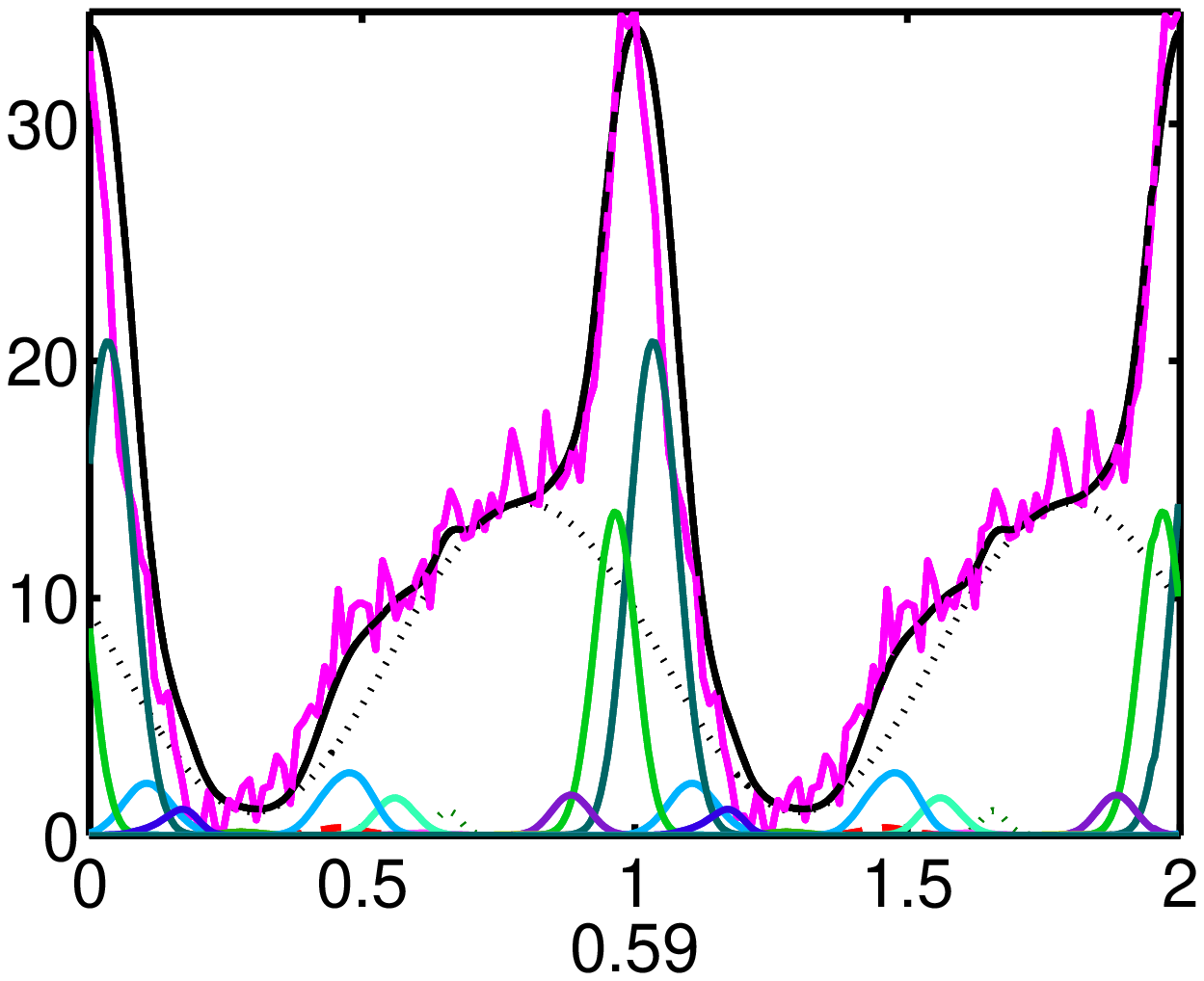}
\caption{181 Ginga}
\label{f181}
\end{figure}

\begin{figure}
%\begin{minipage}[b]{0.5\linewidth}
%\centering
\includegraphics[width=0.24\textwidth]{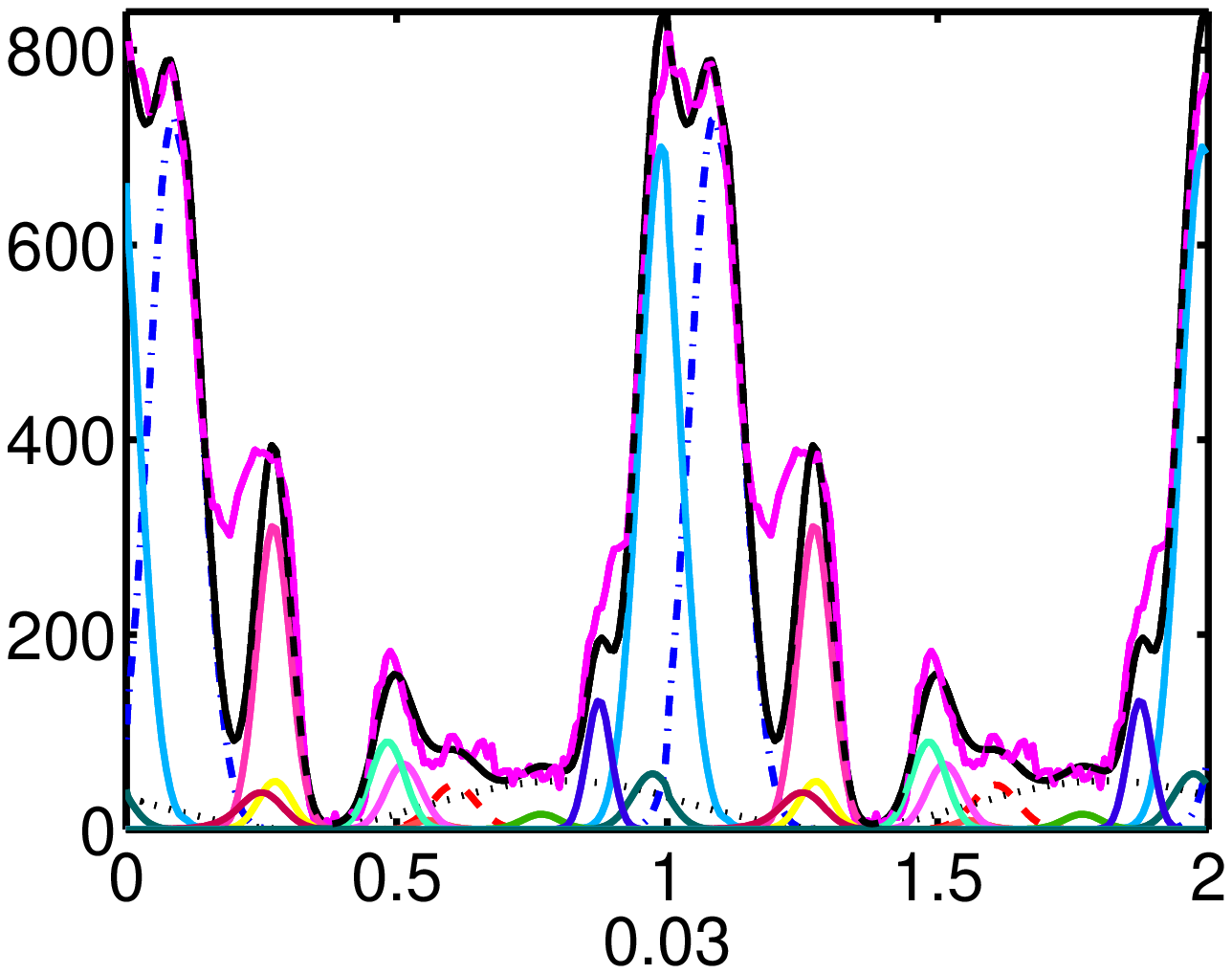}
\caption{252 RXTE}
\label{f252}
%\end{minipage}
\end{figure}

\begin{figure}
%\hspace{0.5cm}
%\begin{minipage}[b]{0.5\linewidth}
%\centering
\includegraphics[width=0.24\textwidth]{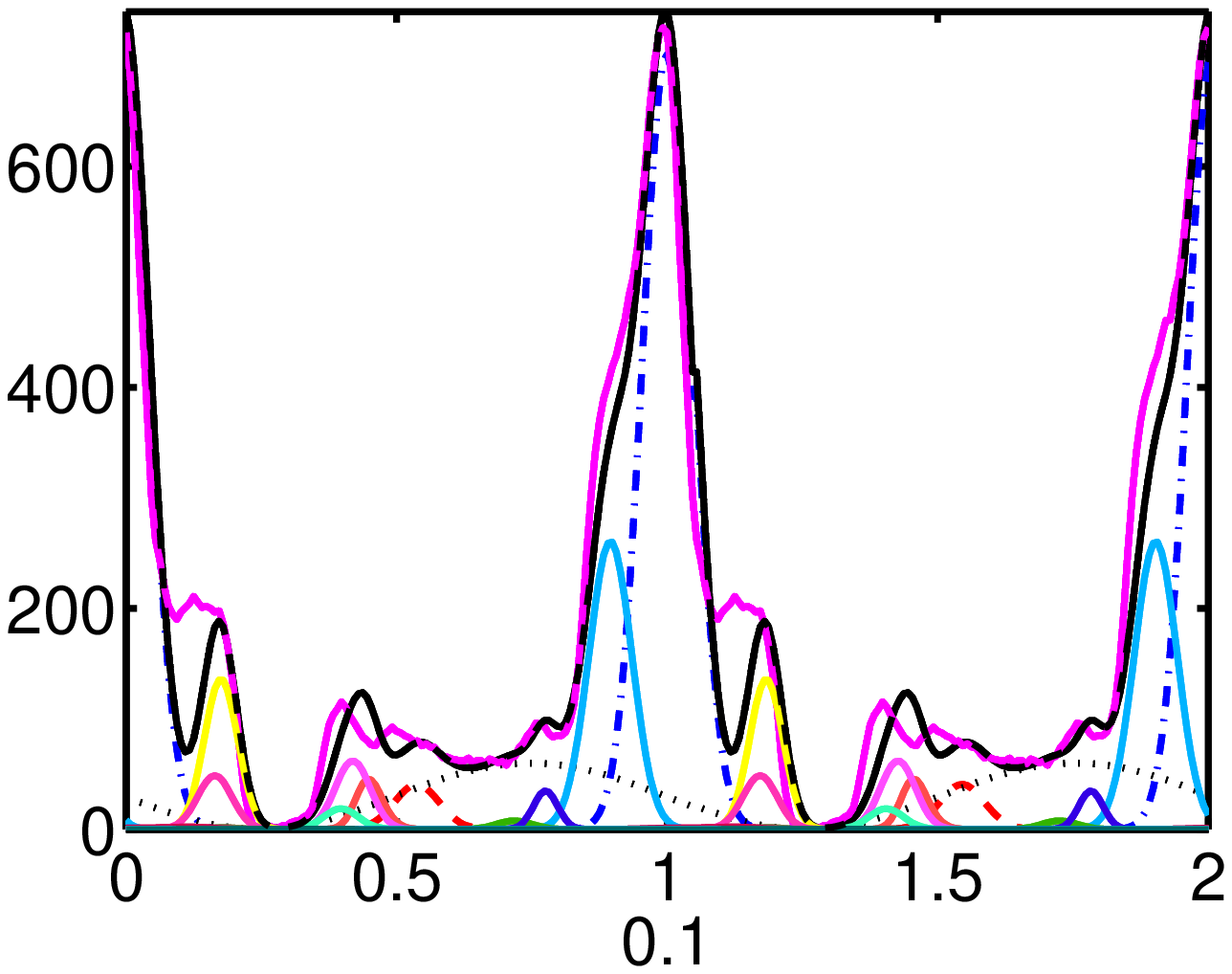}
\caption{257 RXTE}
\label{f257}
%\end{minipage}
\end{figure}

\begin{figure}
%\centering
\includegraphics[width=0.24\textwidth]{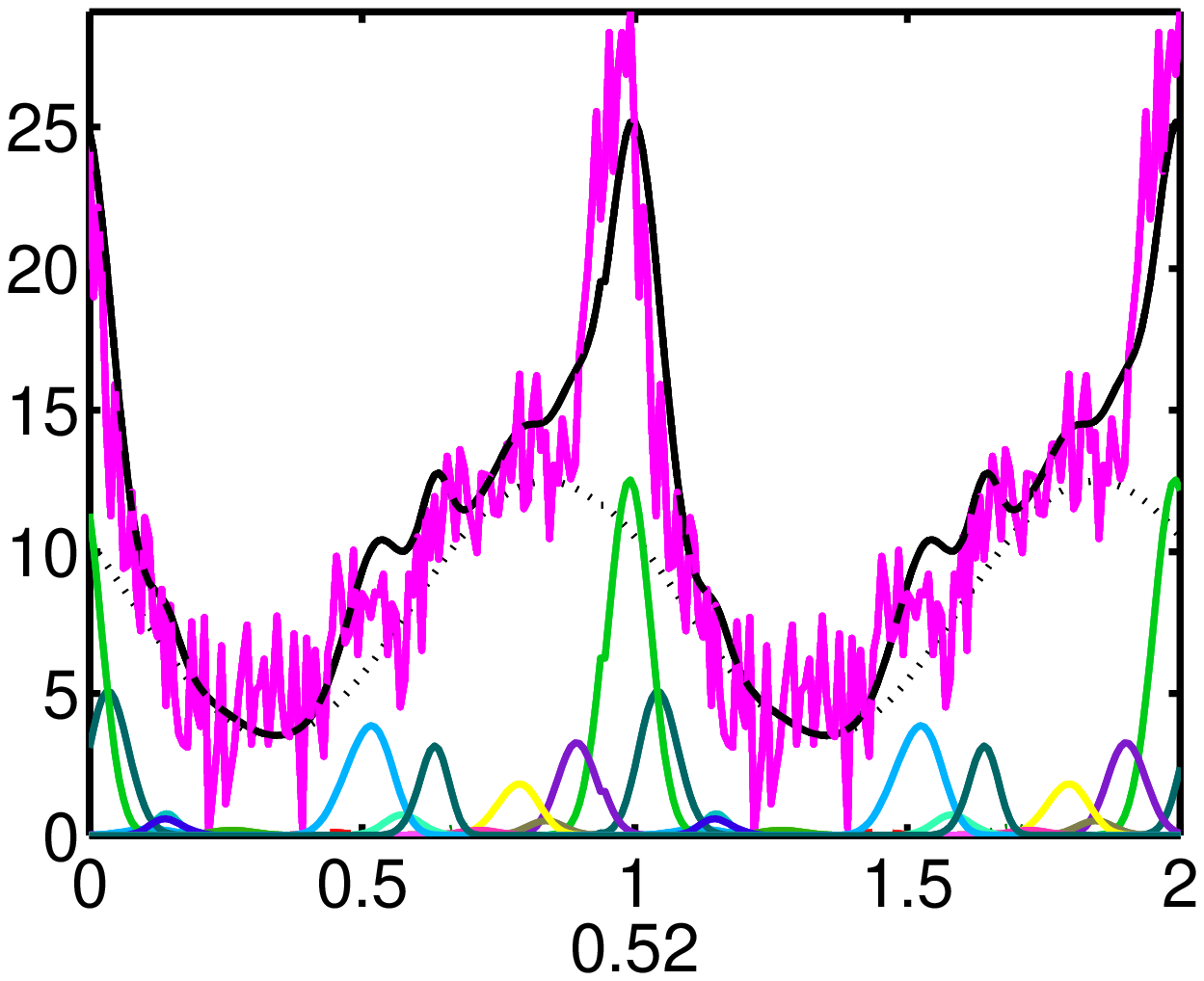}
\includegraphics[width=0.24\textwidth]{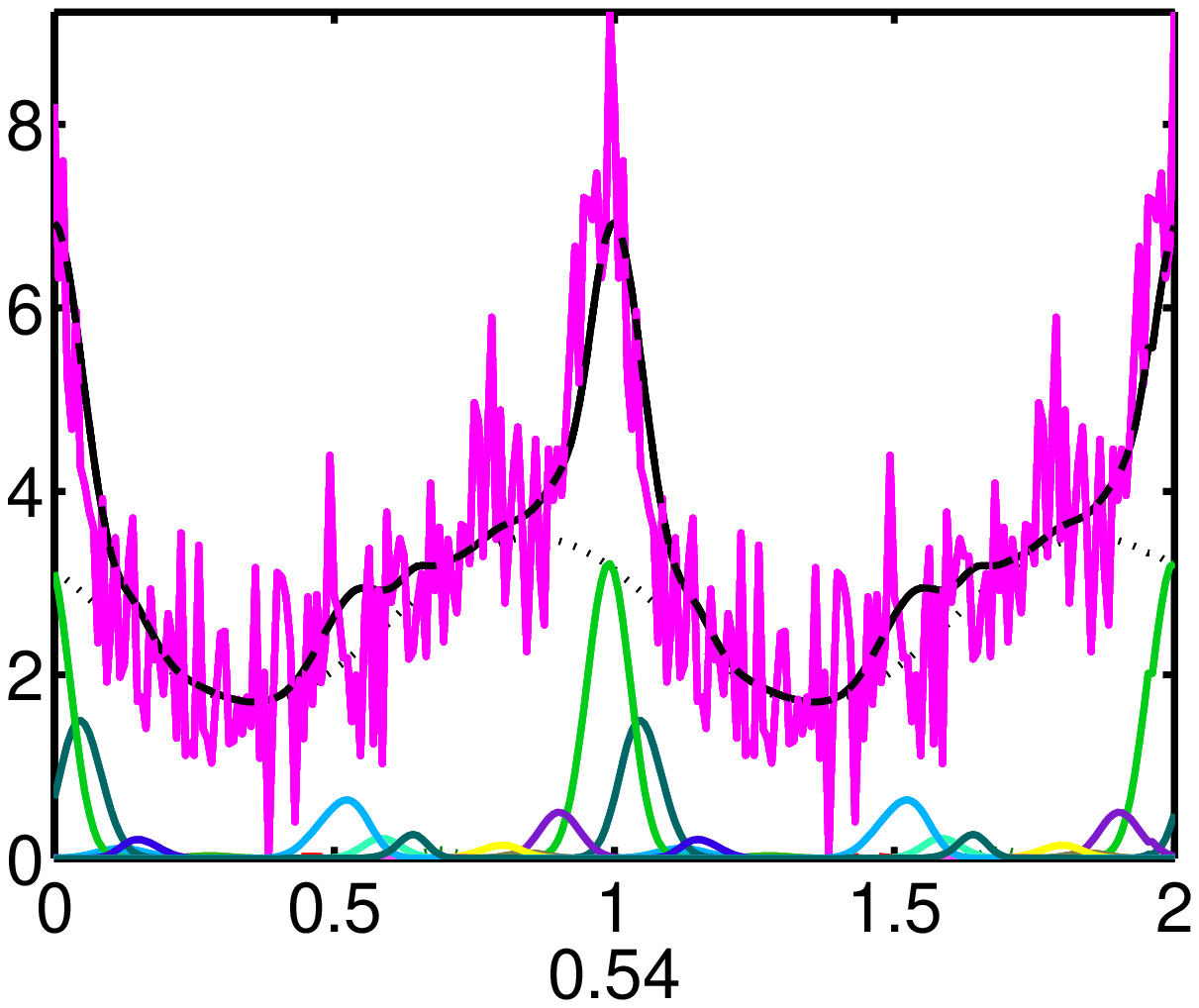}
\includegraphics[width=0.24\textwidth]{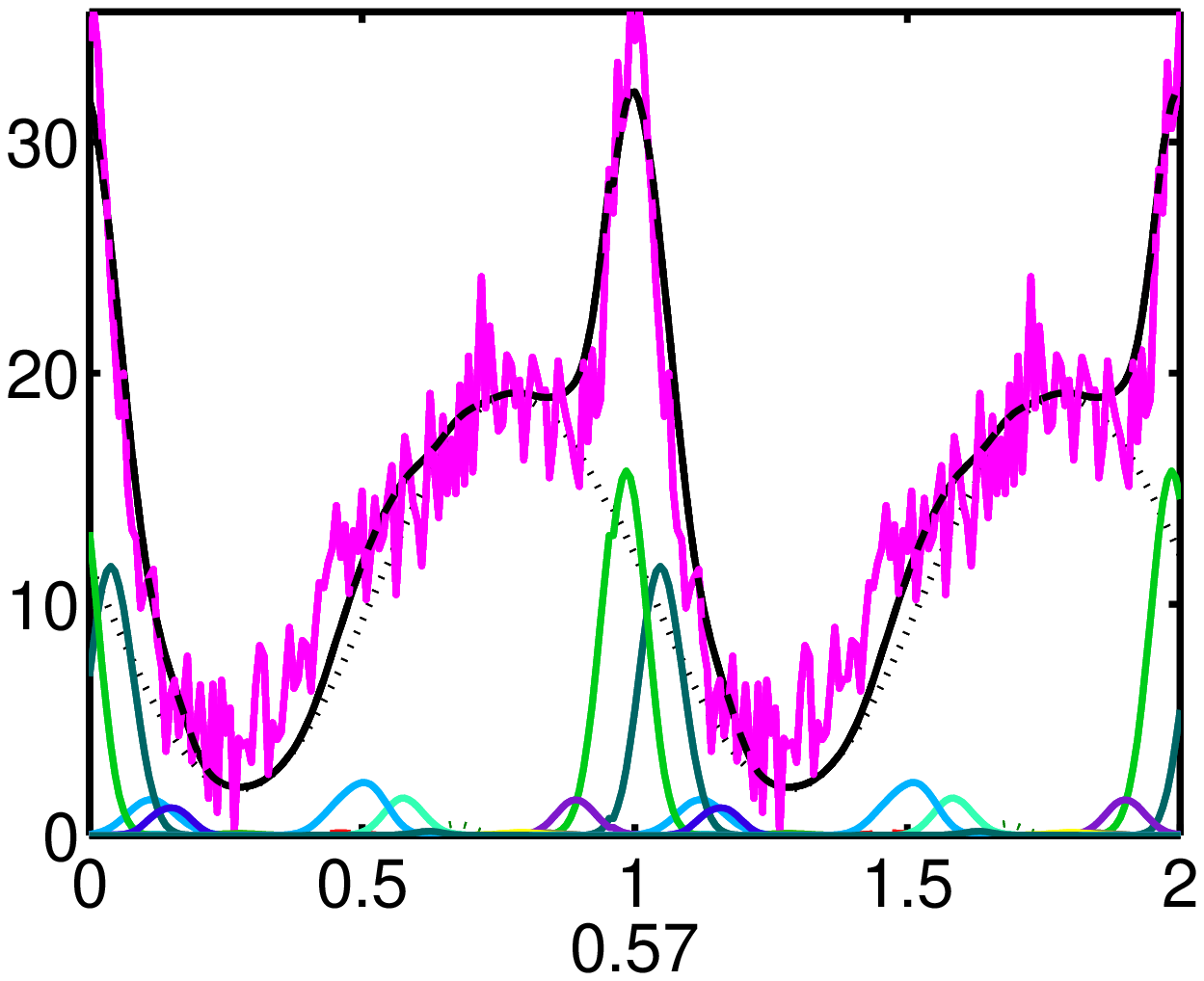}
\caption{258 RXTE}
\label{f258}
\end{figure}

\begin{figure}
%\centering
\includegraphics[width=0.24\textwidth]{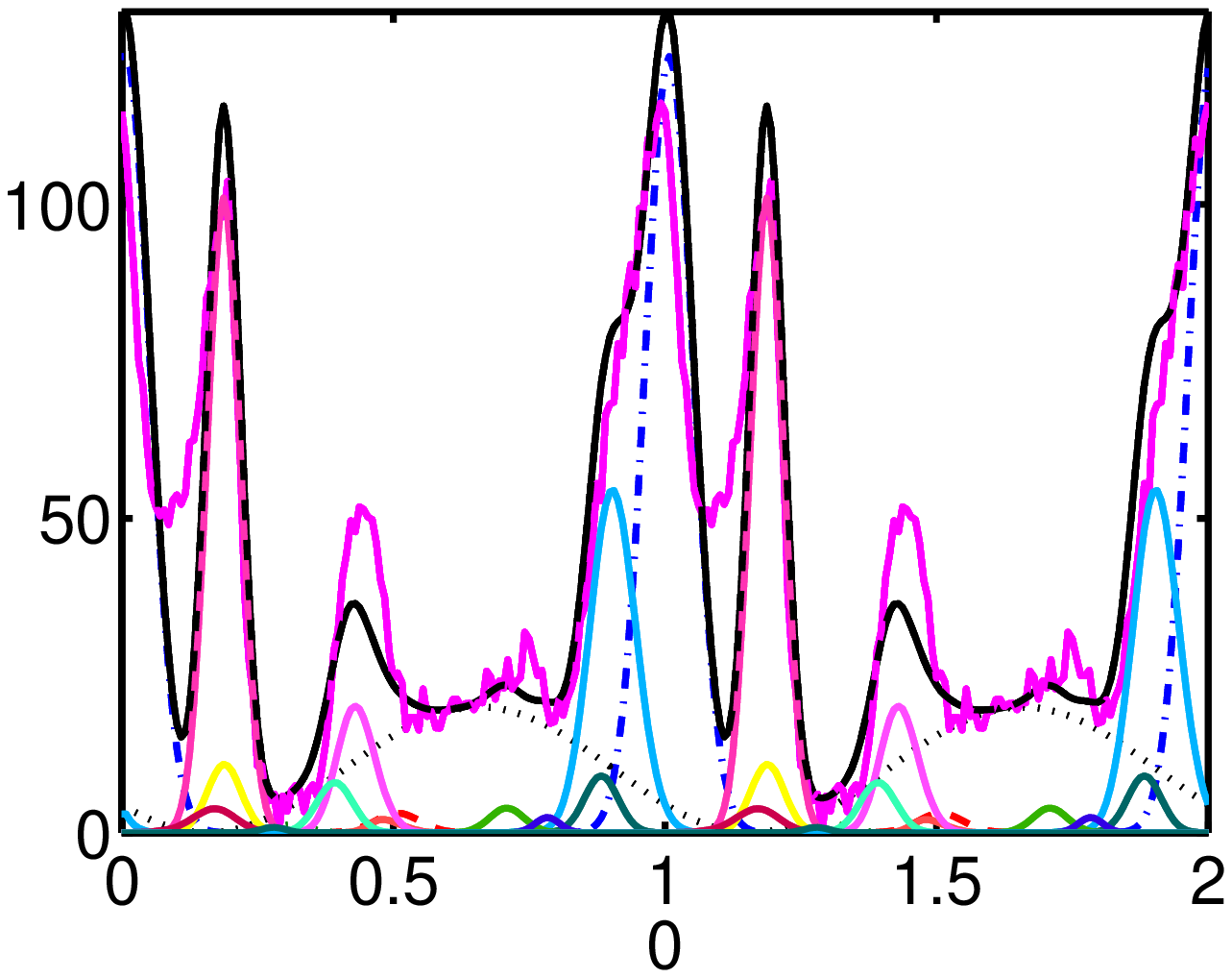}
\includegraphics[width=0.24\textwidth]{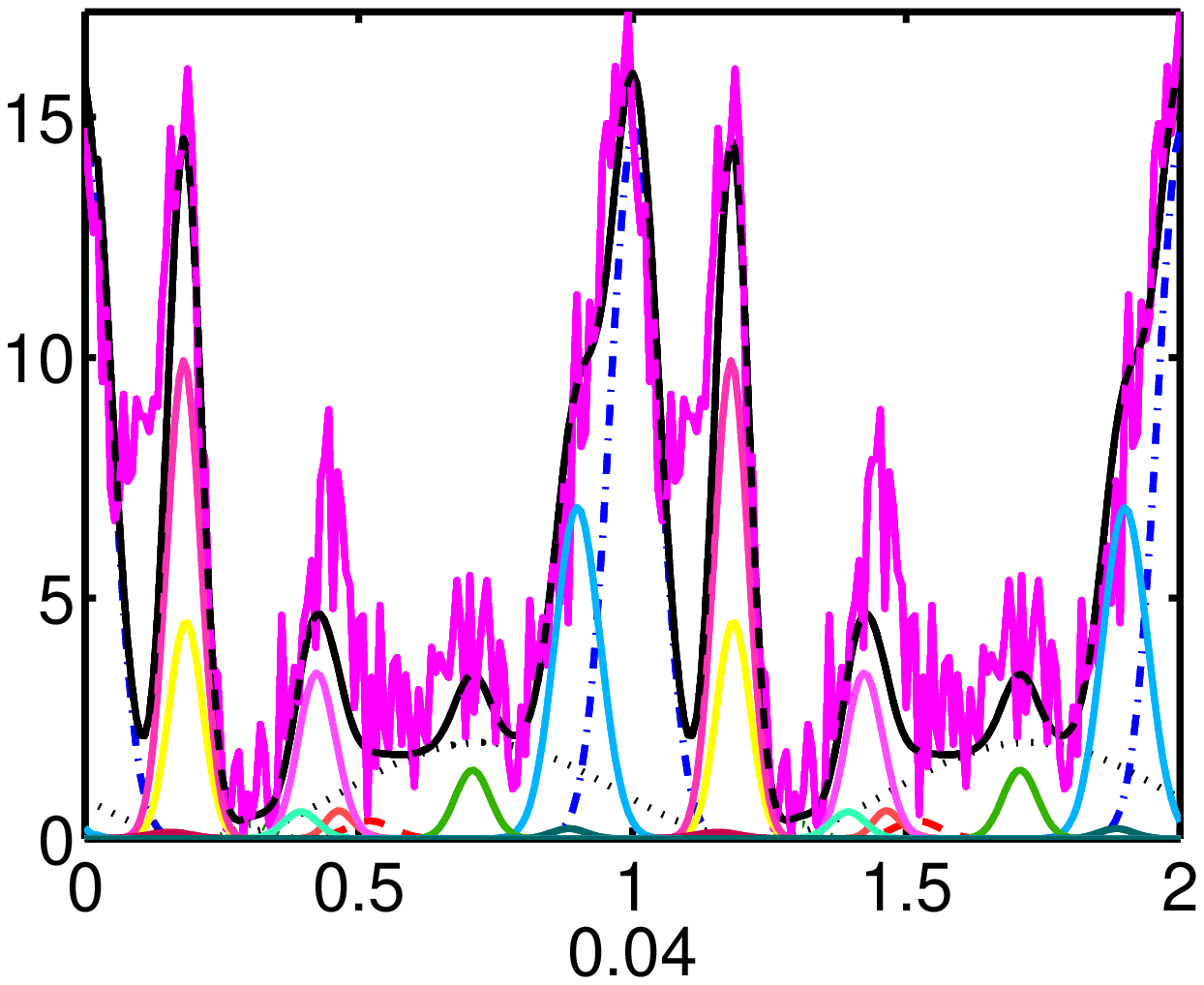}
\includegraphics[width=0.24\textwidth]{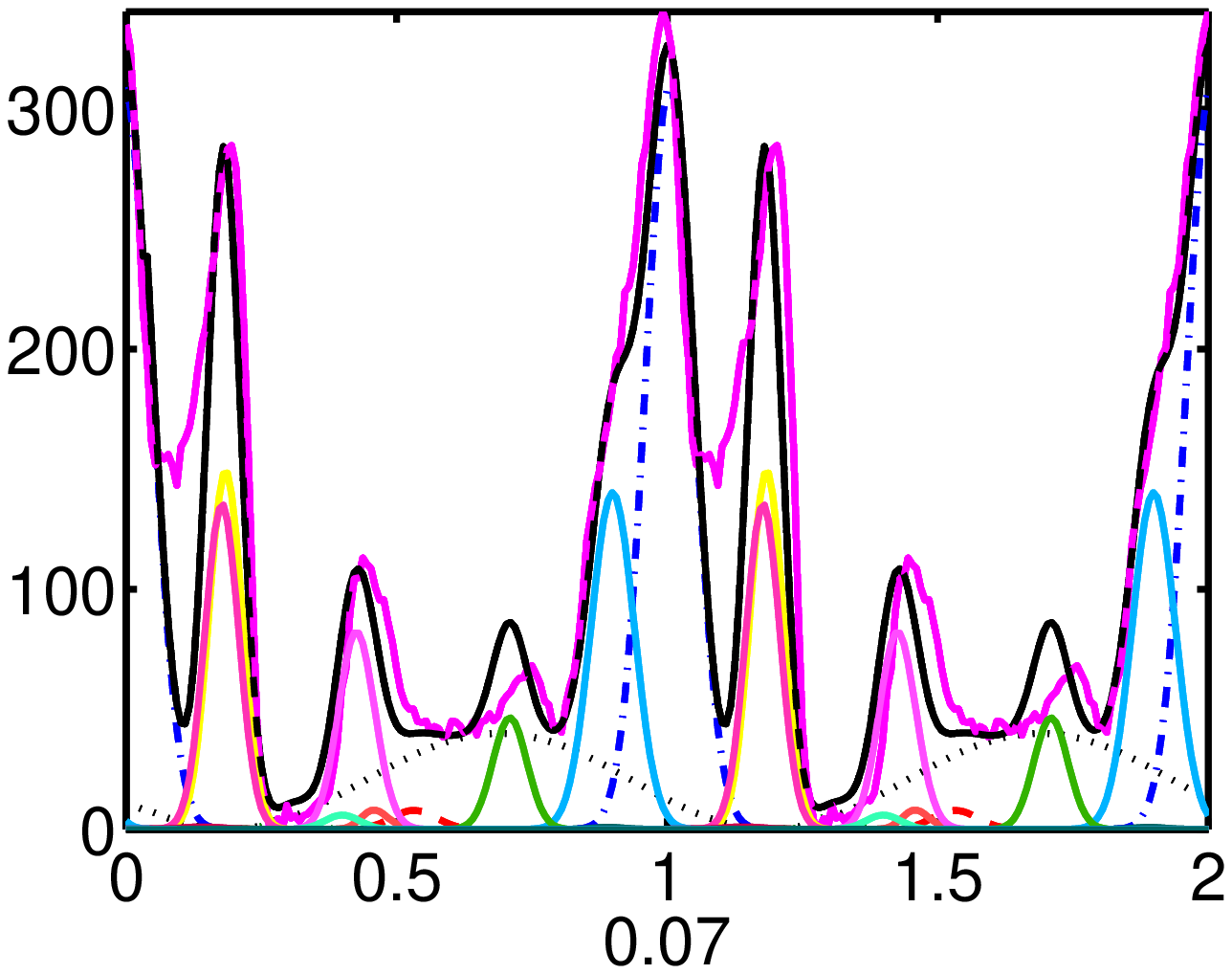}
\includegraphics[width=0.24\textwidth]{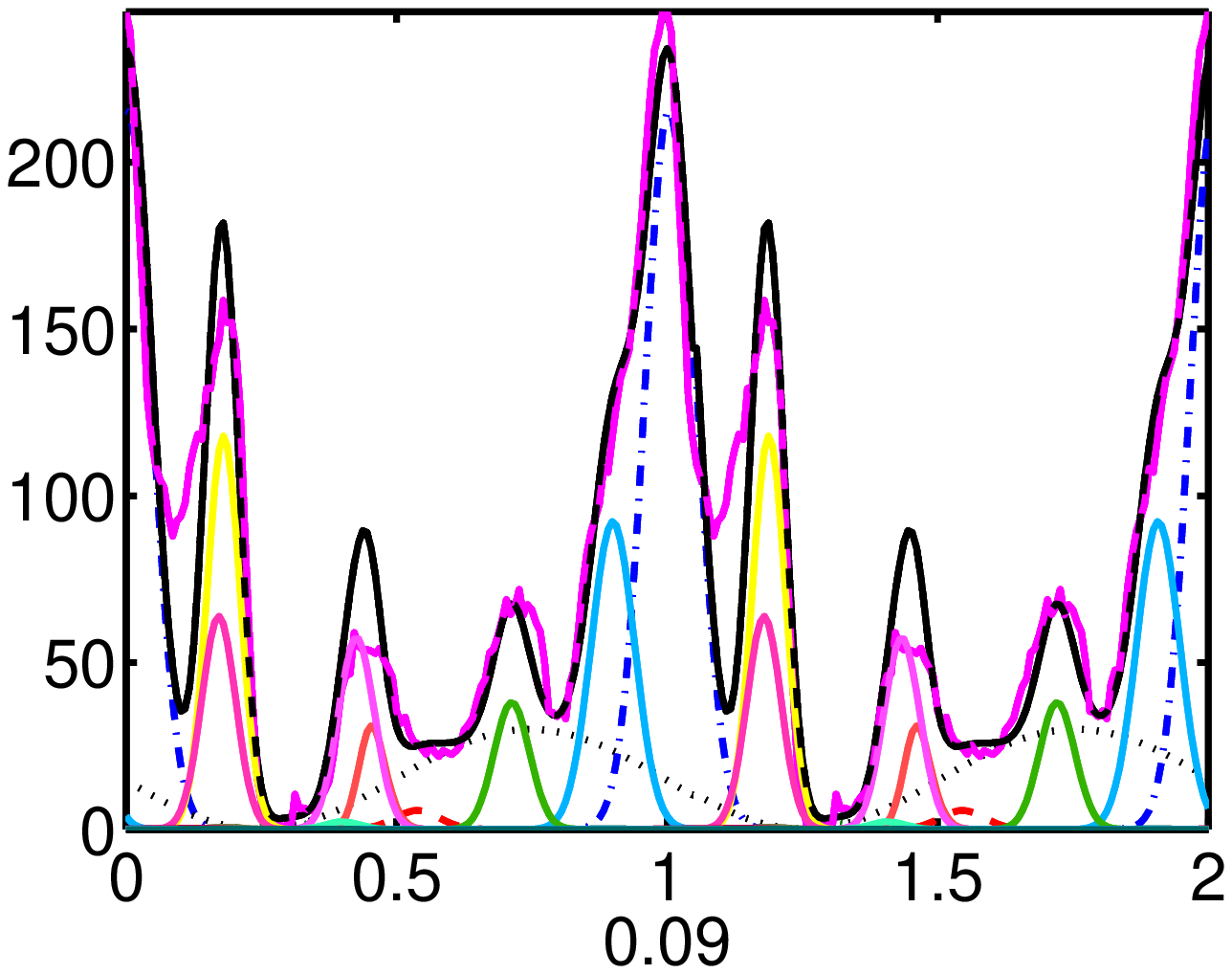}
\includegraphics[width=0.24\textwidth]{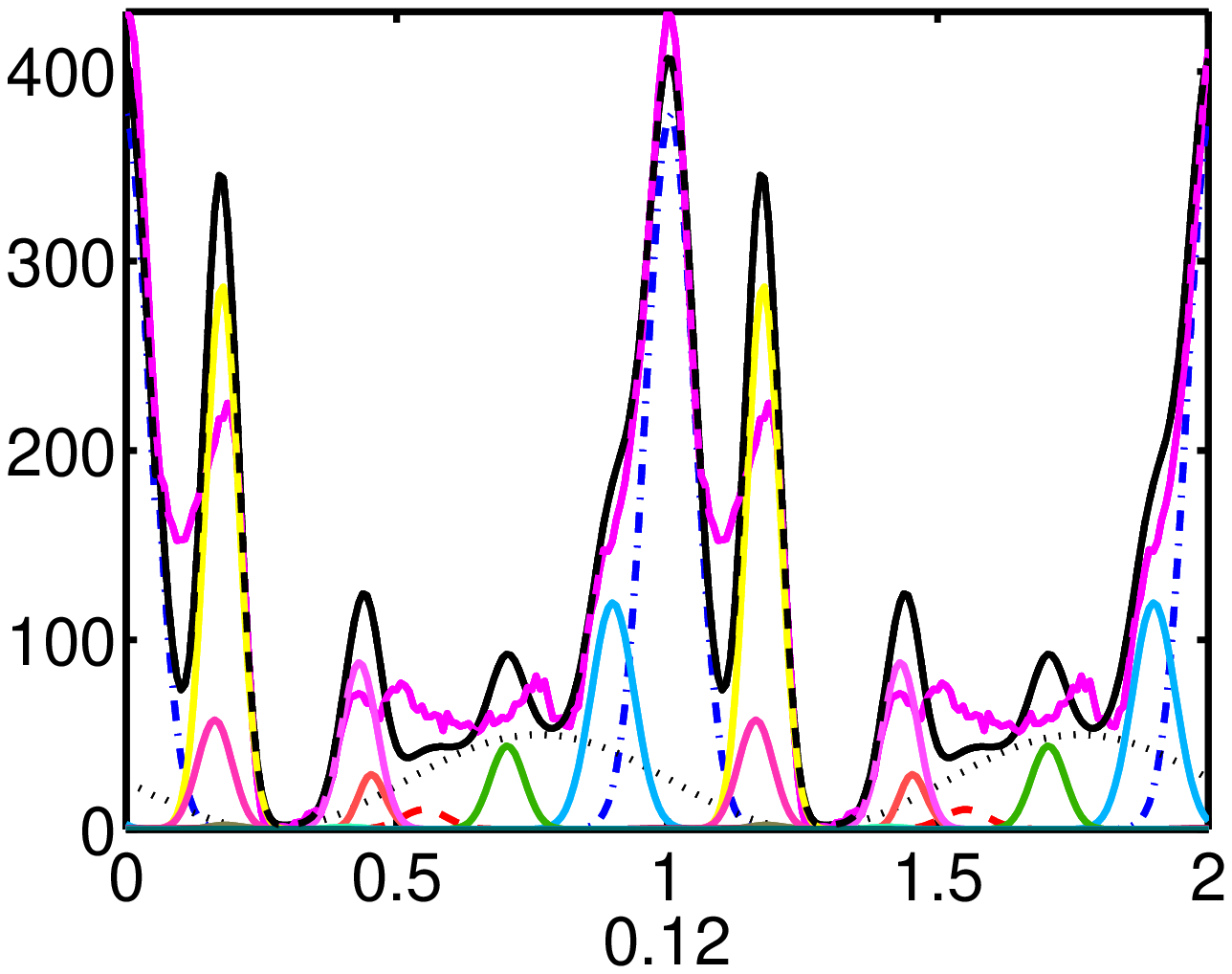}
\includegraphics[width=0.24\textwidth]{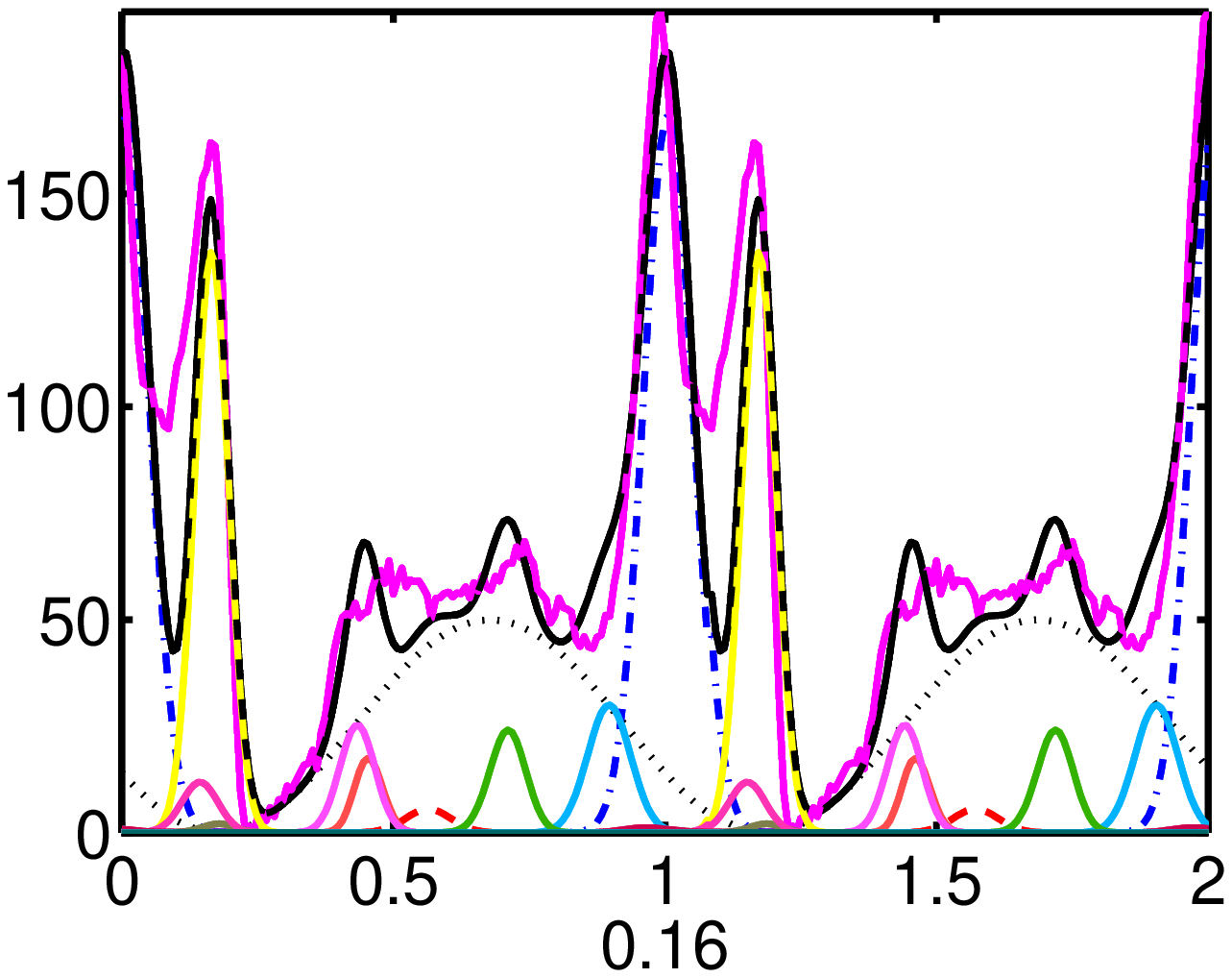}
\includegraphics[width=0.24\textwidth]{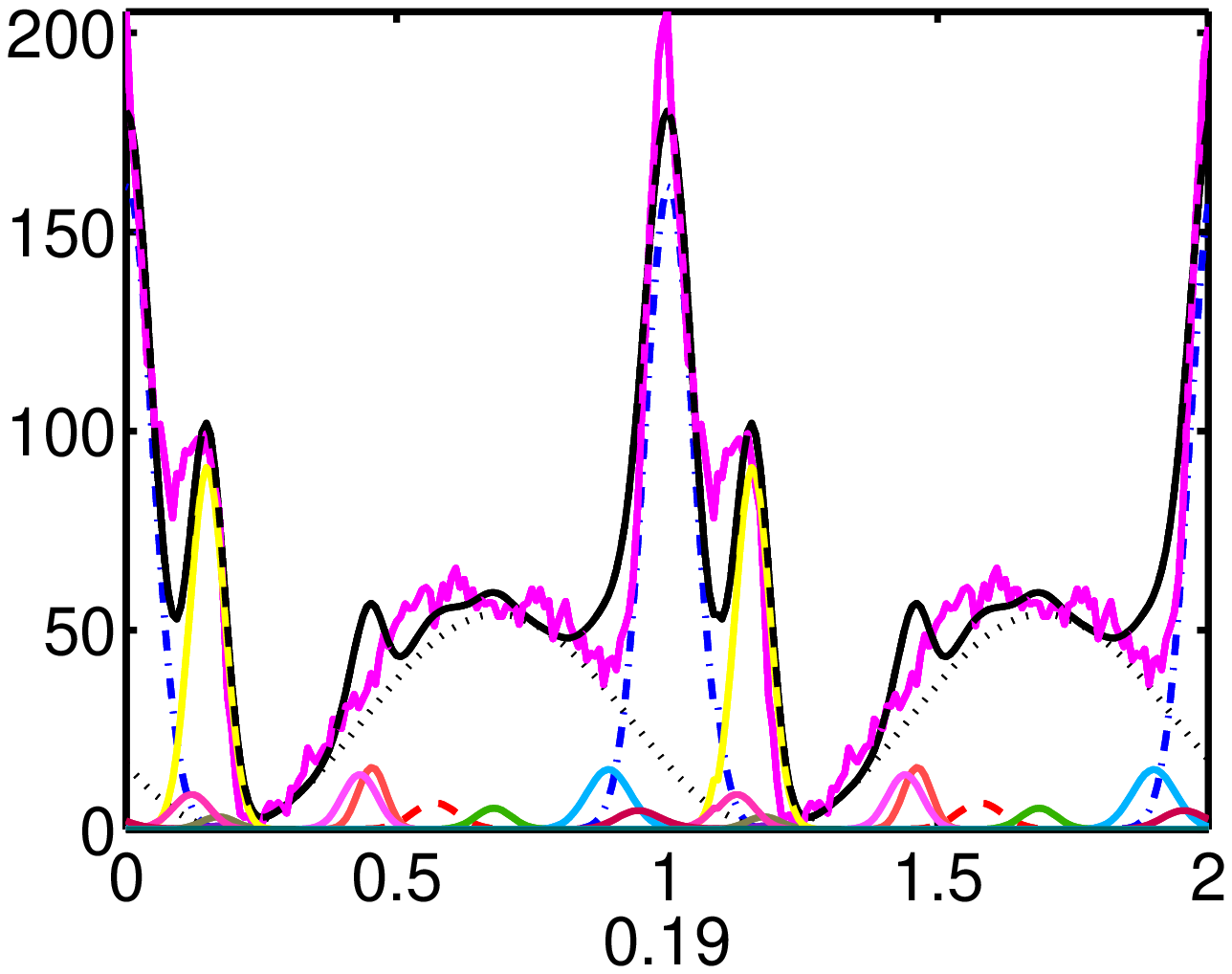}
\includegraphics[width=0.24\textwidth]{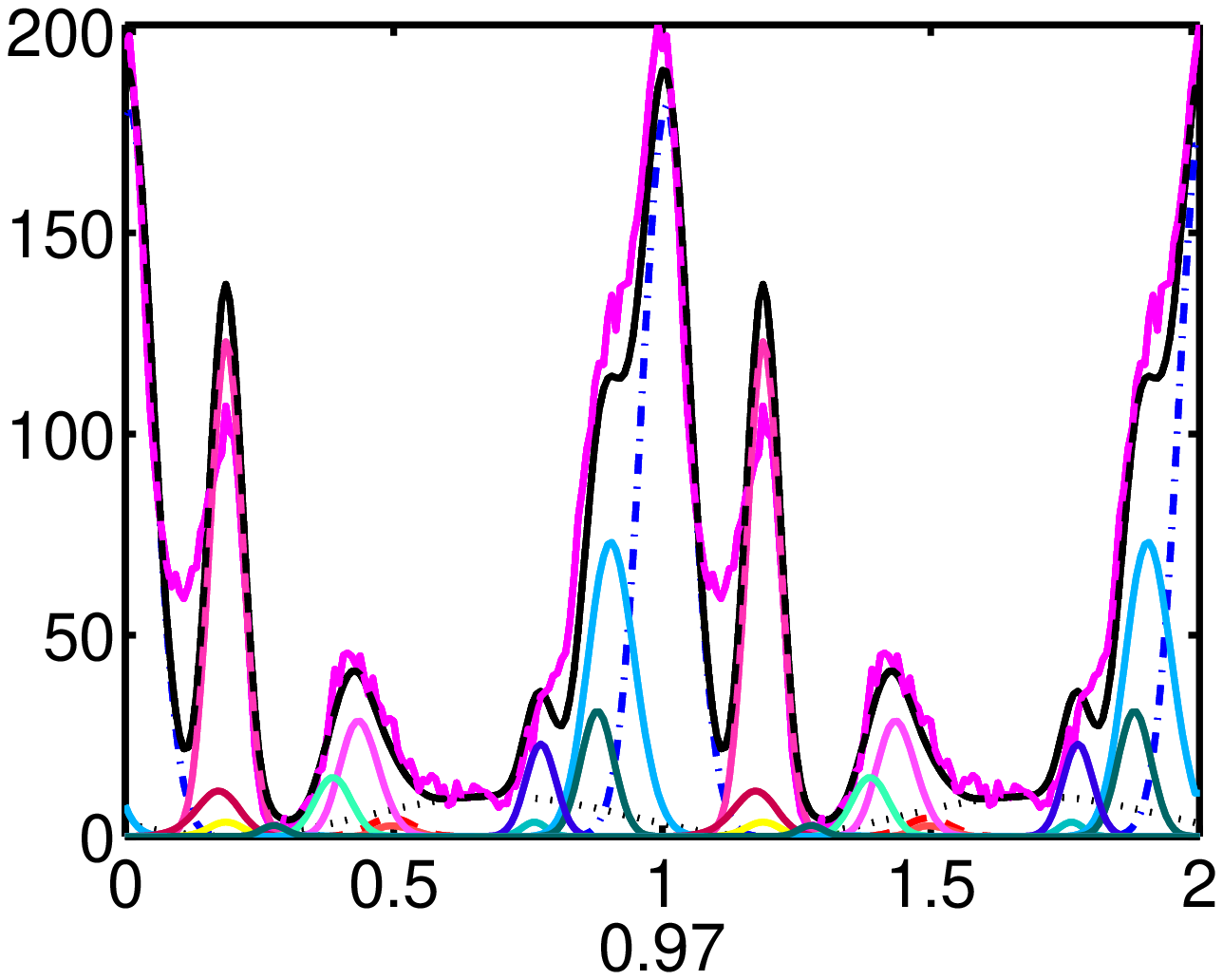}
\caption{259 RXTE}
\label{f259}
\end{figure}

\begin{figure}
%\centering
\includegraphics[width=0.24\textwidth]{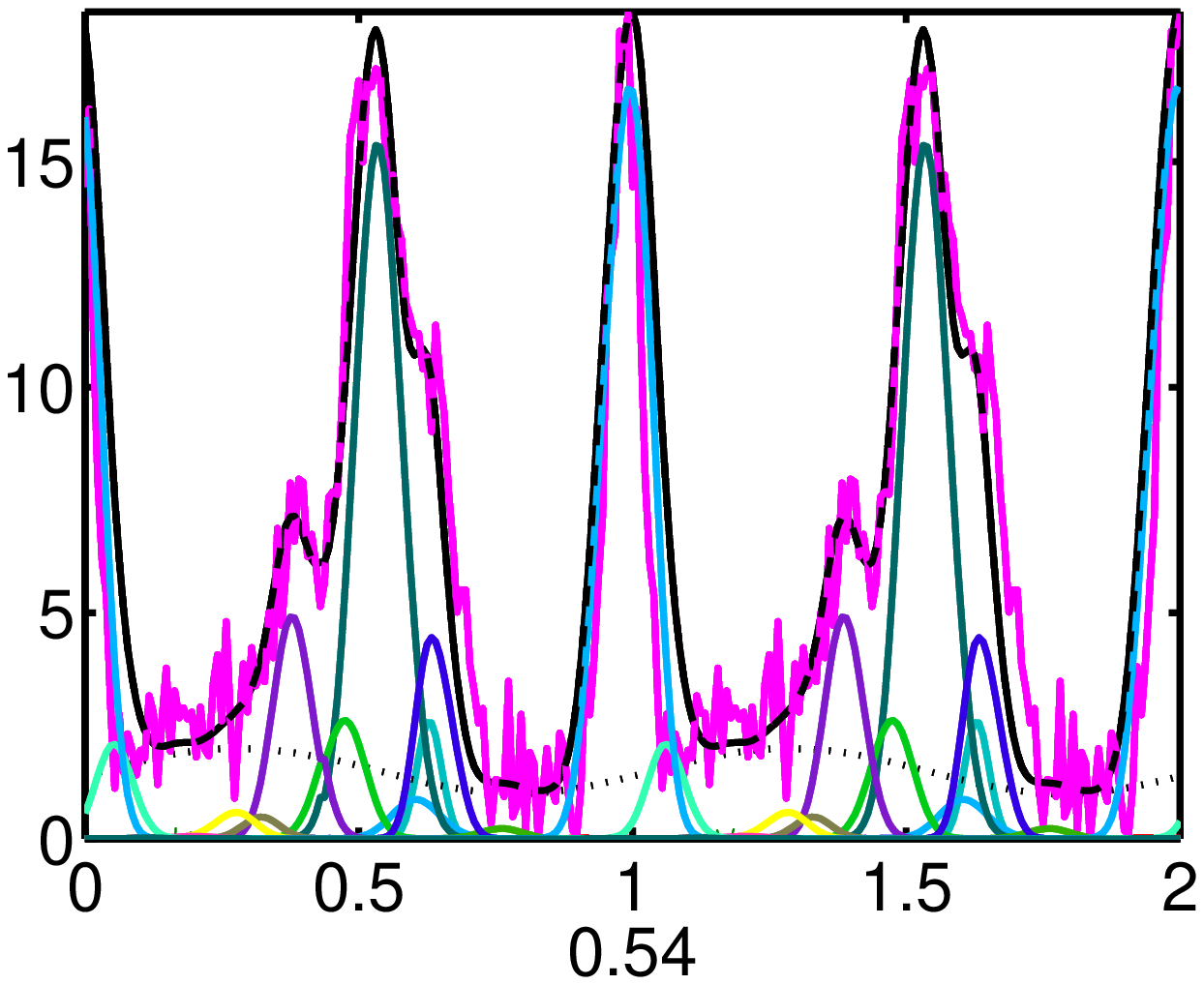}
\includegraphics[width=0.24\textwidth]{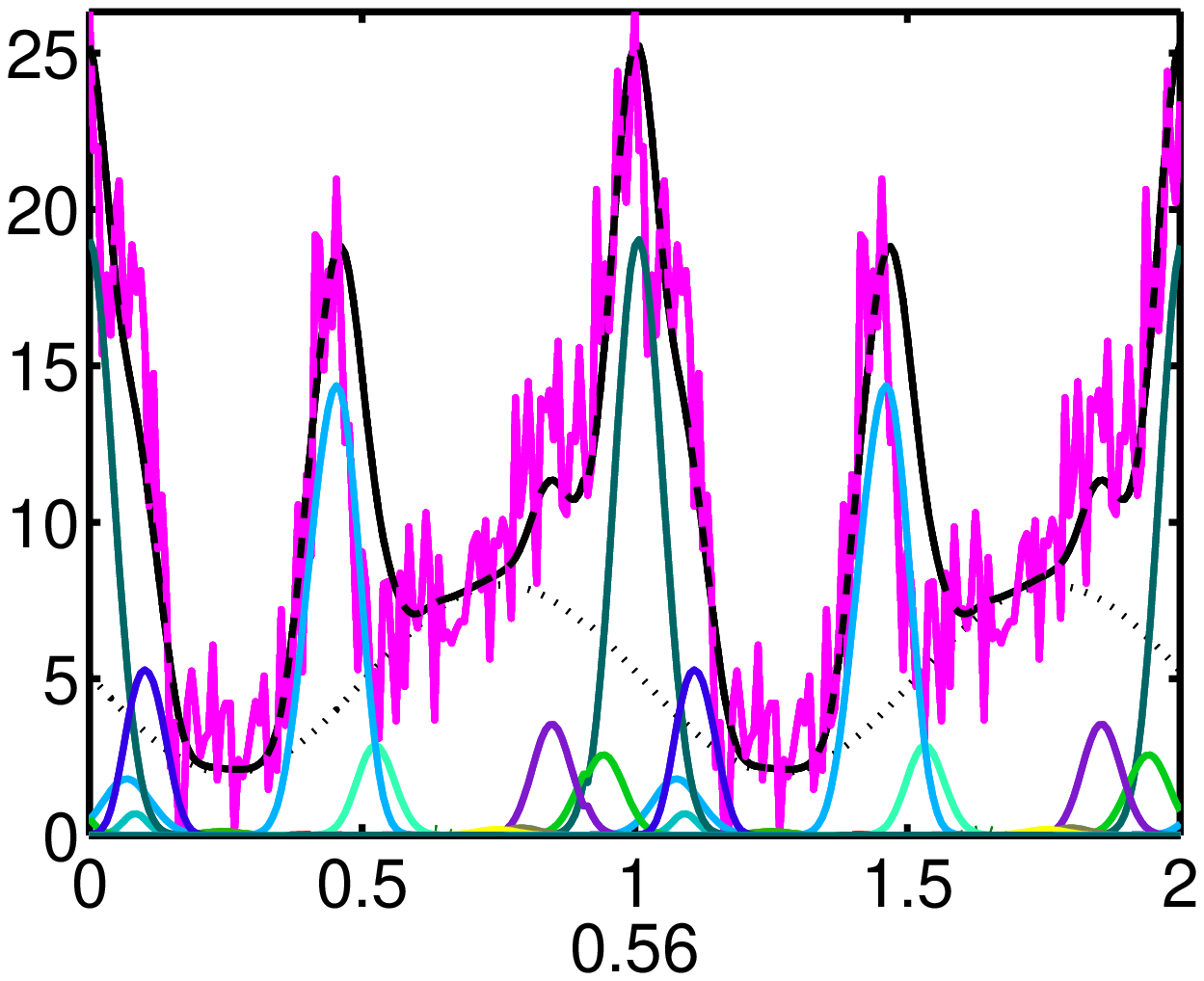}
\includegraphics[width=0.24\textwidth]{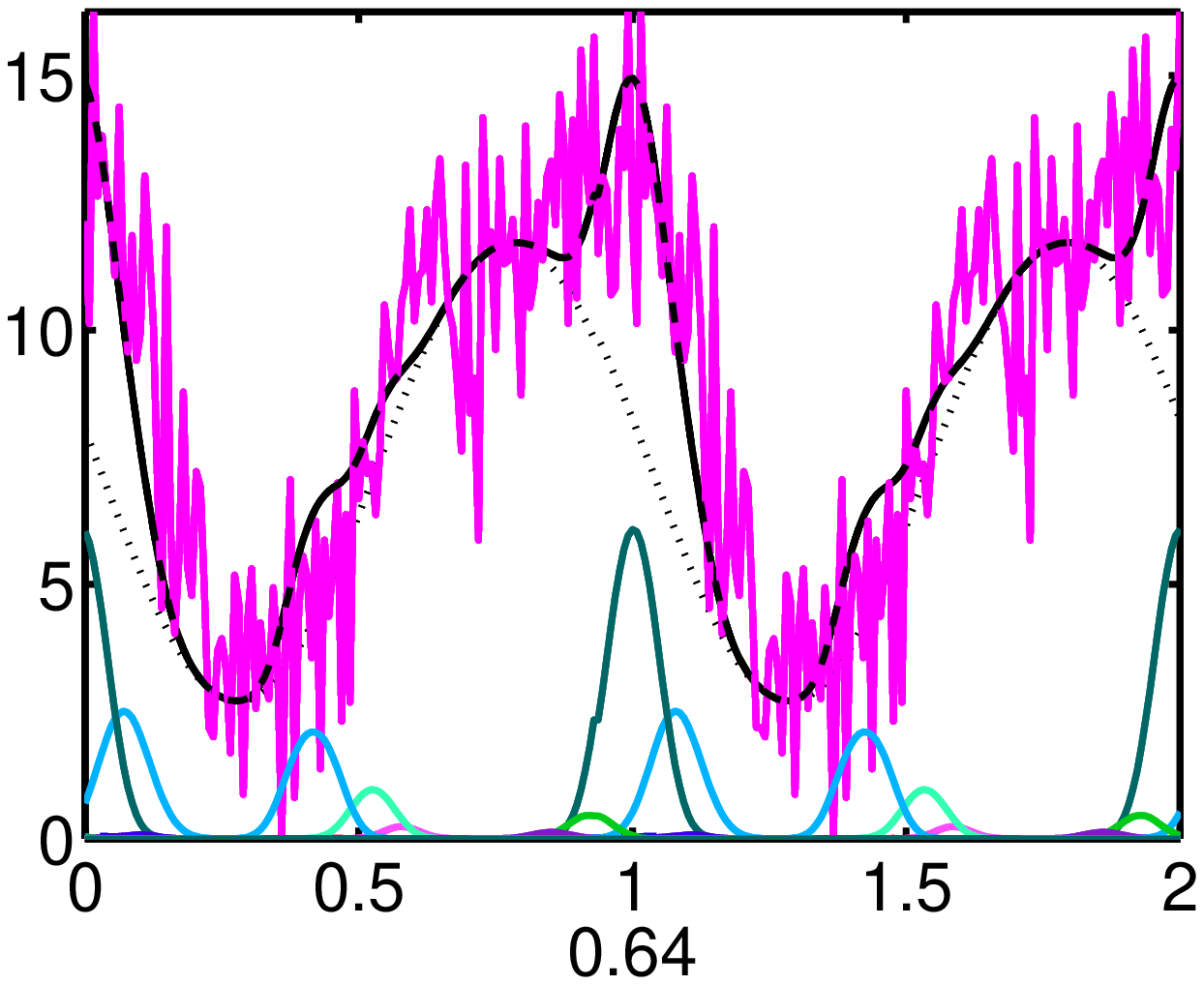}
\caption{268 RXTE}
\label{f268}
\end{figure}

\begin{figure}
%\centering
\includegraphics[width=0.24\textwidth]{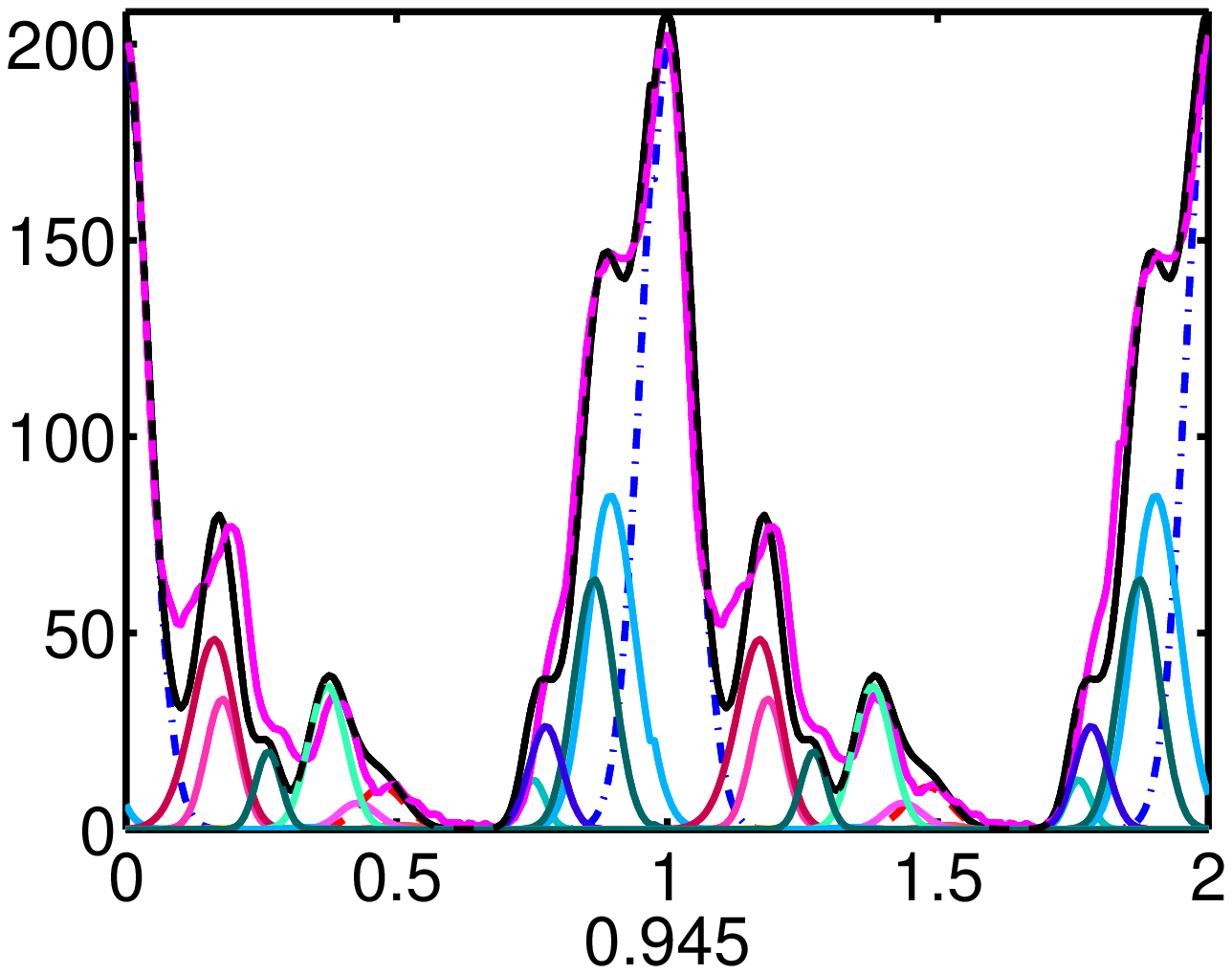}
\includegraphics[width=0.24\textwidth]{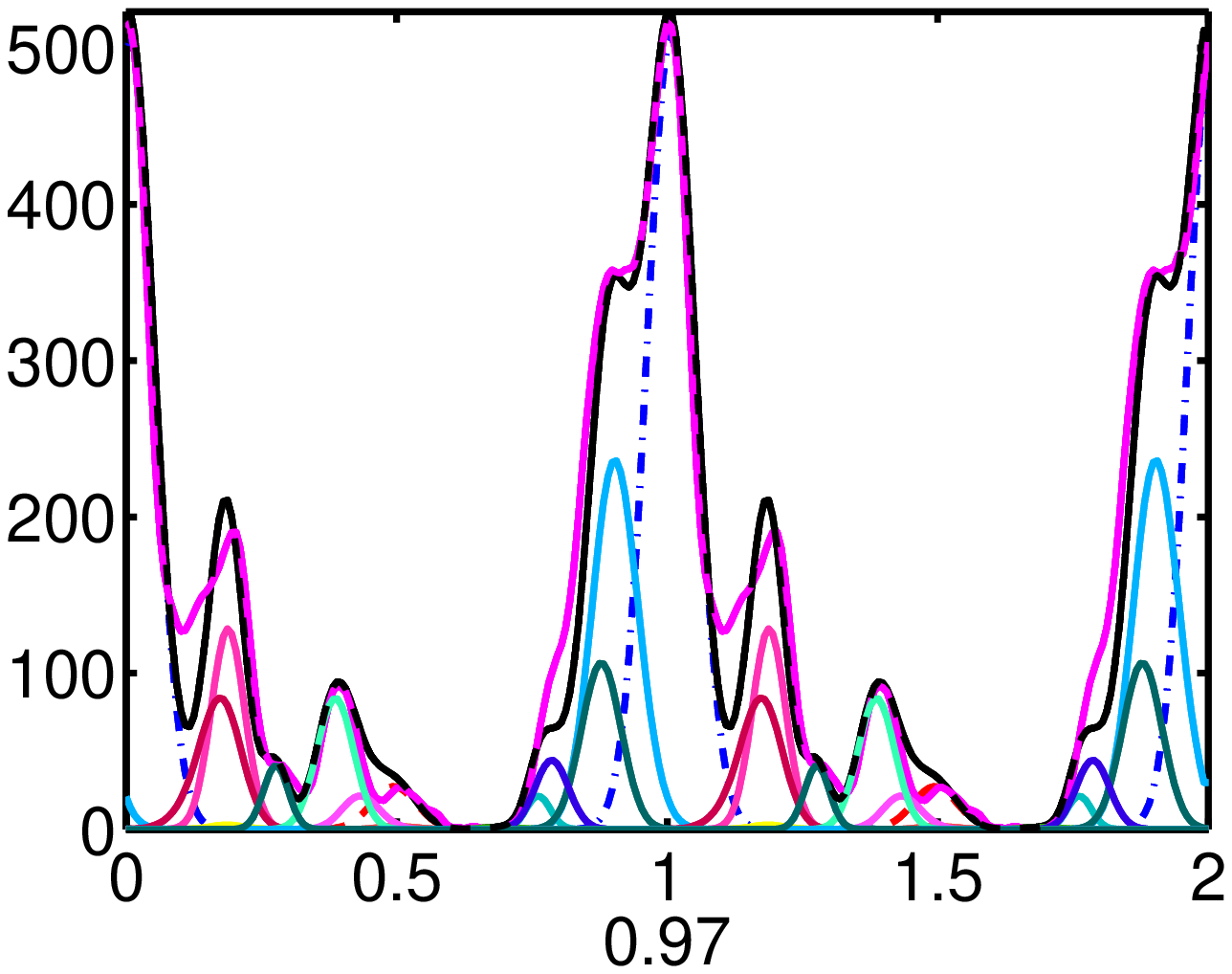}
\includegraphics[width=0.24\textwidth]{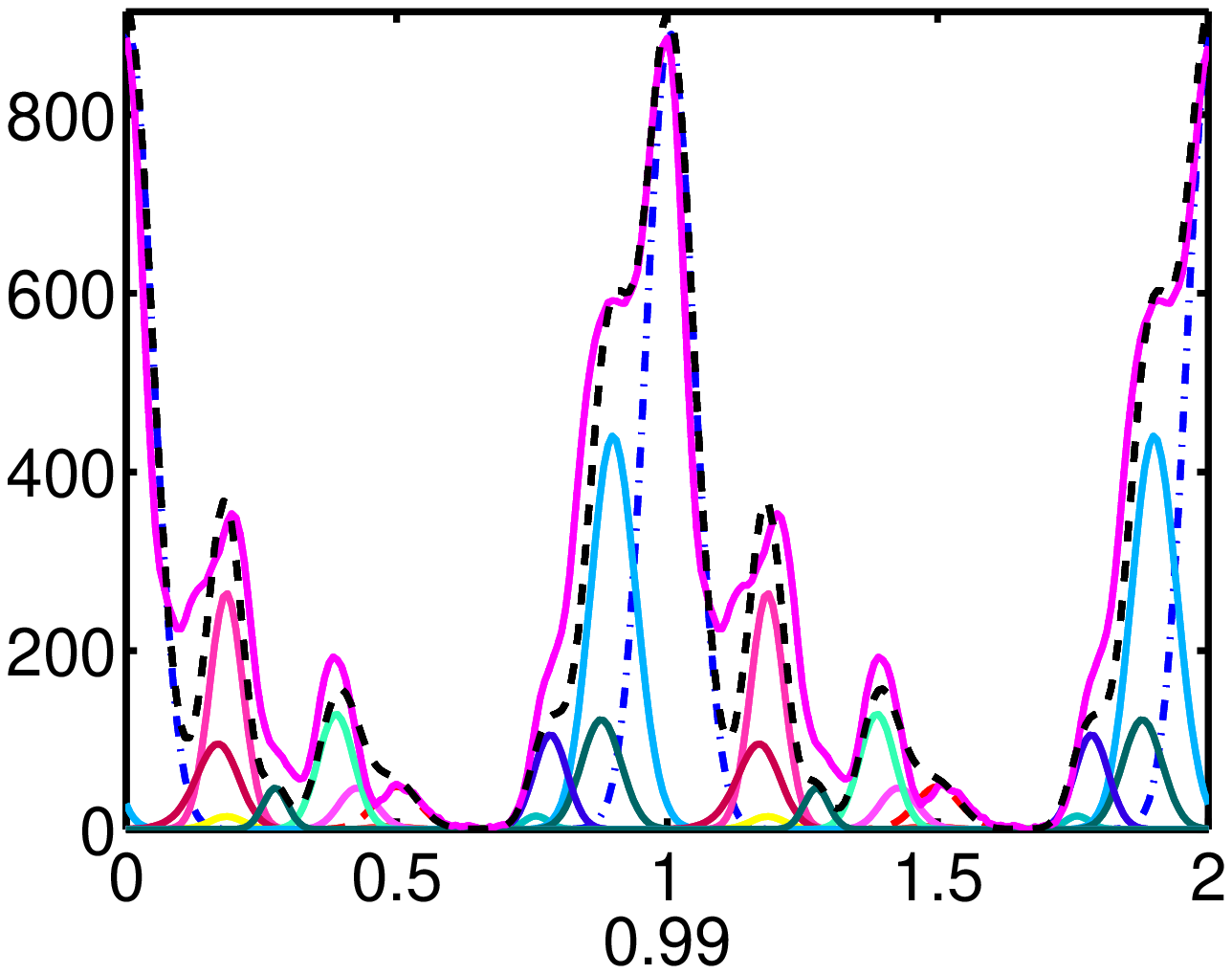}
\includegraphics[width=0.24\textwidth]{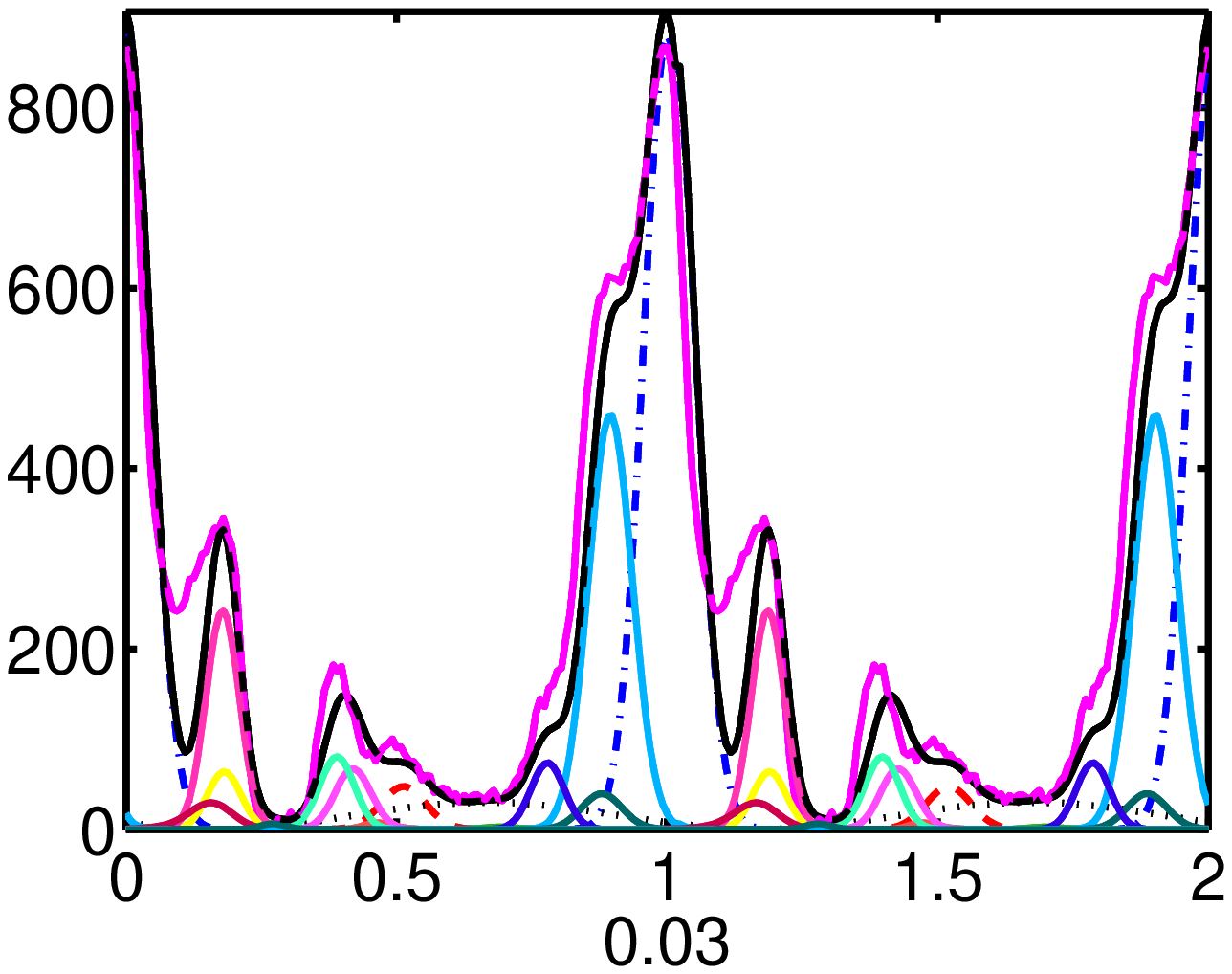}
\includegraphics[width=0.24\textwidth]{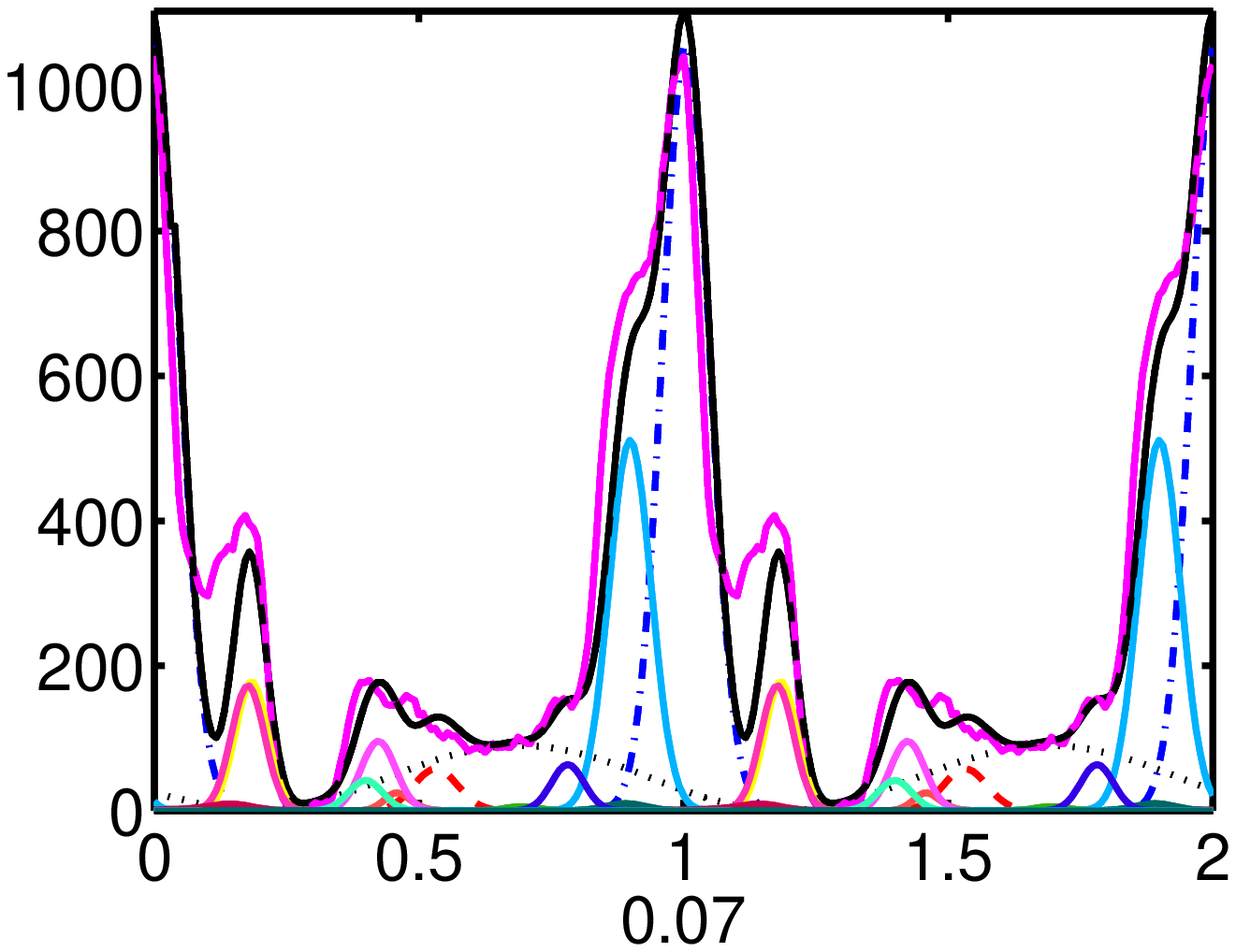}
\includegraphics[width=0.24\textwidth]{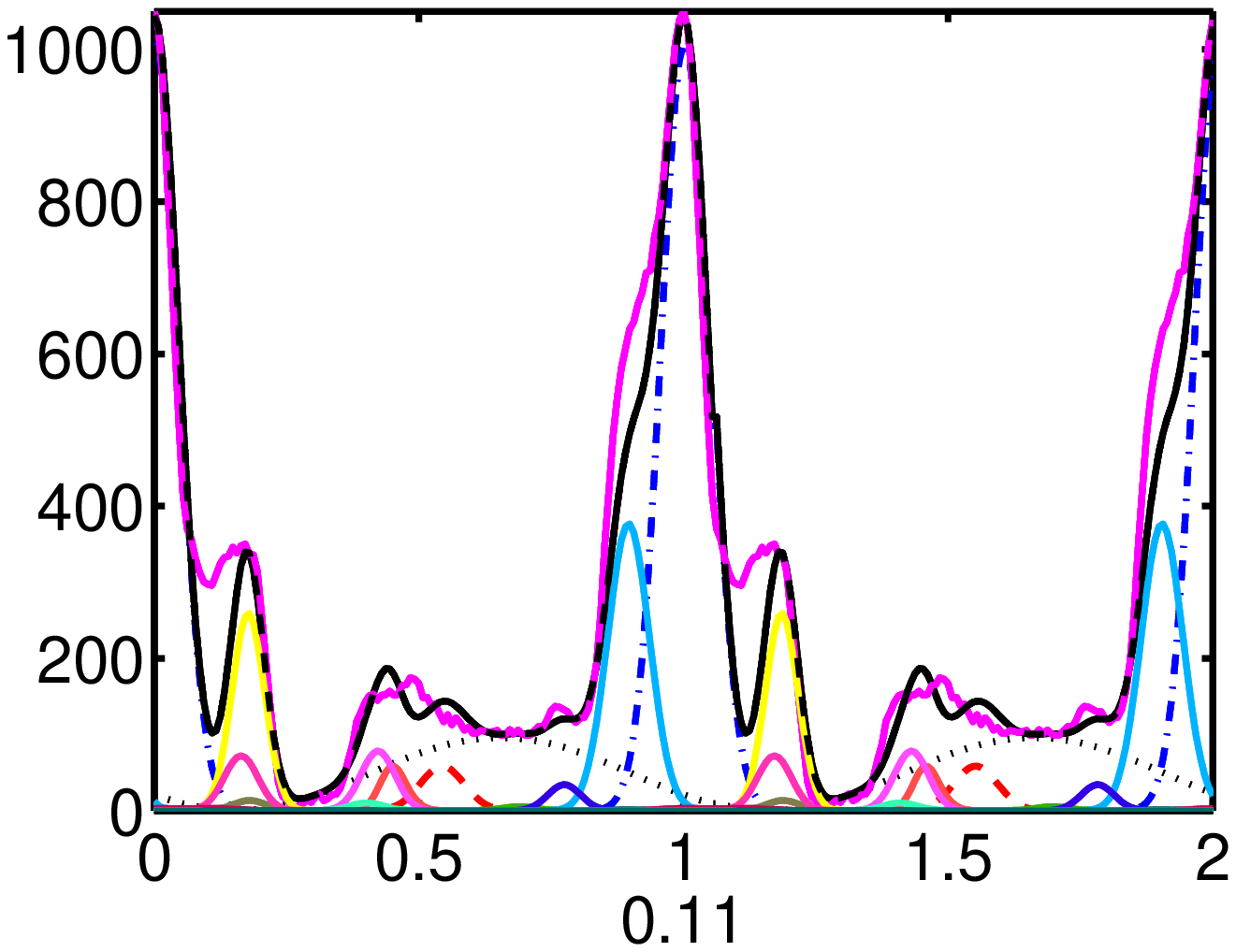}
\includegraphics[width=0.24\textwidth]{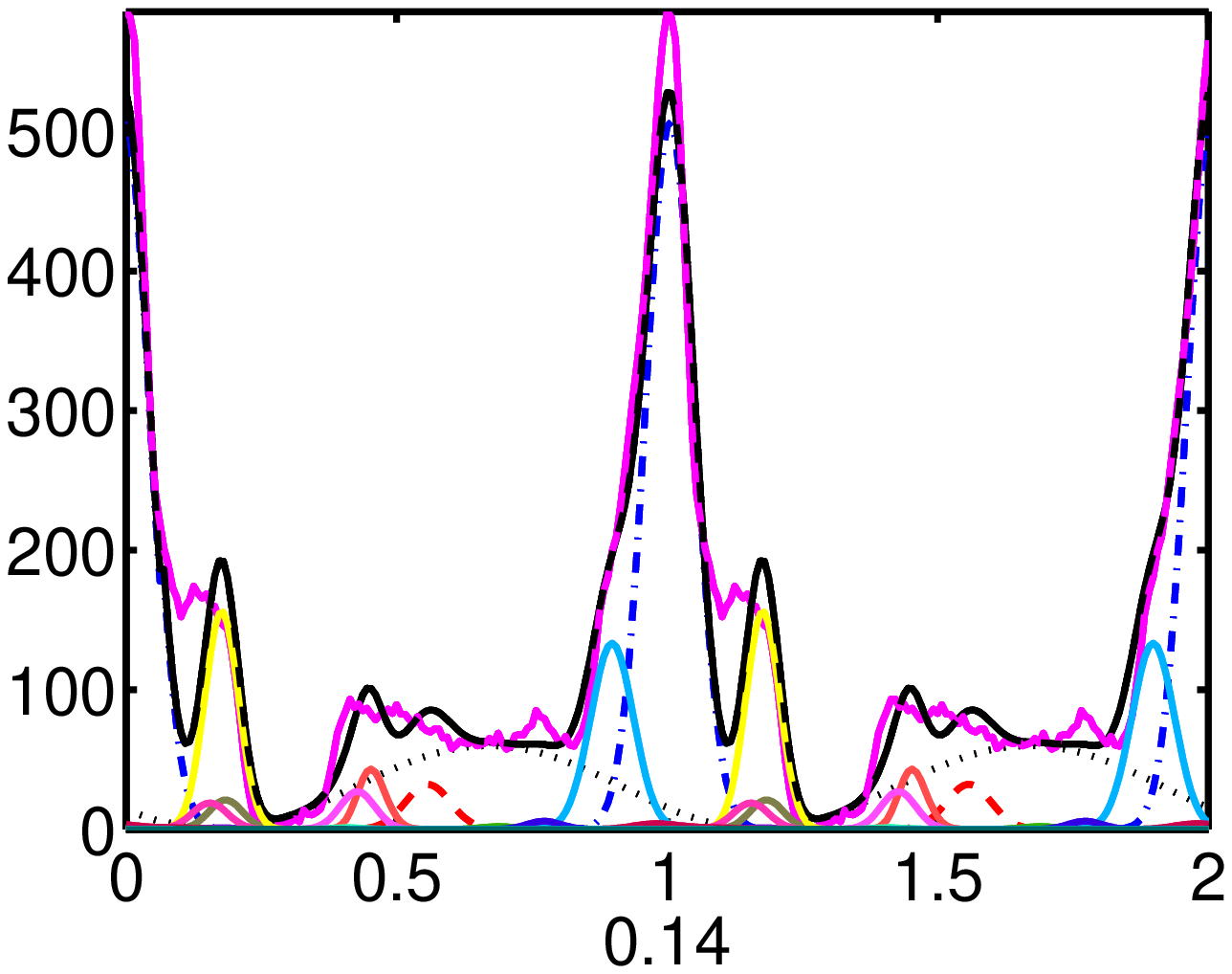}
\includegraphics[width=0.24\textwidth]{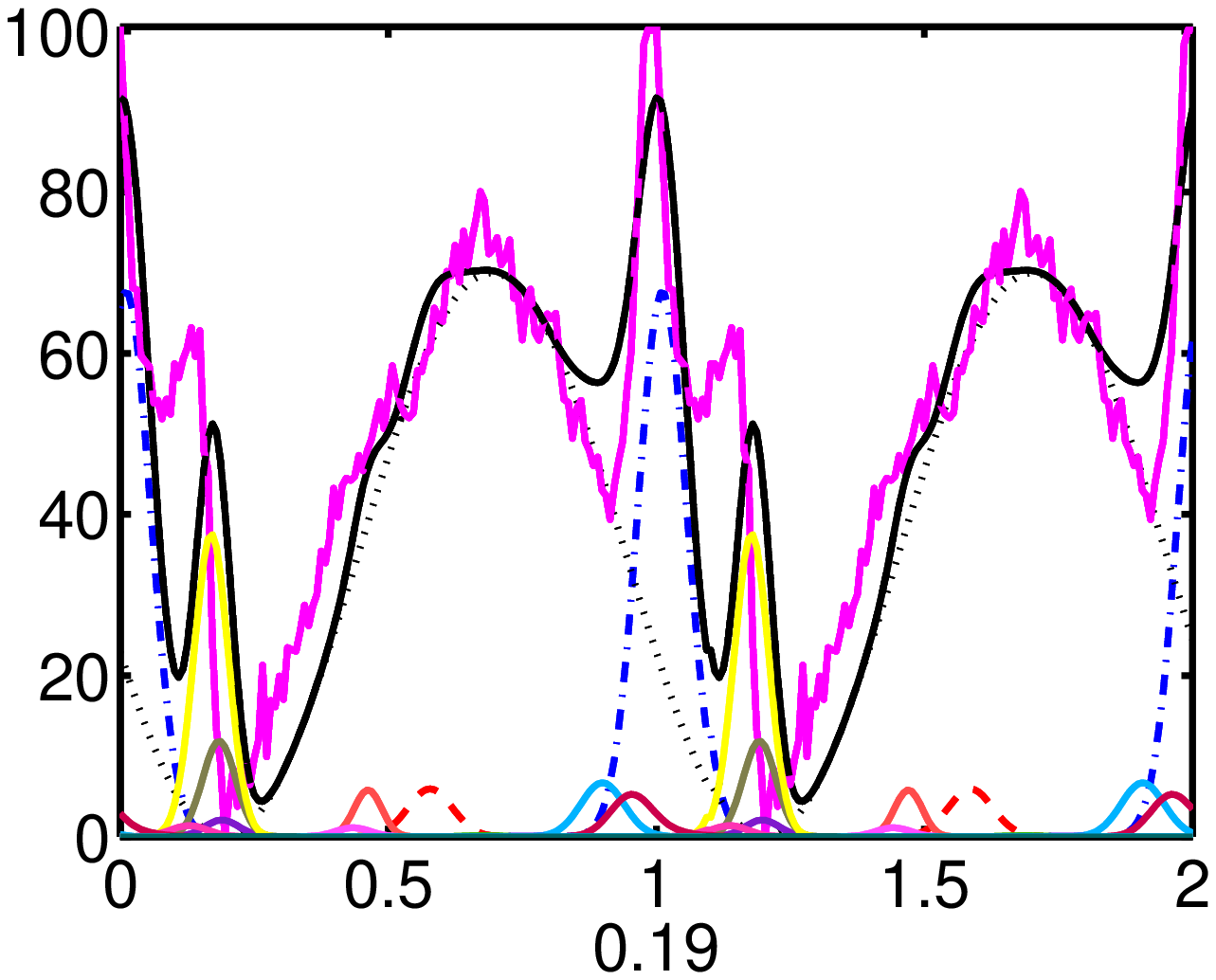}
\caption{269 RXTE}
\label{f269}
\end{figure}

\begin{figure}
%\centering
\includegraphics[width=0.24\textwidth]{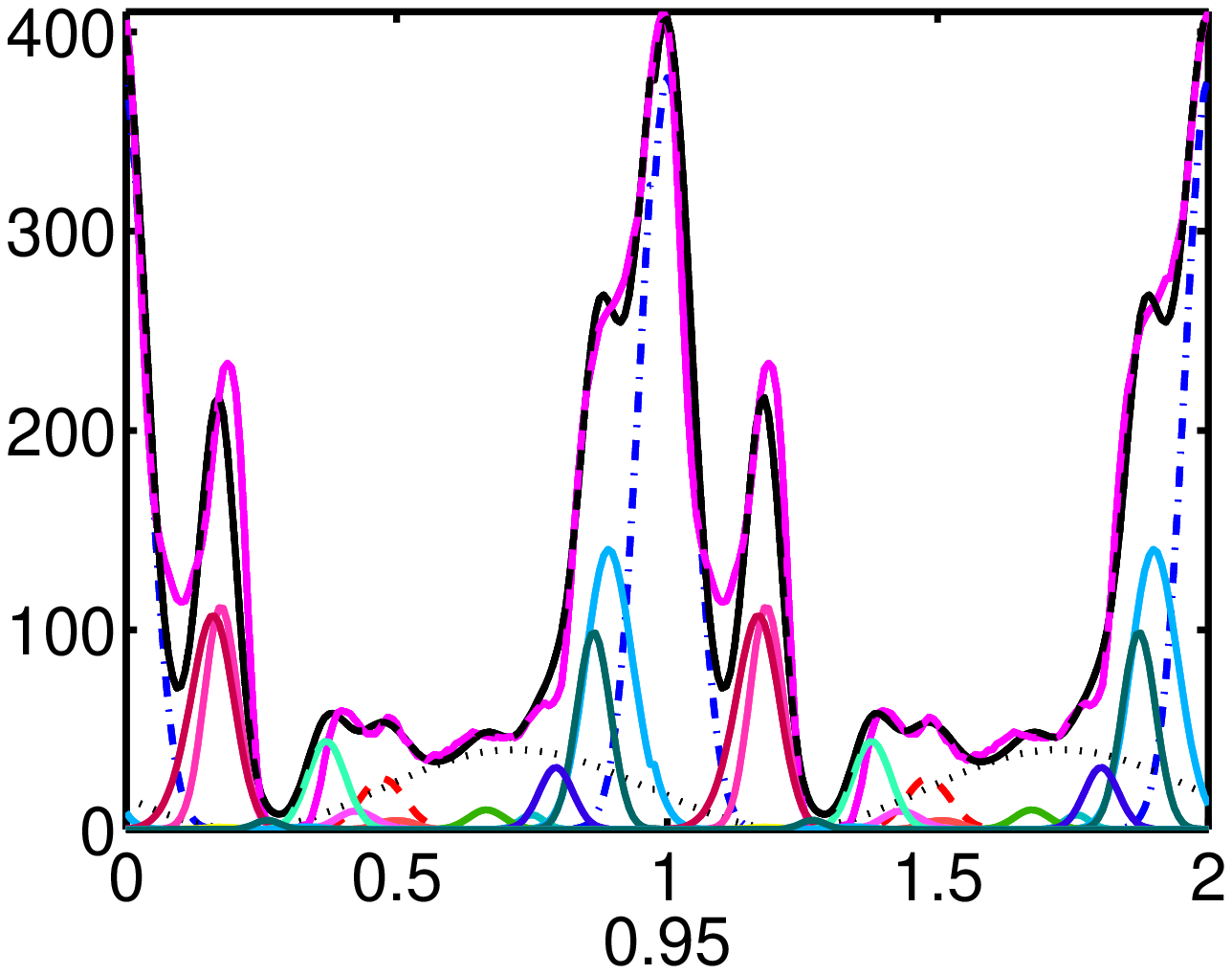}
\caption{271 RXTE}
\label{f271}
\end{figure}

\begin{figure}
%\centering
\includegraphics[width=0.24\textwidth]{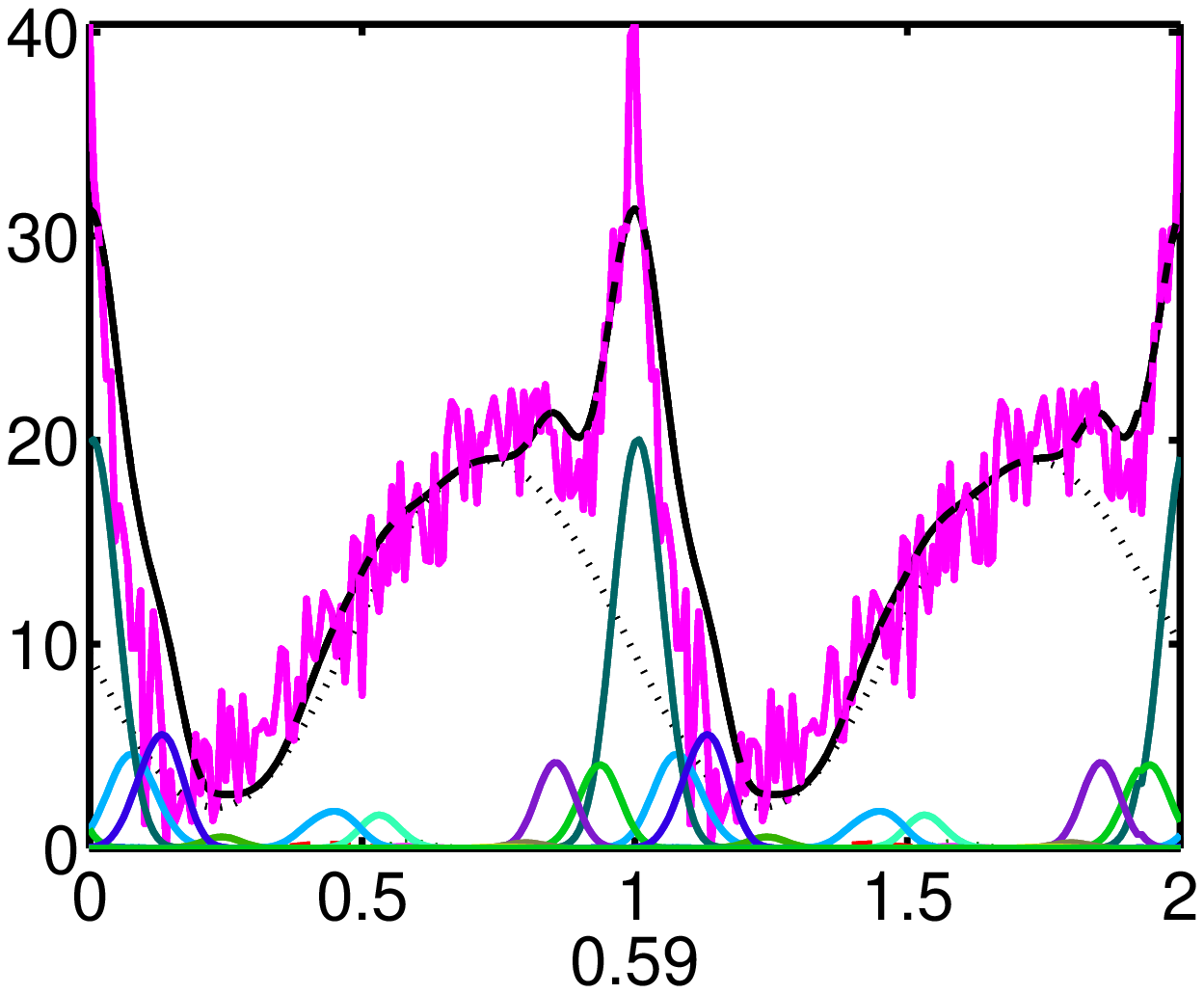}
\includegraphics[width=0.24\textwidth]{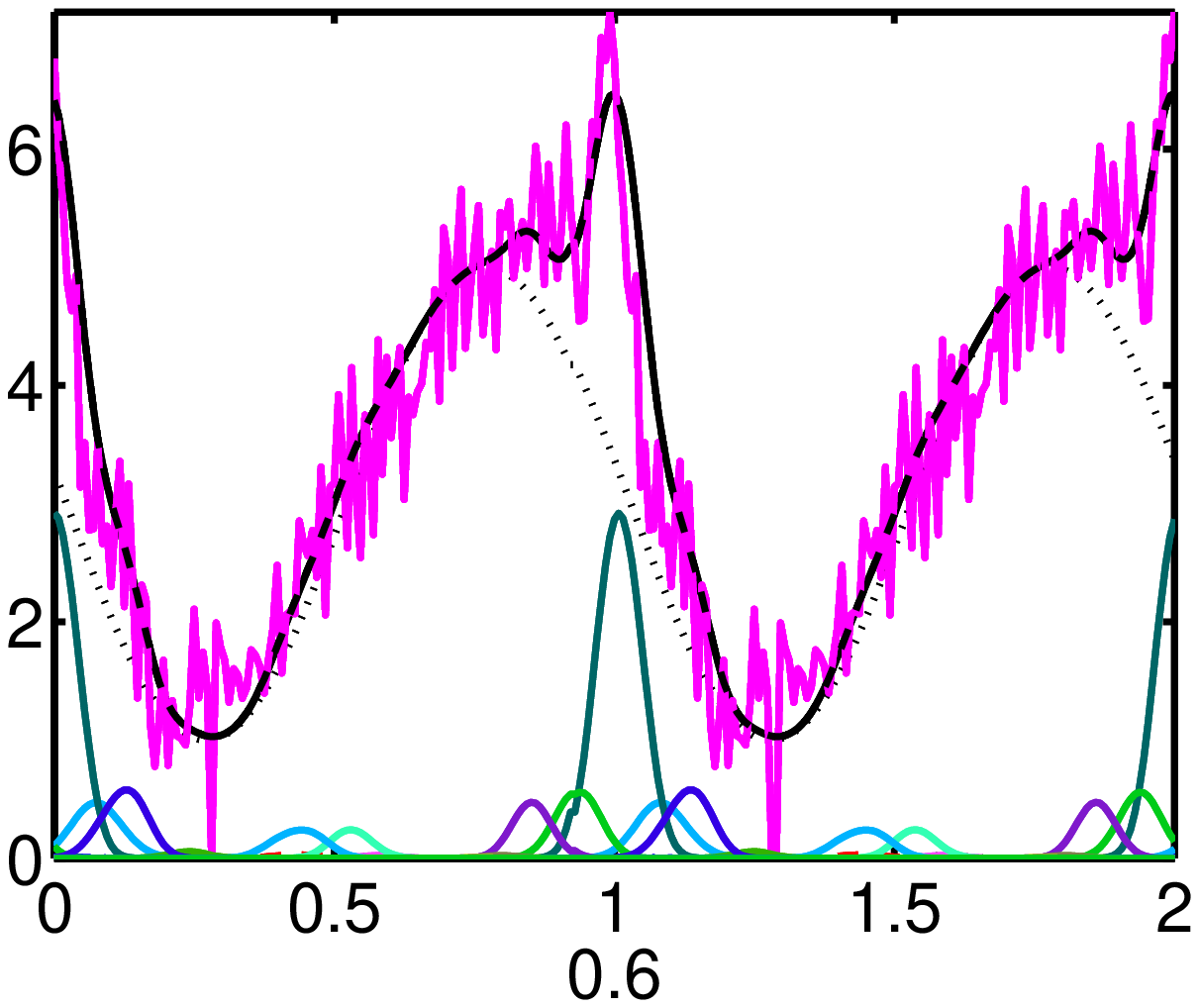}
\includegraphics[width=0.24\textwidth]{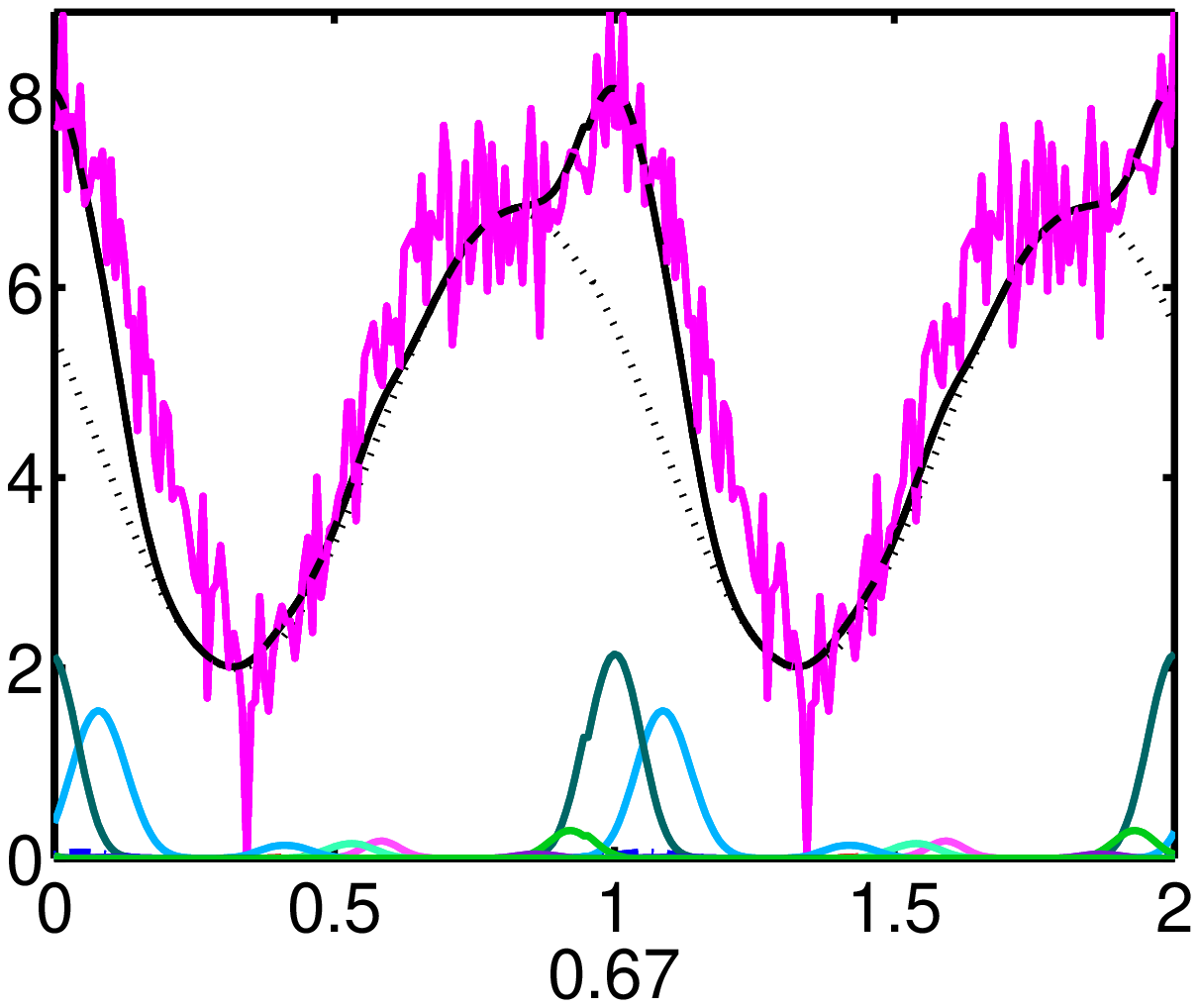}
\caption{277 RXTE}
\label{f277}
\end{figure}

\begin{figure}
%\centering
\includegraphics[width=0.24\textwidth]{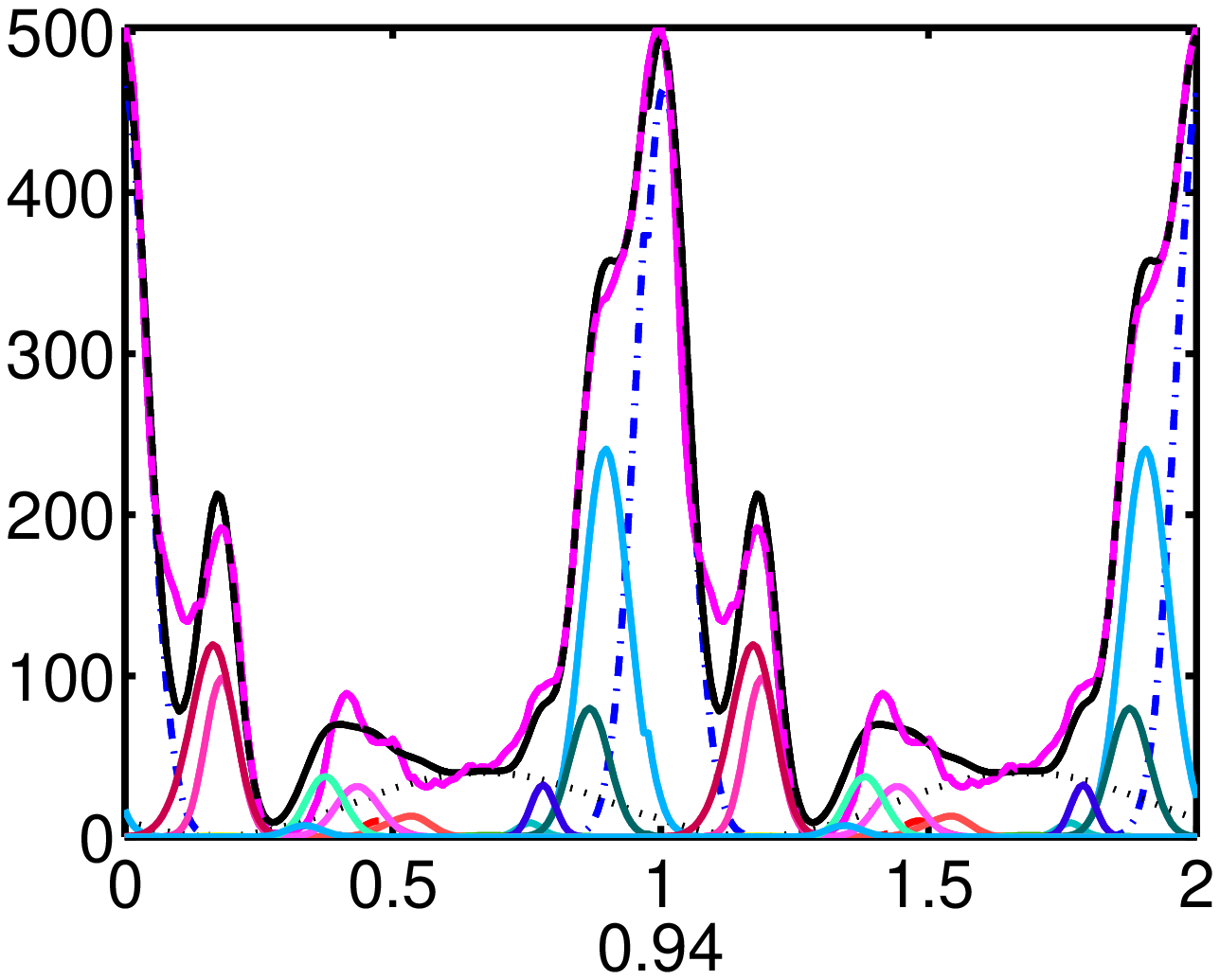}
\includegraphics[width=0.24\textwidth]{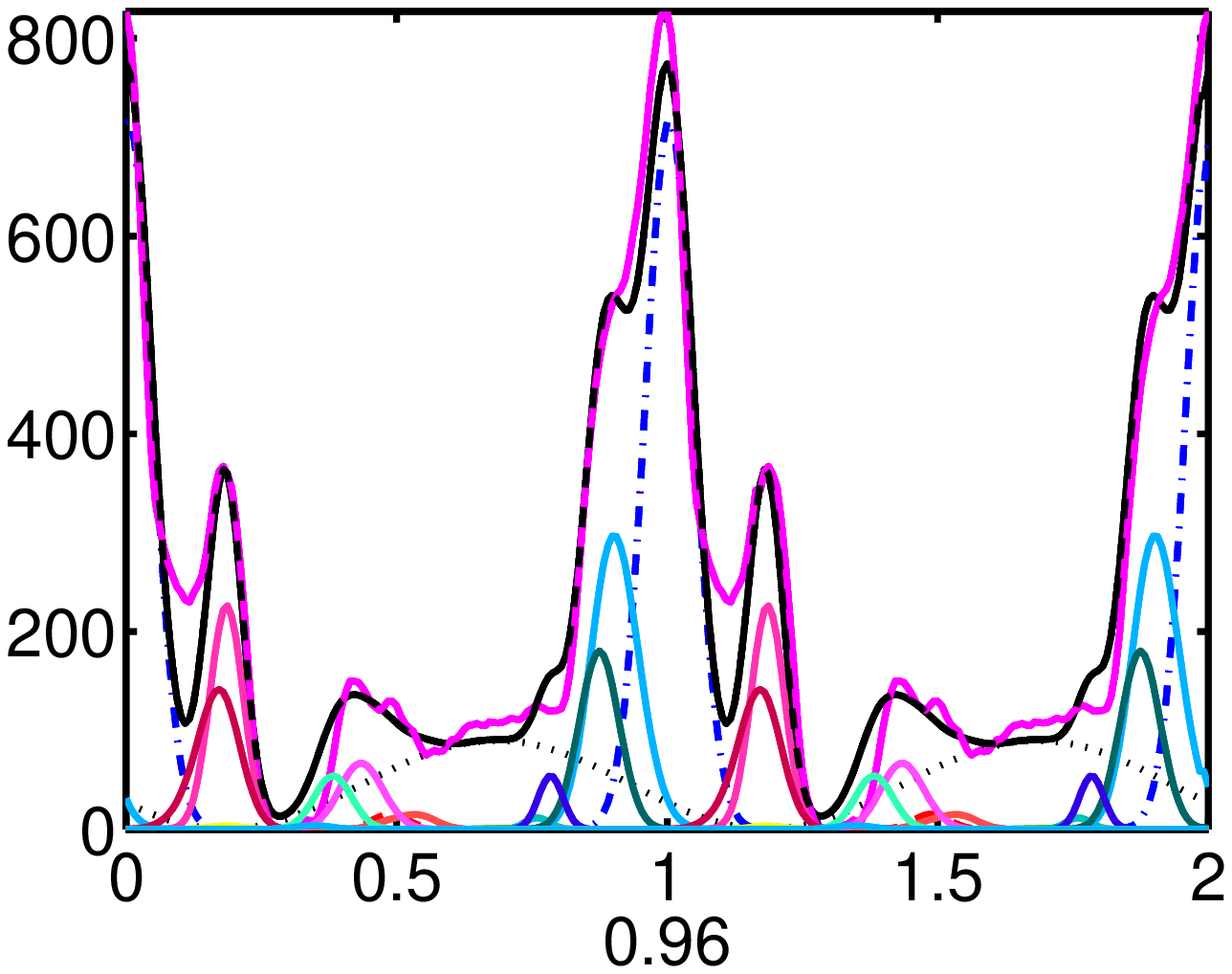}
\includegraphics[width=0.24\textwidth]{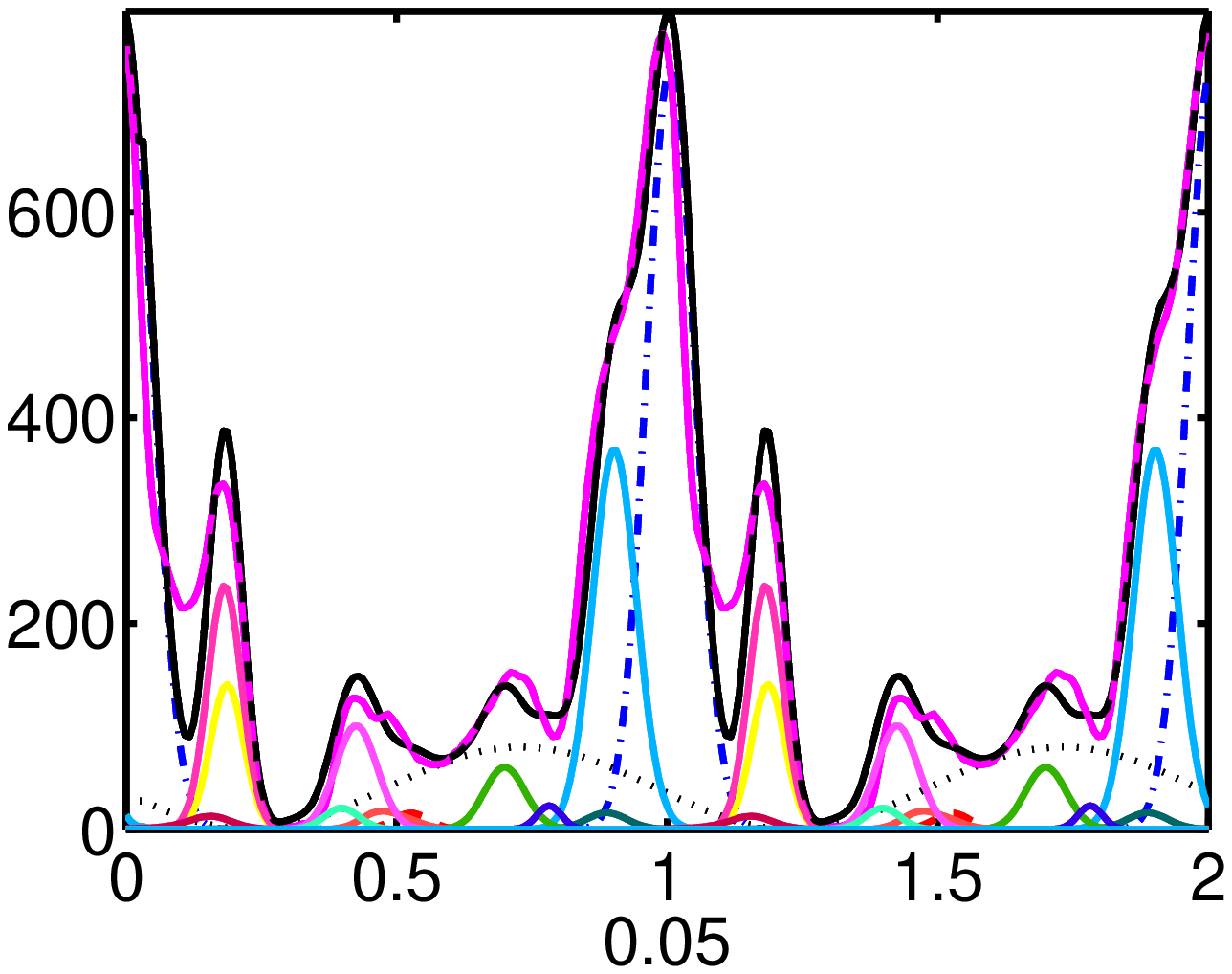}
\includegraphics[width=0.24\textwidth]{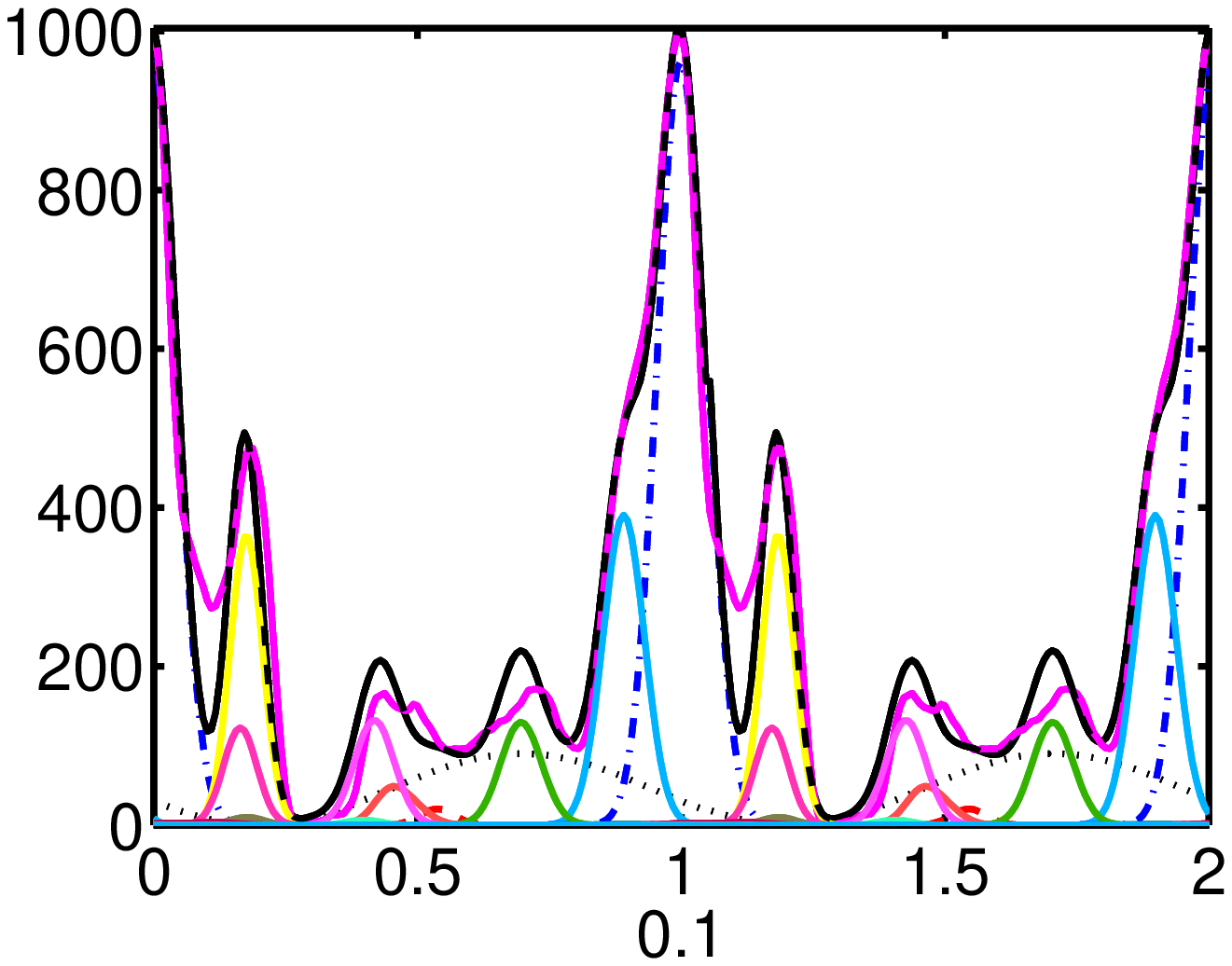}
\includegraphics[width=0.24\textwidth]{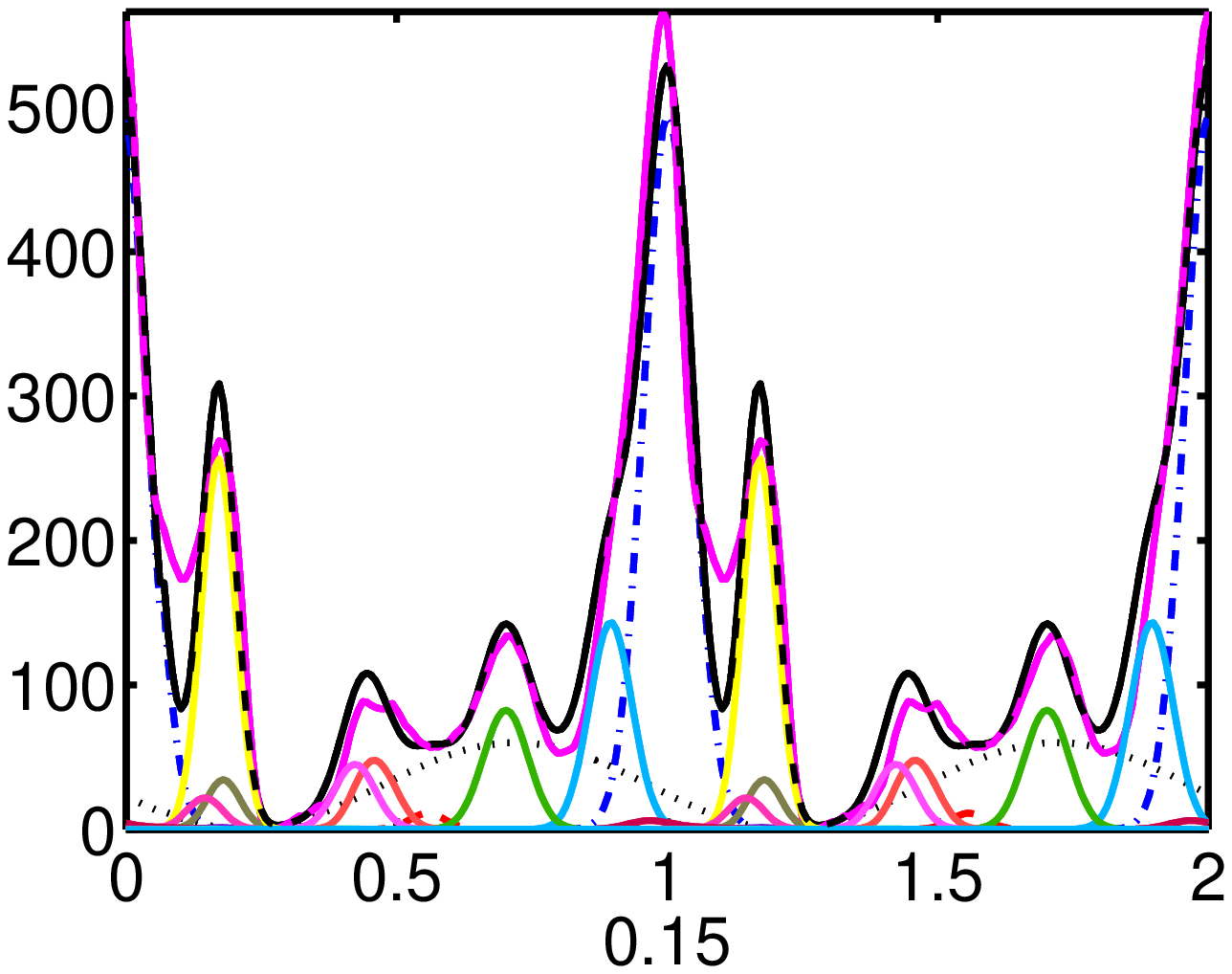}
\includegraphics[width=0.24\textwidth]{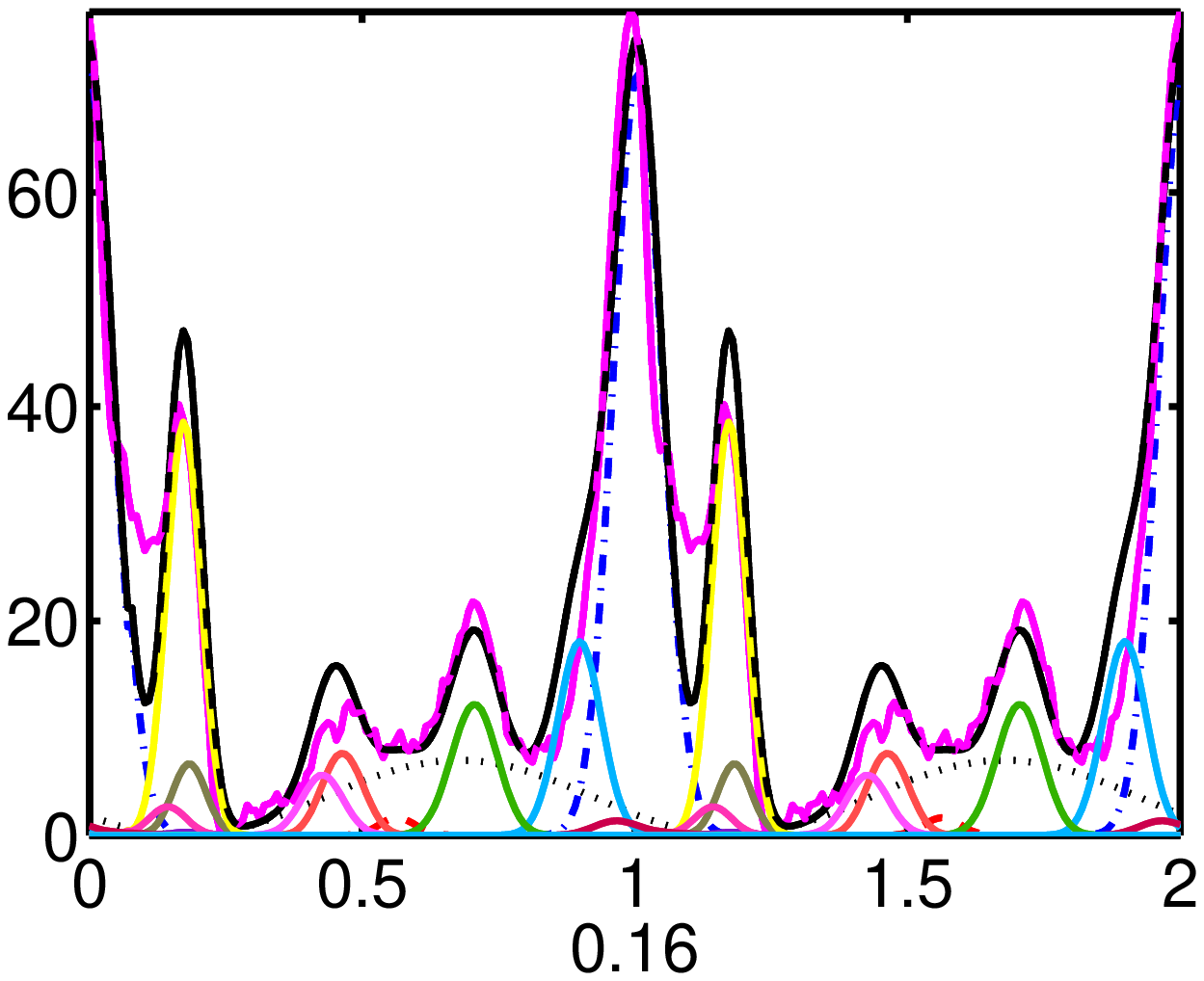}
\caption{303 RXTE}
\label{f303}
\end{figure}

\begin{figure}
%\centering
\includegraphics[width=0.24\textwidth]{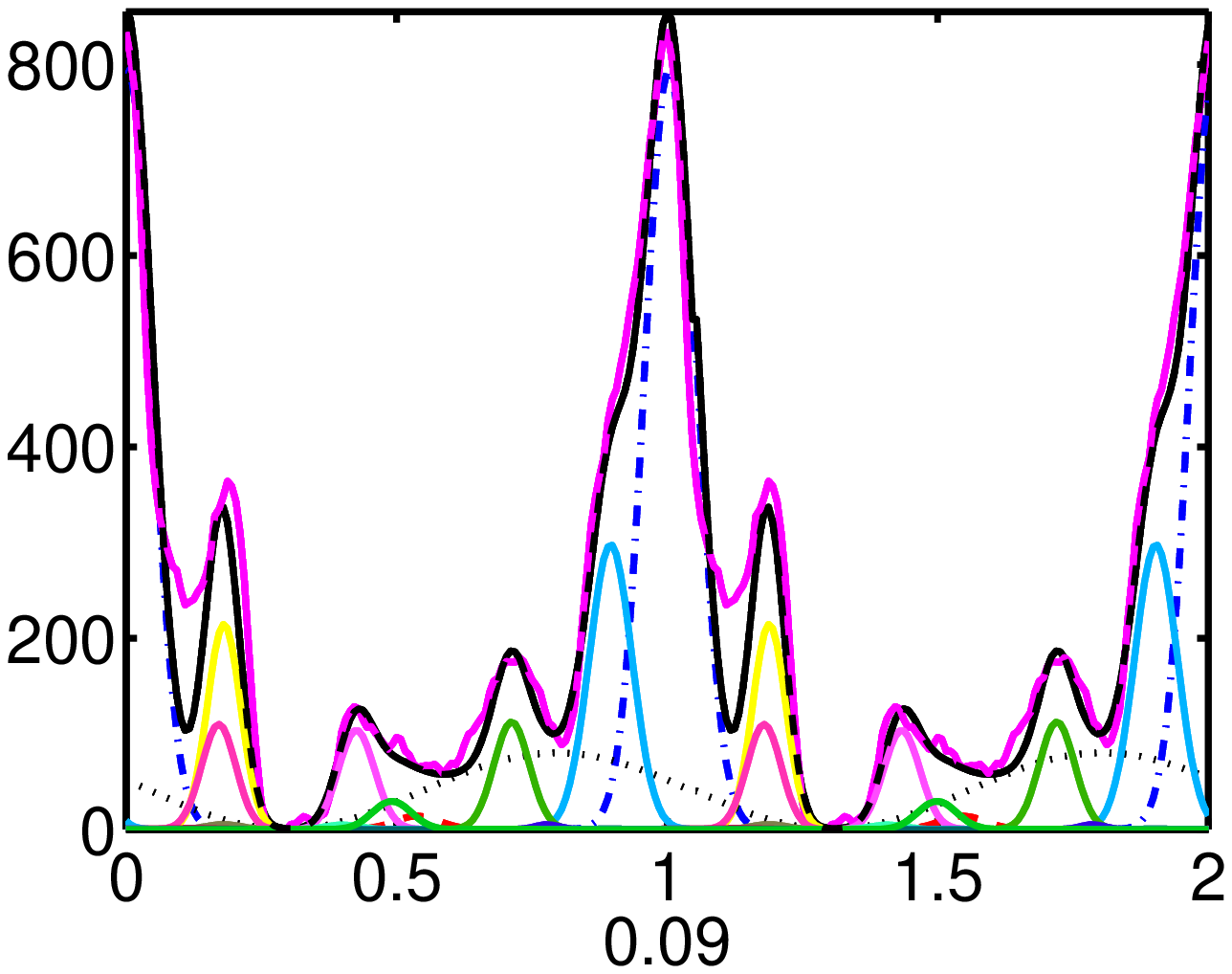}
\caption{304 RXTE}
\label{f304}
\end{figure}

\begin{figure}
%\centering
\includegraphics[width=0.24\textwidth]{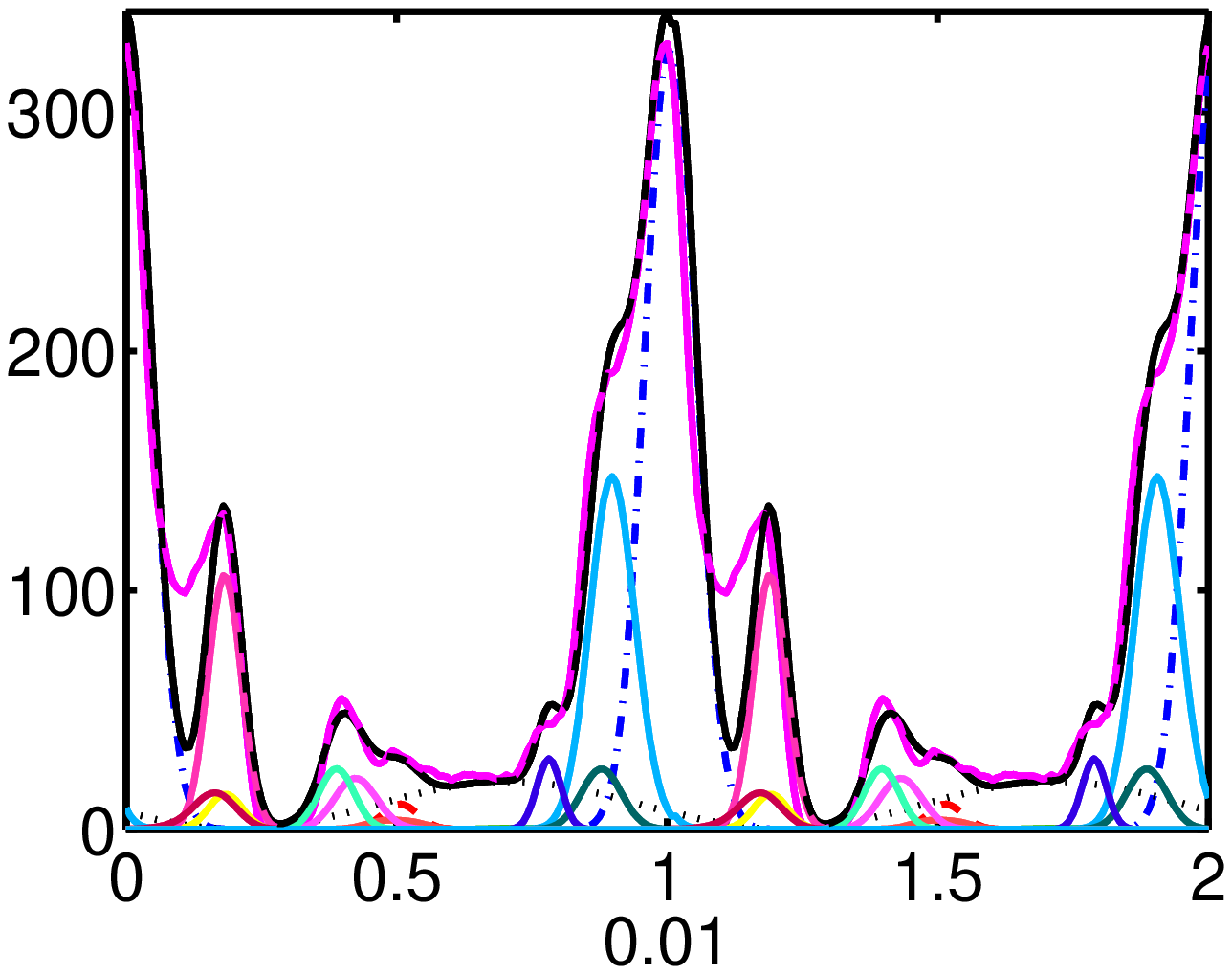}
\includegraphics[width=0.24\textwidth]{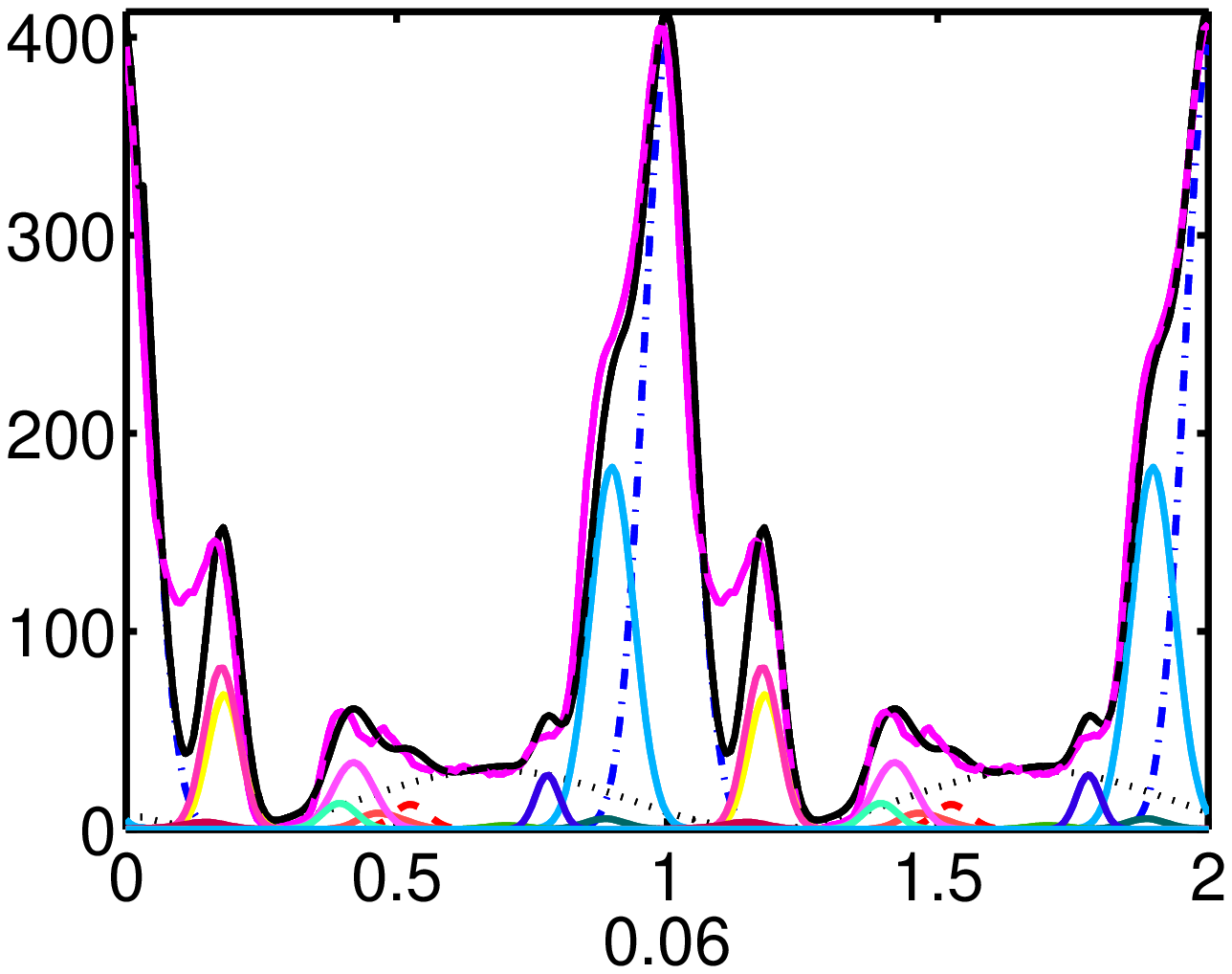}
\includegraphics[width=0.24\textwidth]{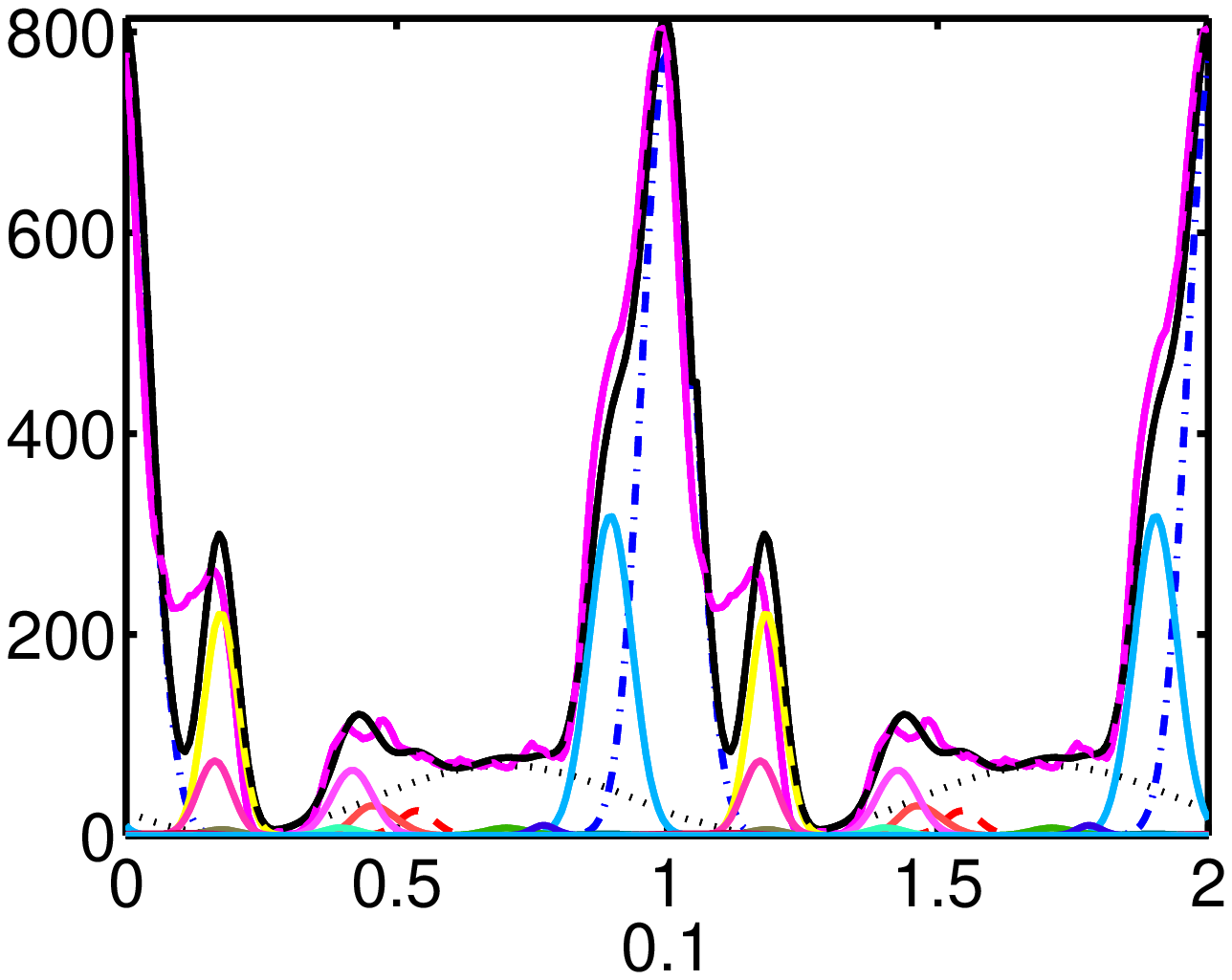}
\includegraphics[width=0.24\textwidth]{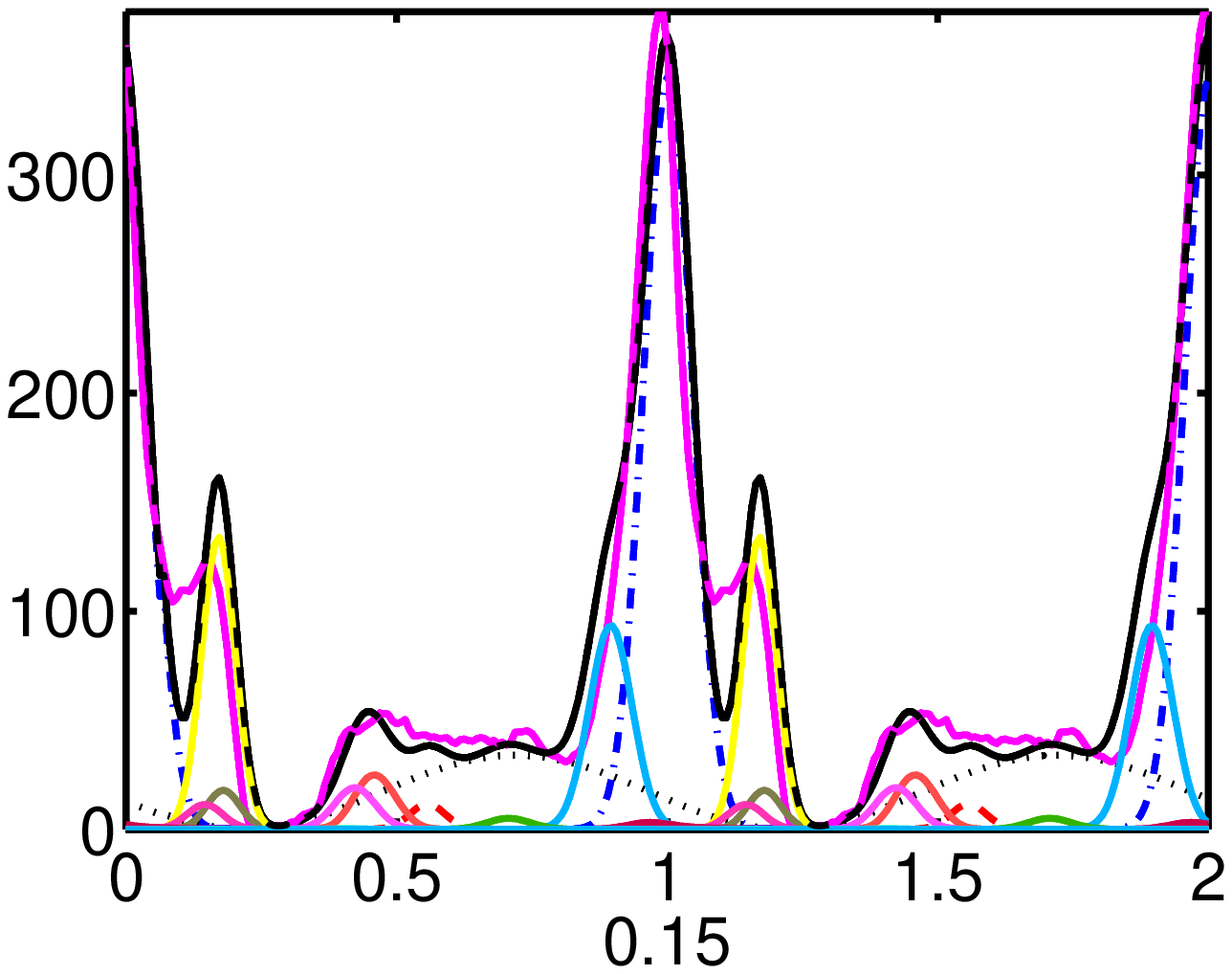}
\caption{307 RXTE}
\label{f307}
\end{figure}

\begin{figure}
%\centering
\includegraphics[width=0.24\textwidth]{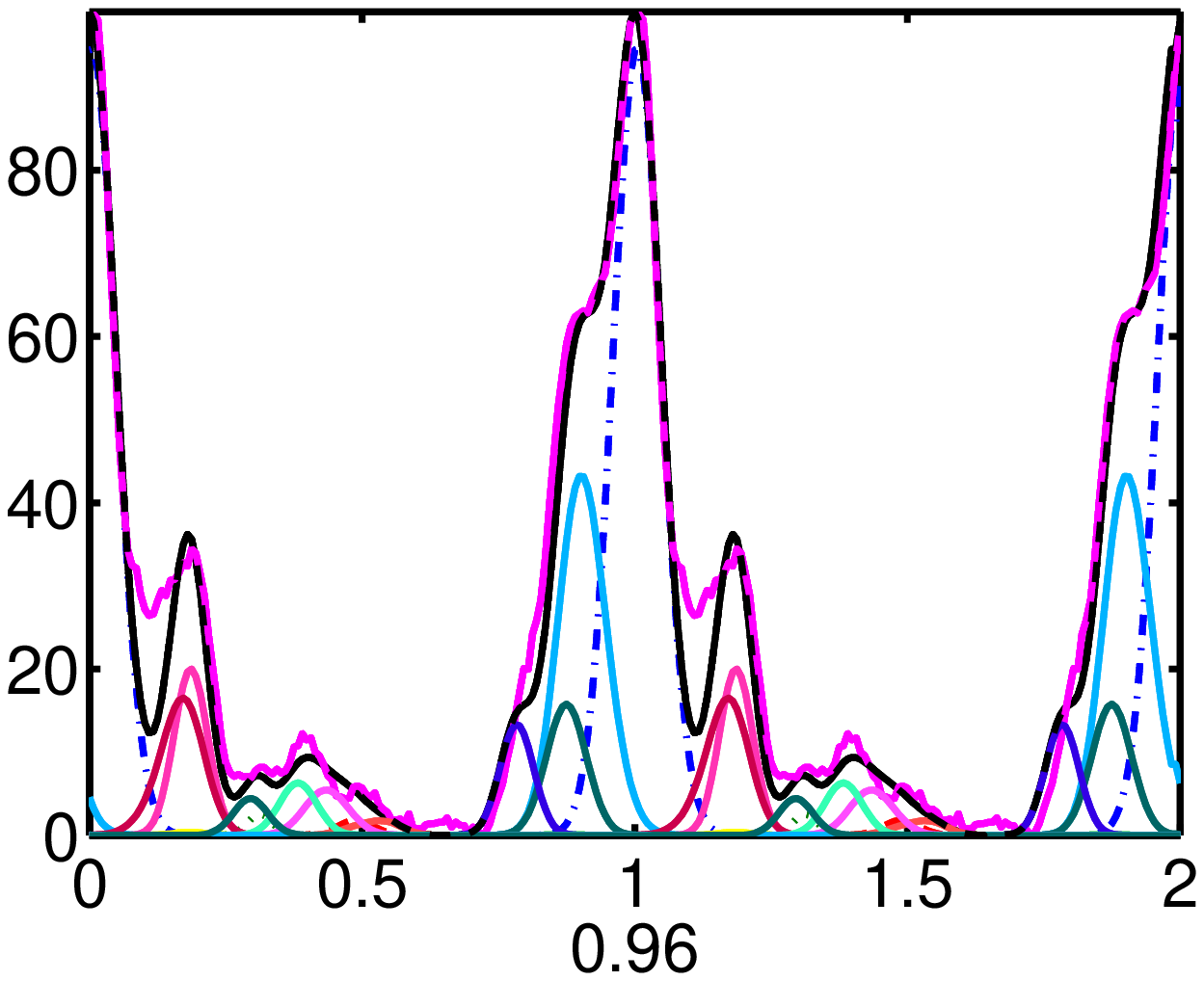}
\includegraphics[width=0.24\textwidth]{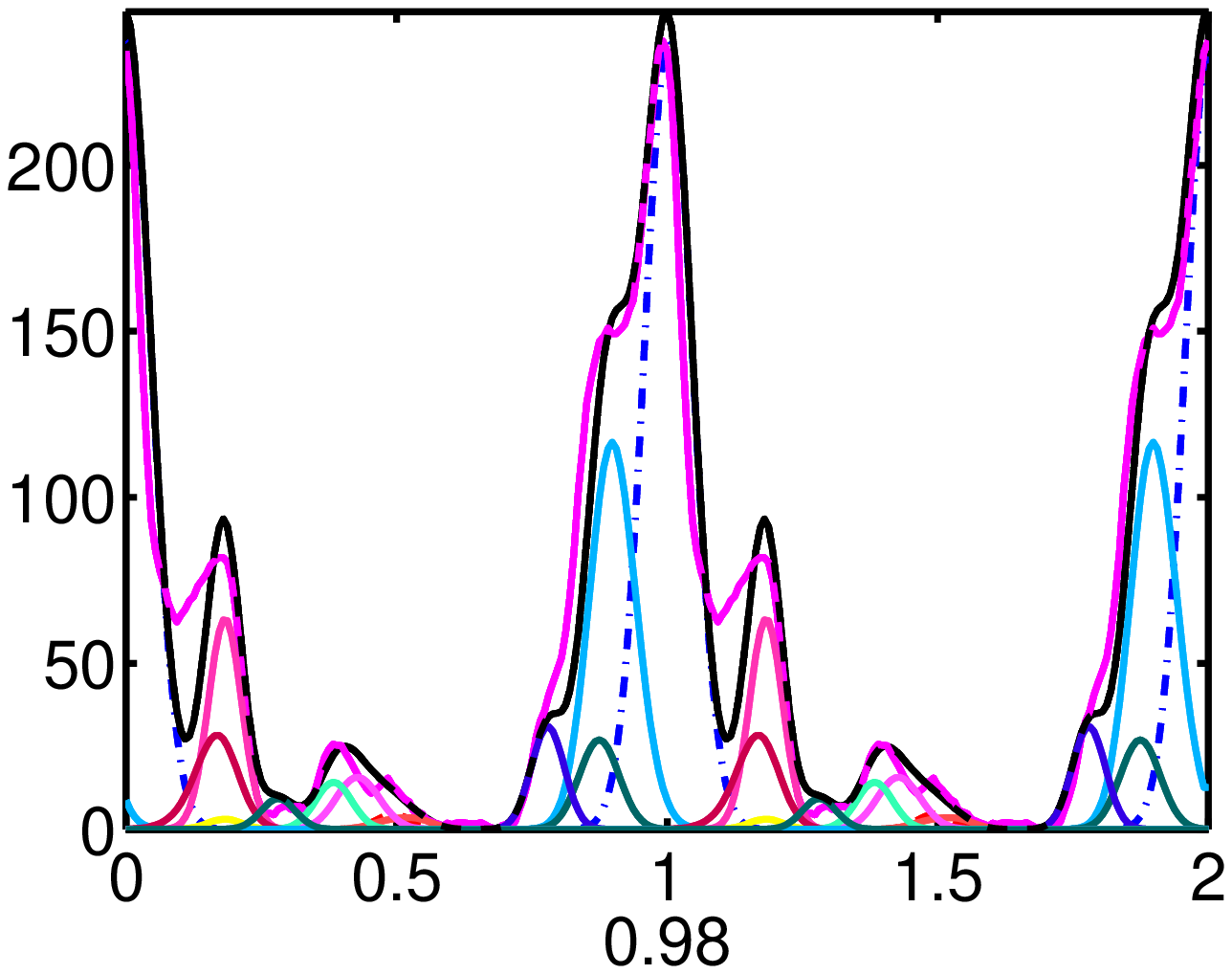}
\includegraphics[width=0.24\textwidth]{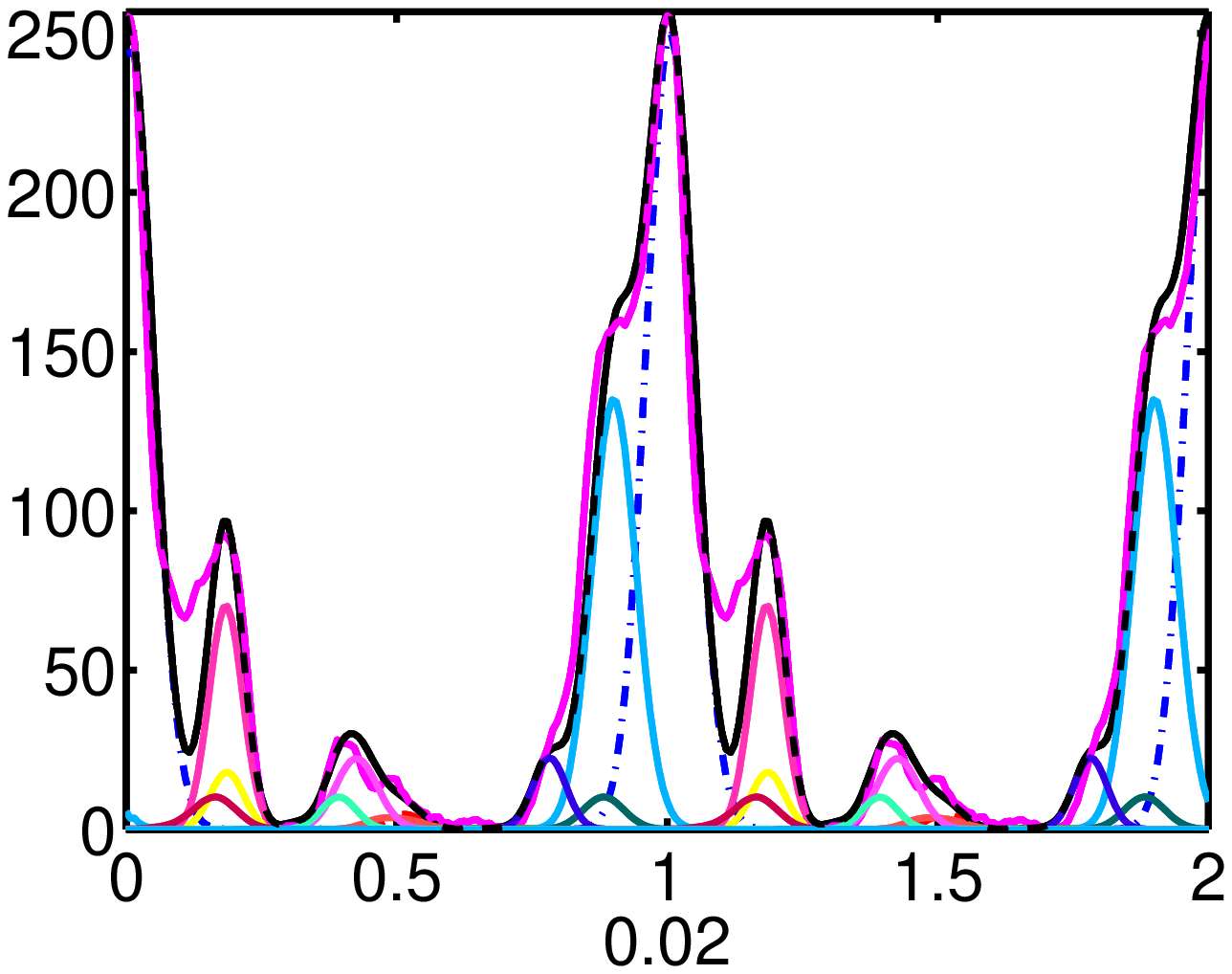}
\includegraphics[width=0.24\textwidth]{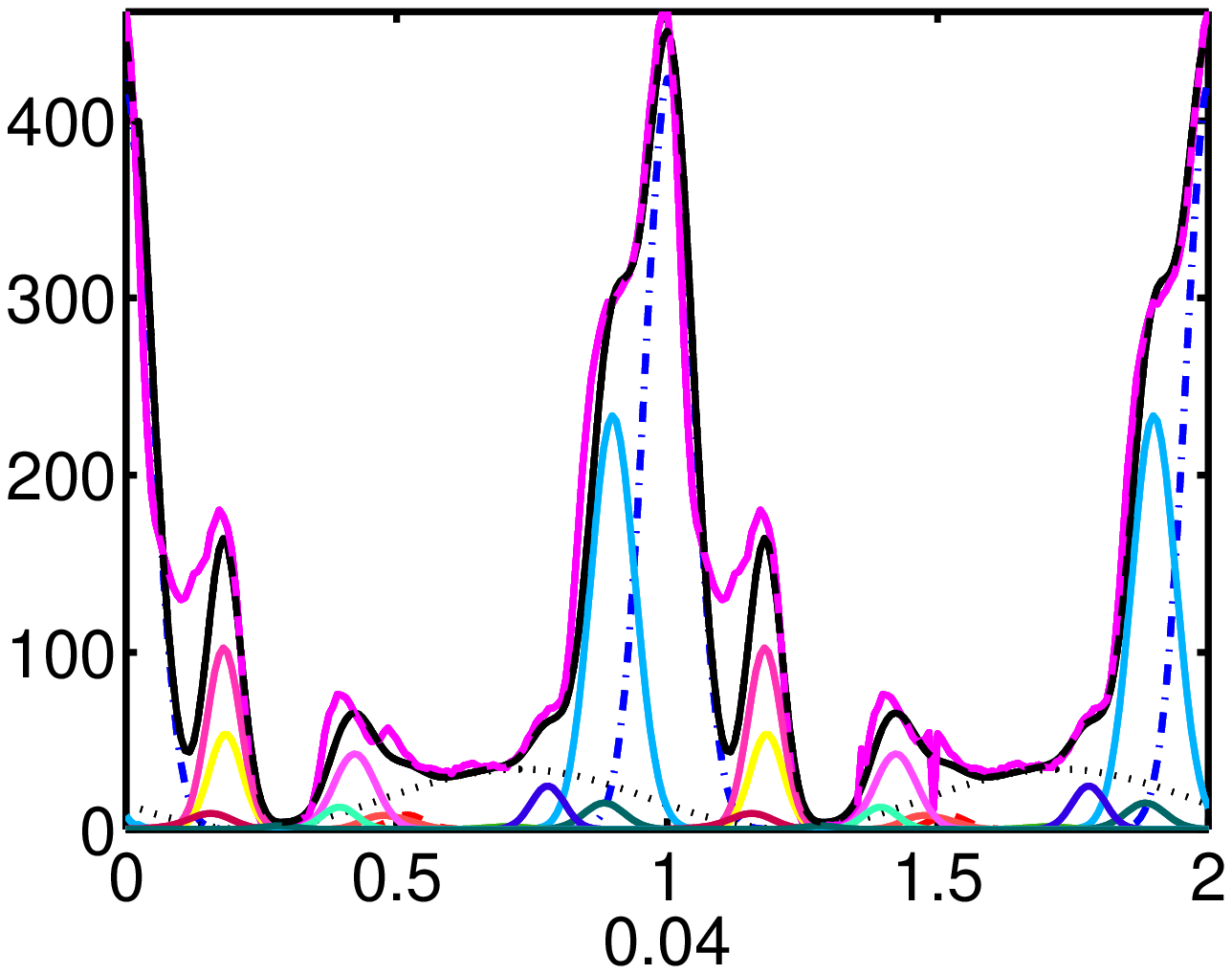}
\includegraphics[width=0.24\textwidth]{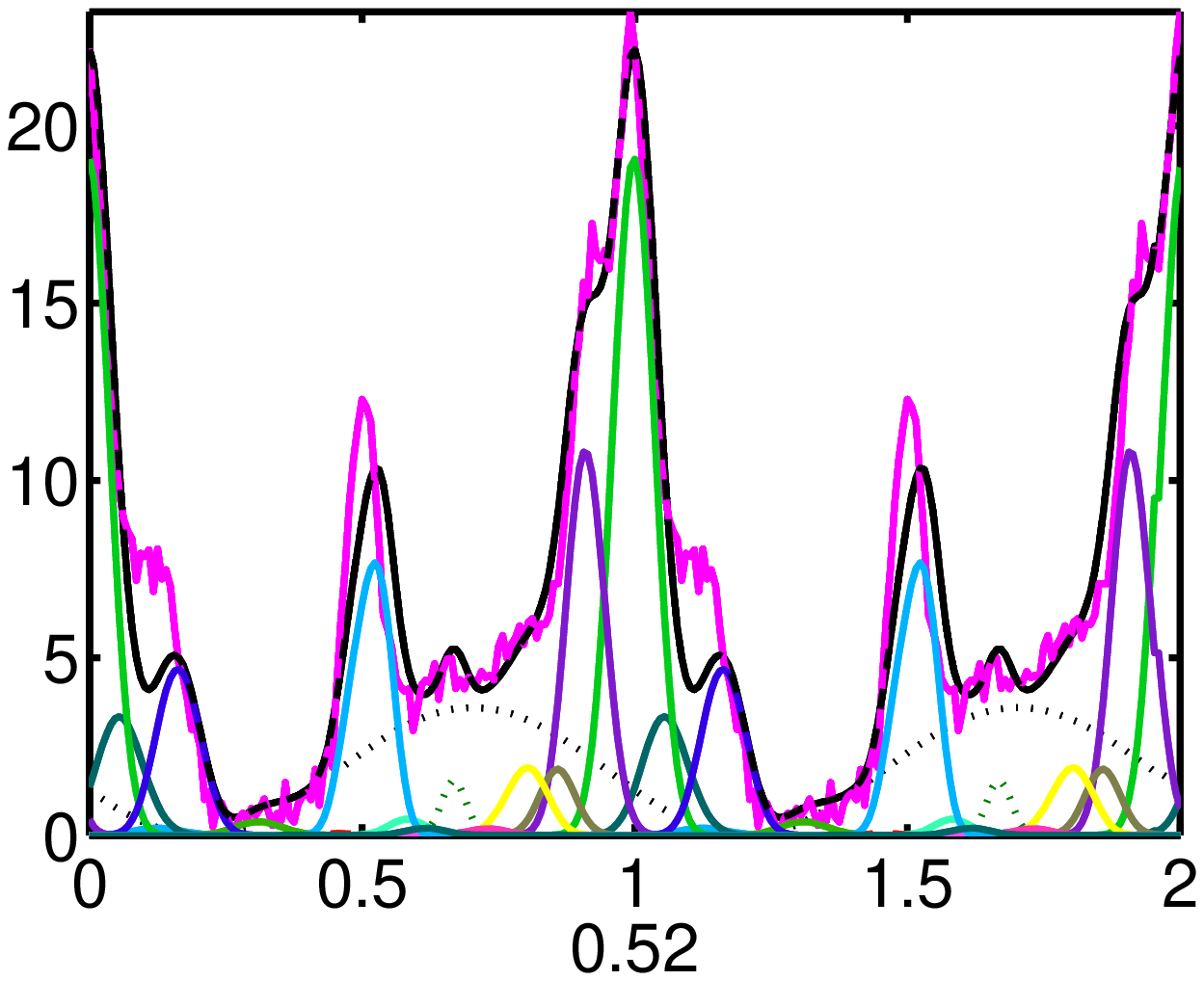}
\includegraphics[width=0.24\textwidth]{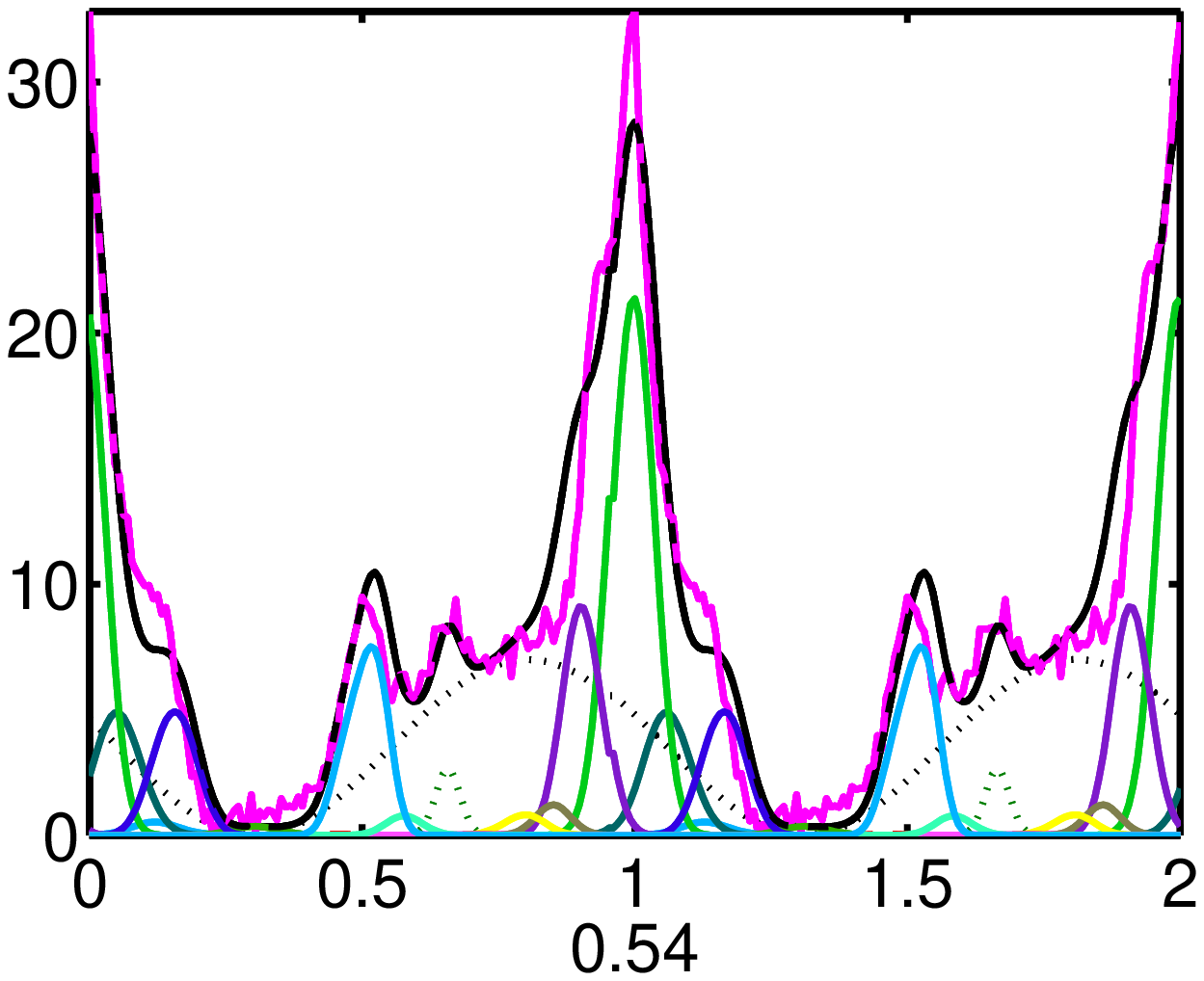}
\includegraphics[width=0.24\textwidth]{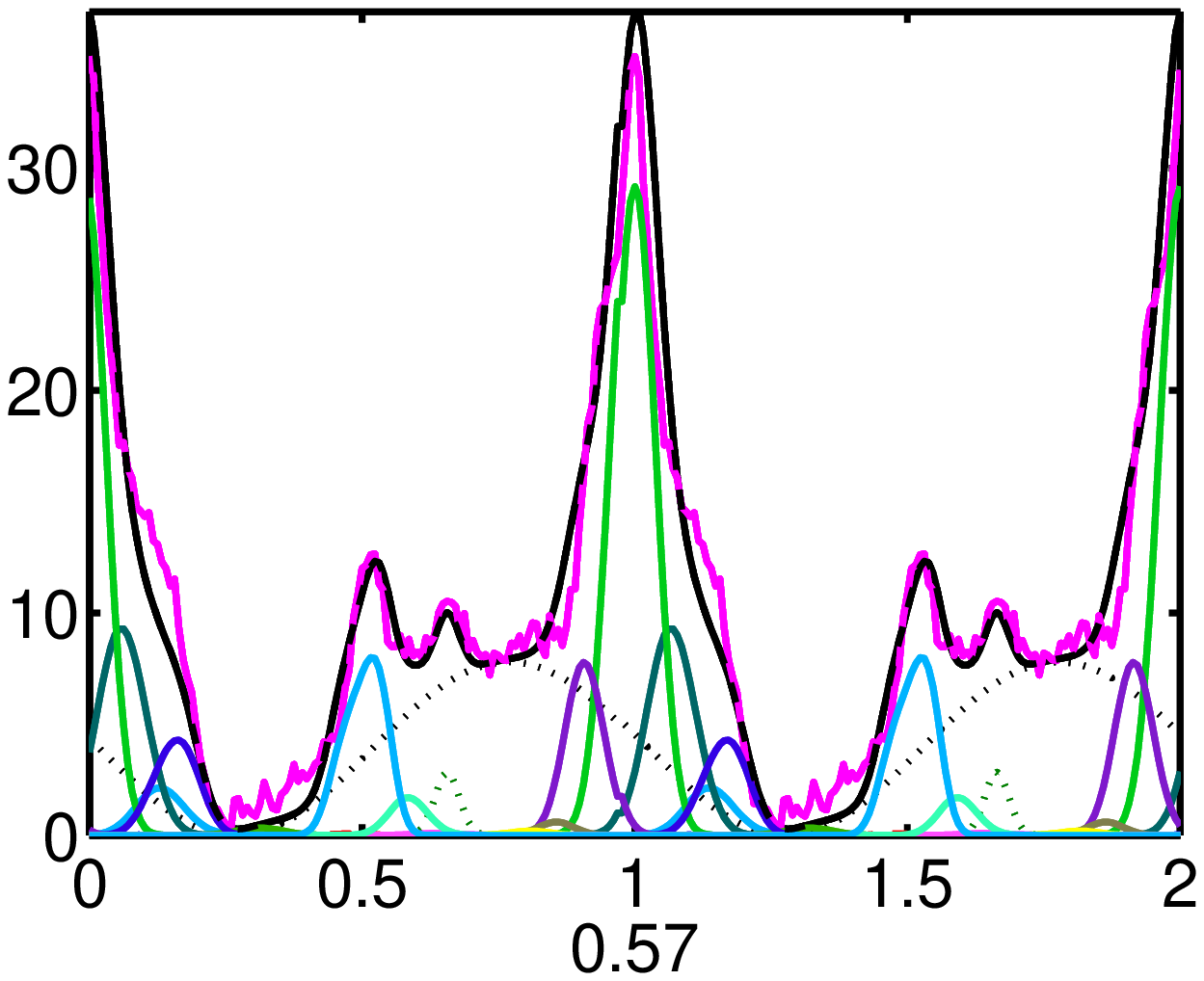}
\includegraphics[width=0.24\textwidth]{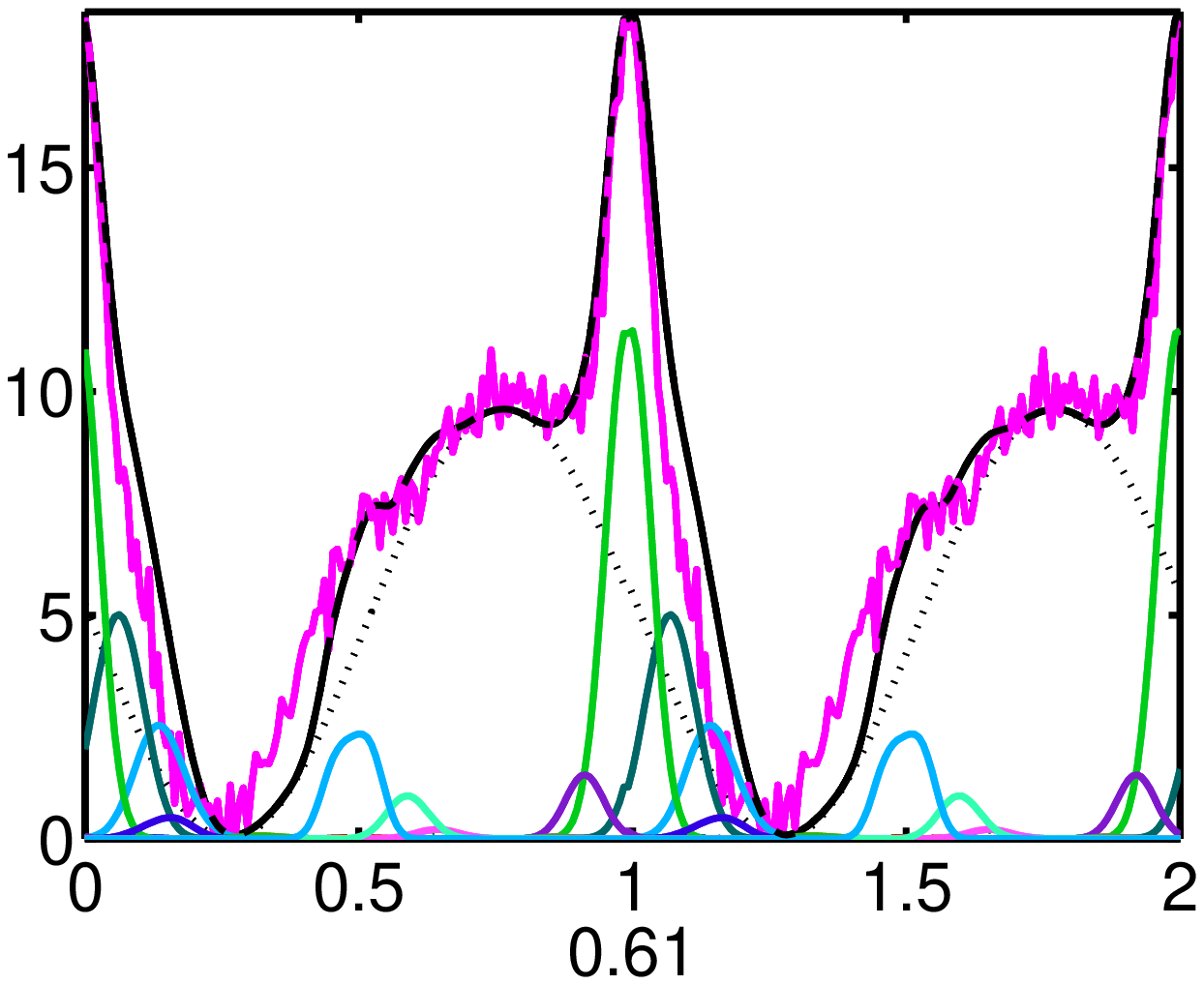}
\caption{313 RXTE}
\label{f313}
\end{figure}

\begin{figure}
%\centering
\includegraphics[width=0.24\textwidth]{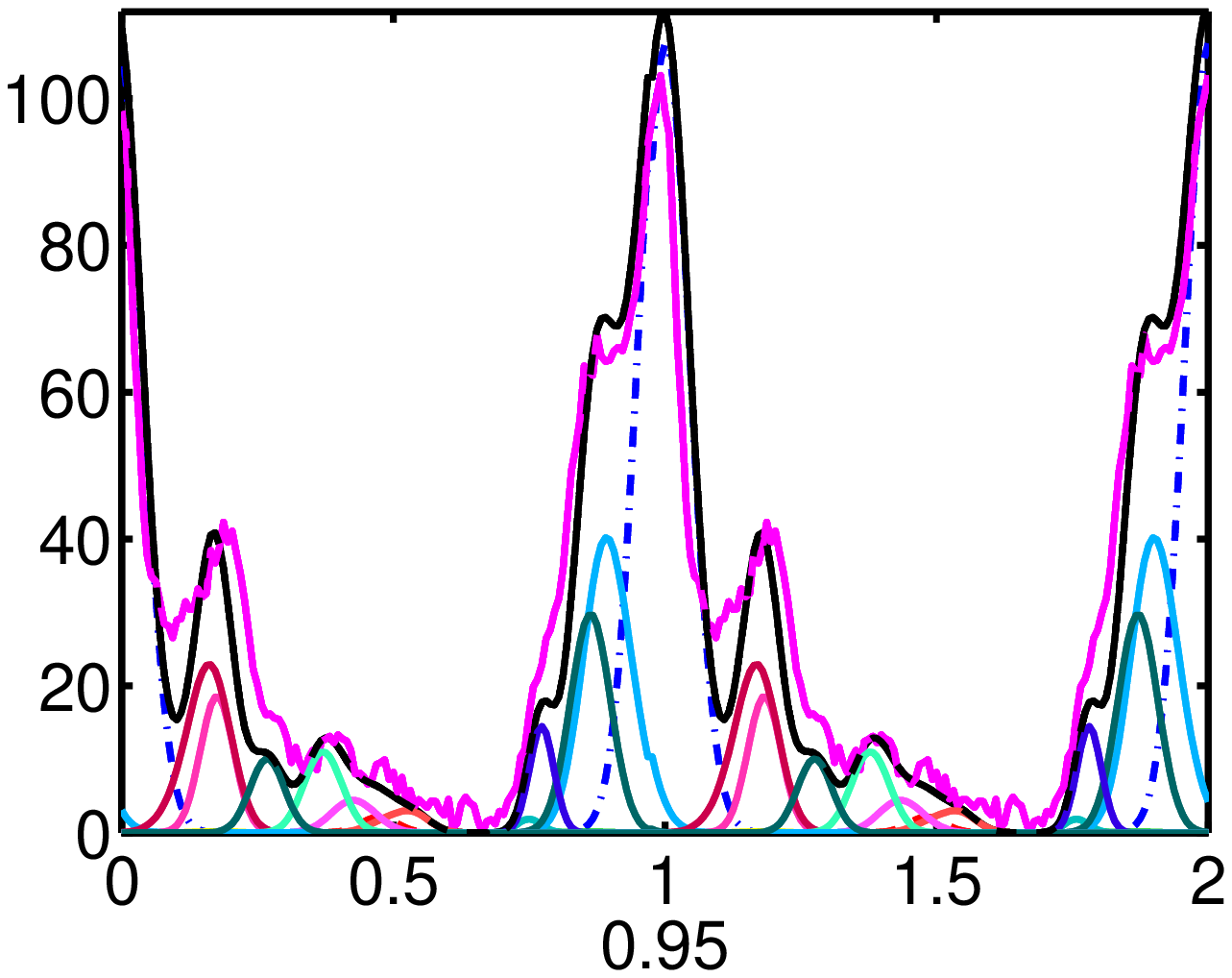}
\includegraphics[width=0.24\textwidth]{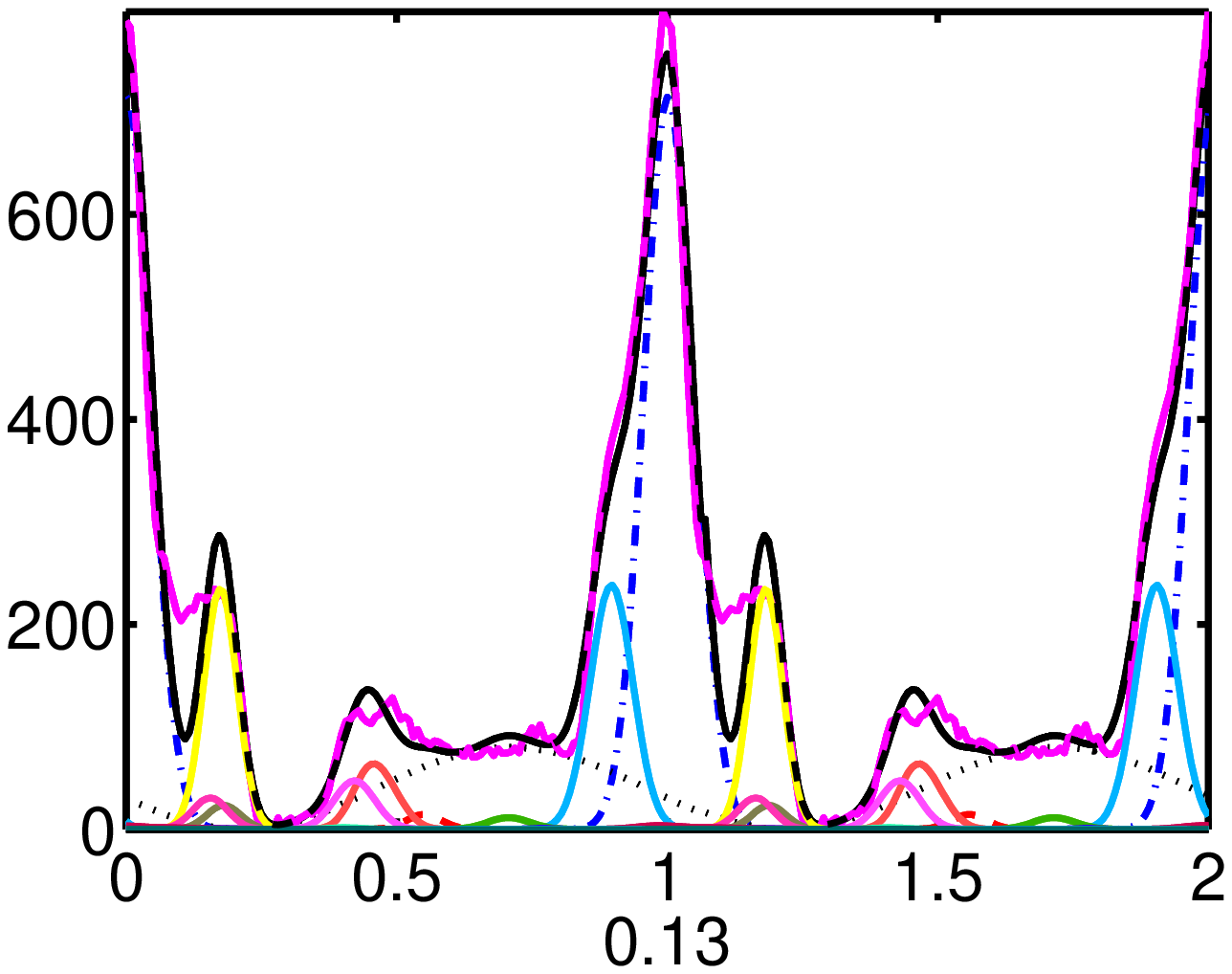}
\caption{319 RXTE}
\label{f319}
\end{figure}

\begin{figure}
%\centering
\includegraphics[width=0.24\textwidth]{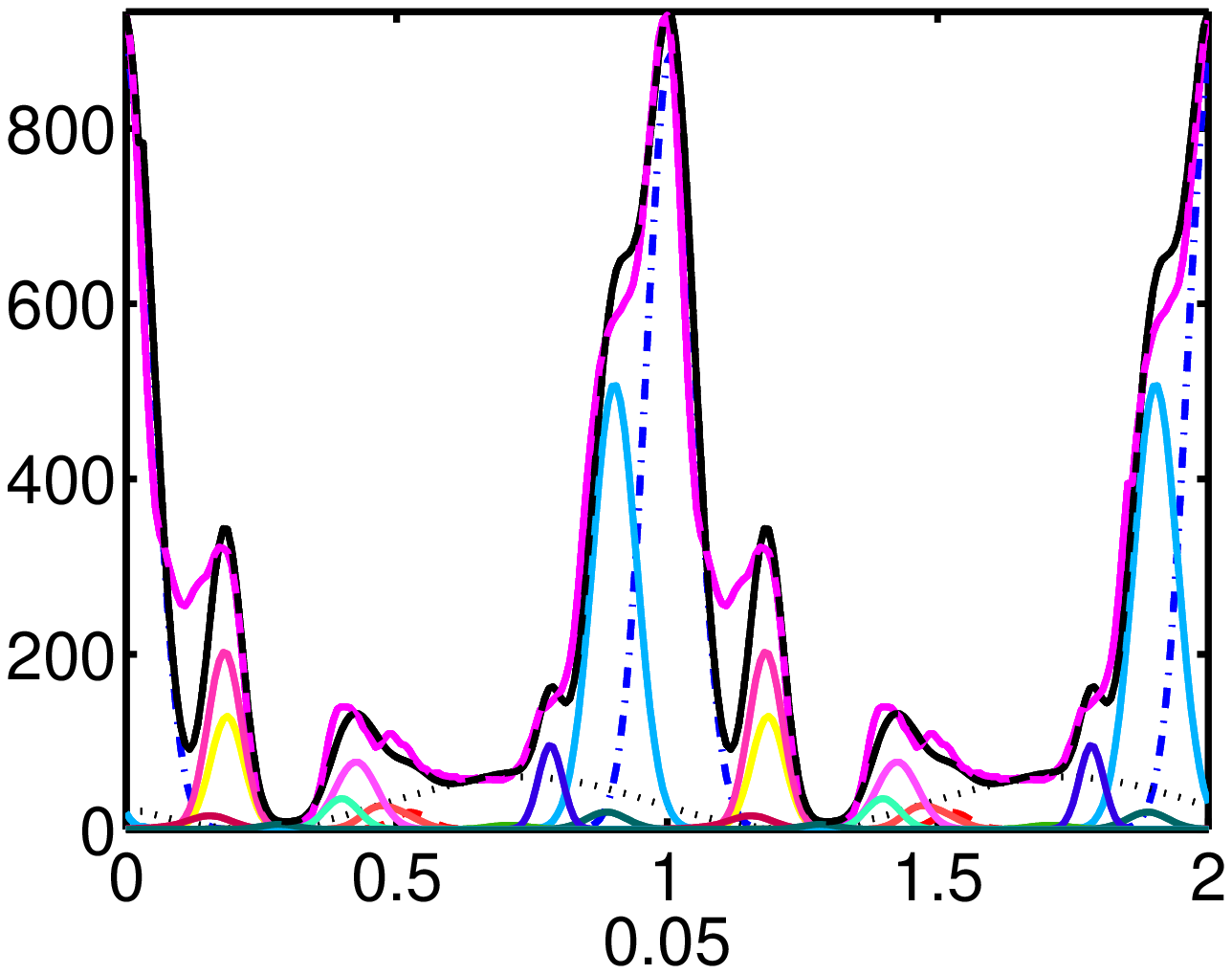}
\caption{320 RXTE }
\label{f320}
\end{figure}

\begin{figure}
%\centering
\includegraphics[width=0.24\textwidth]{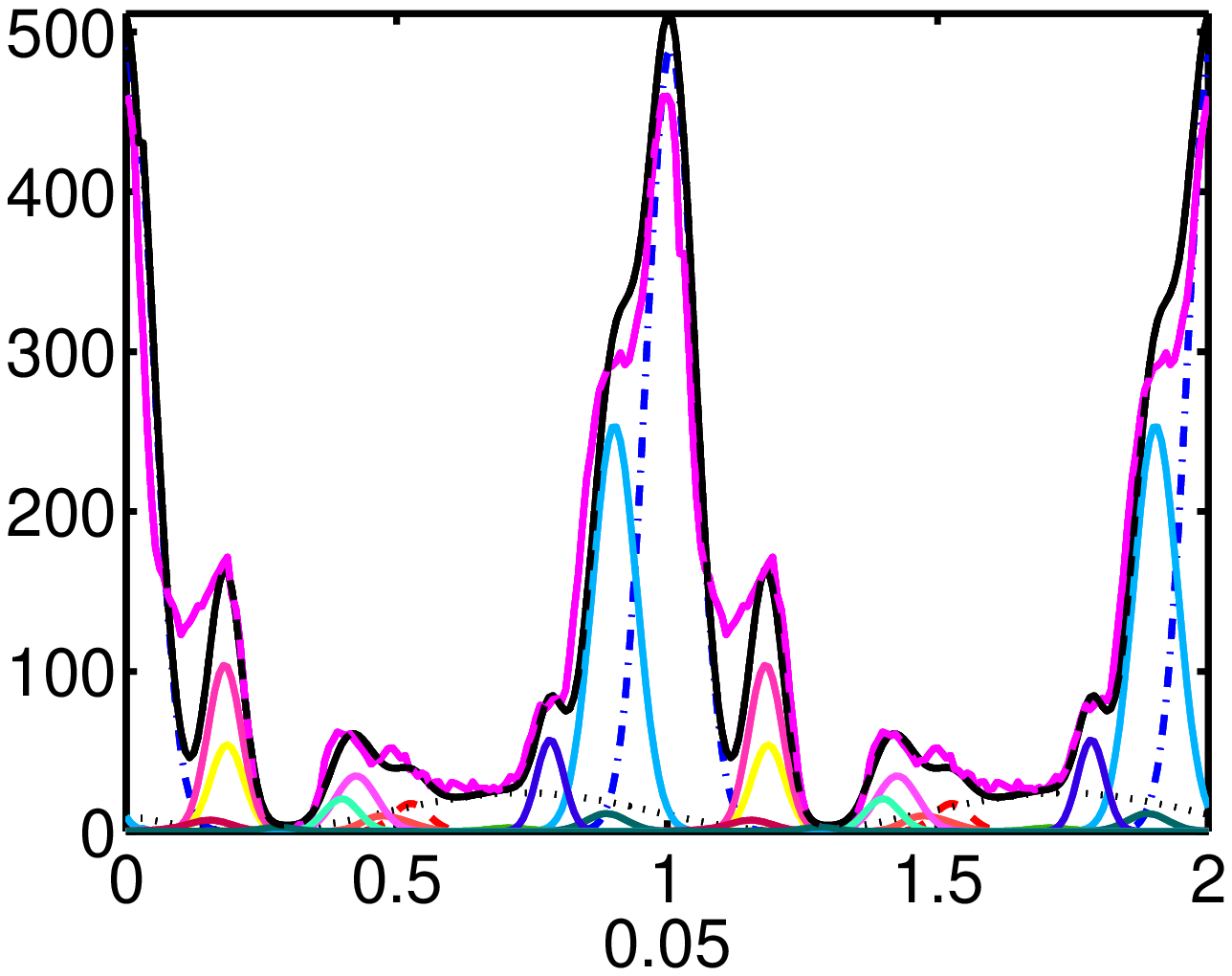}
\includegraphics[width=0.24\textwidth]{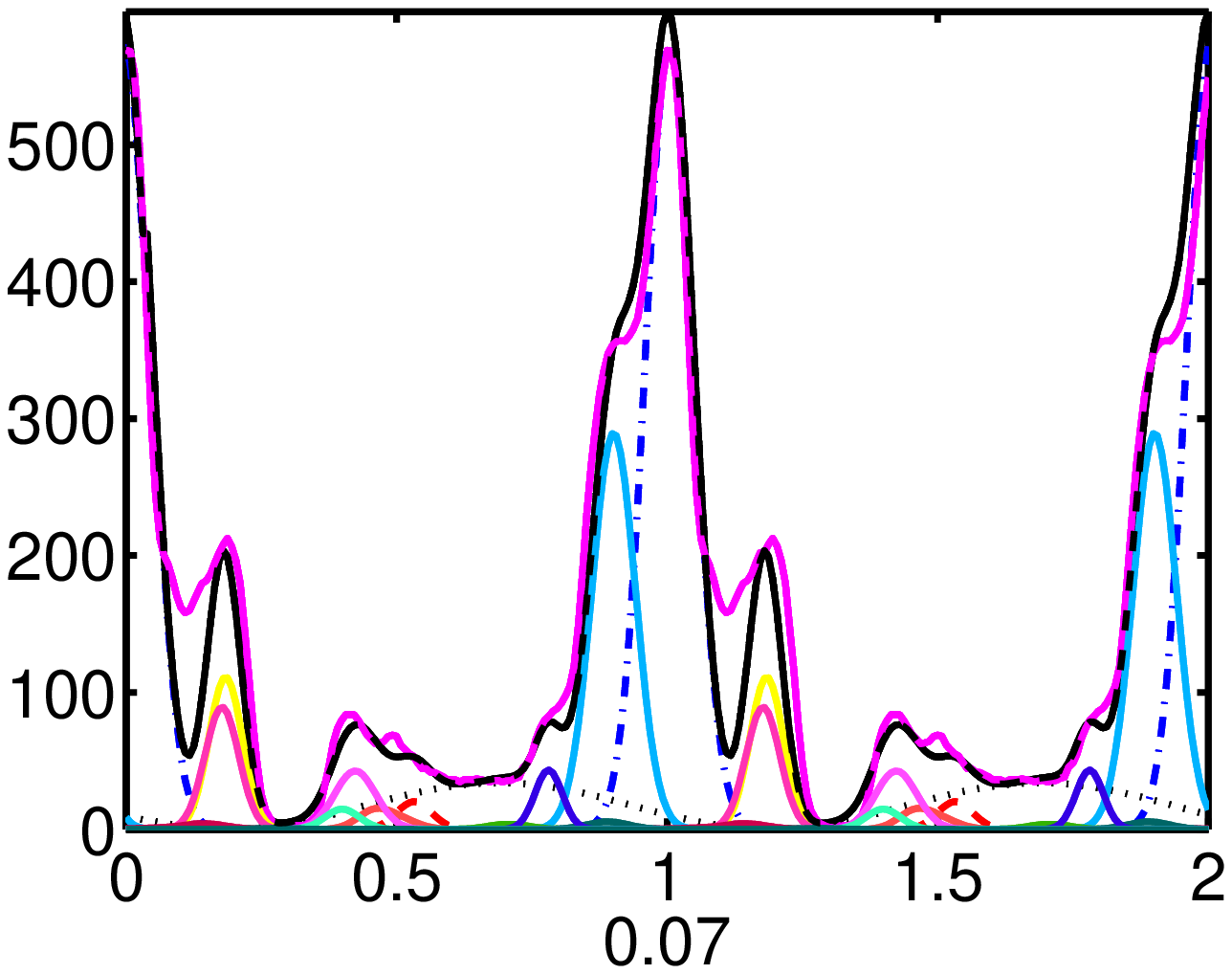}
\includegraphics[width=0.24\textwidth]{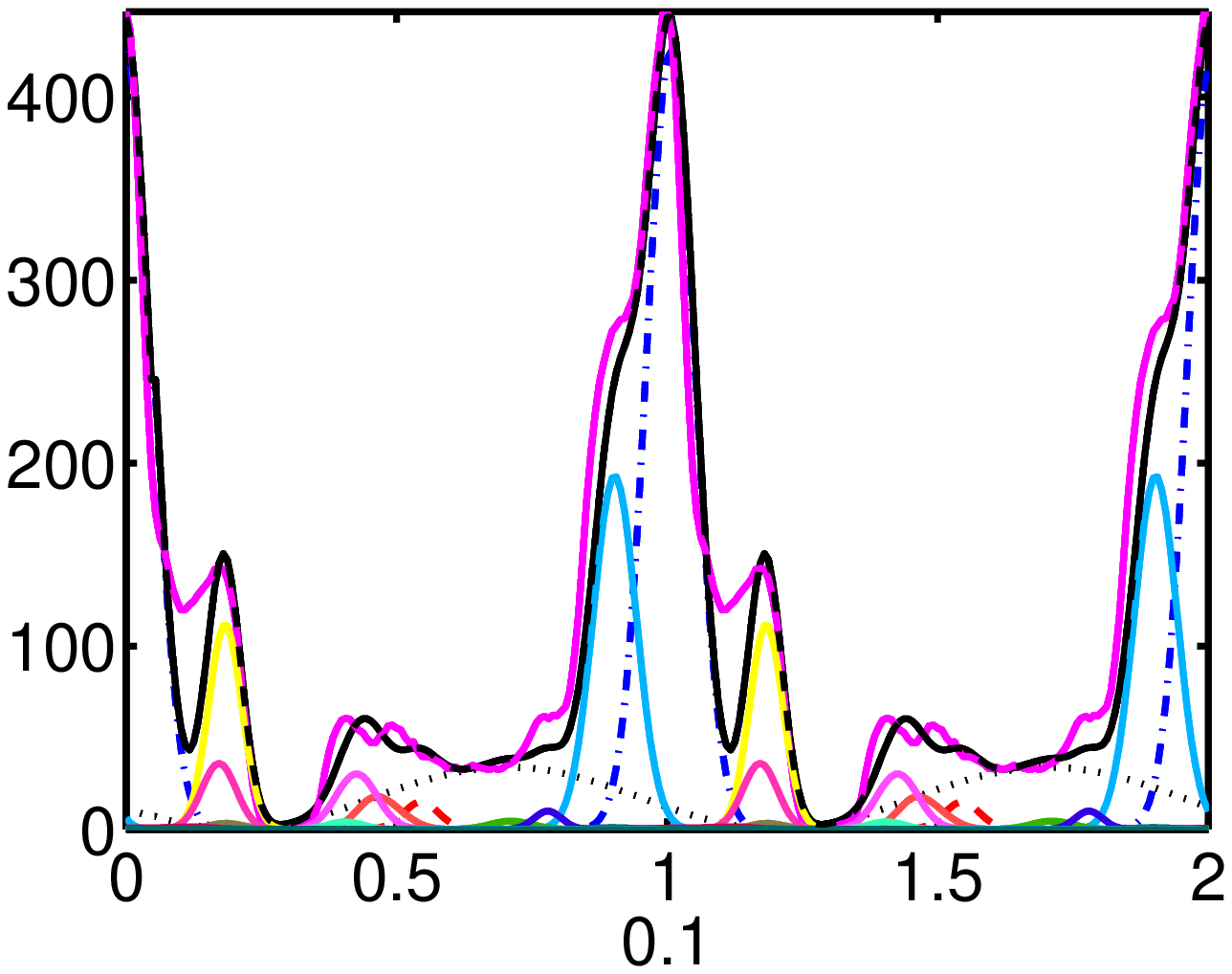}
\includegraphics[width=0.24\textwidth]{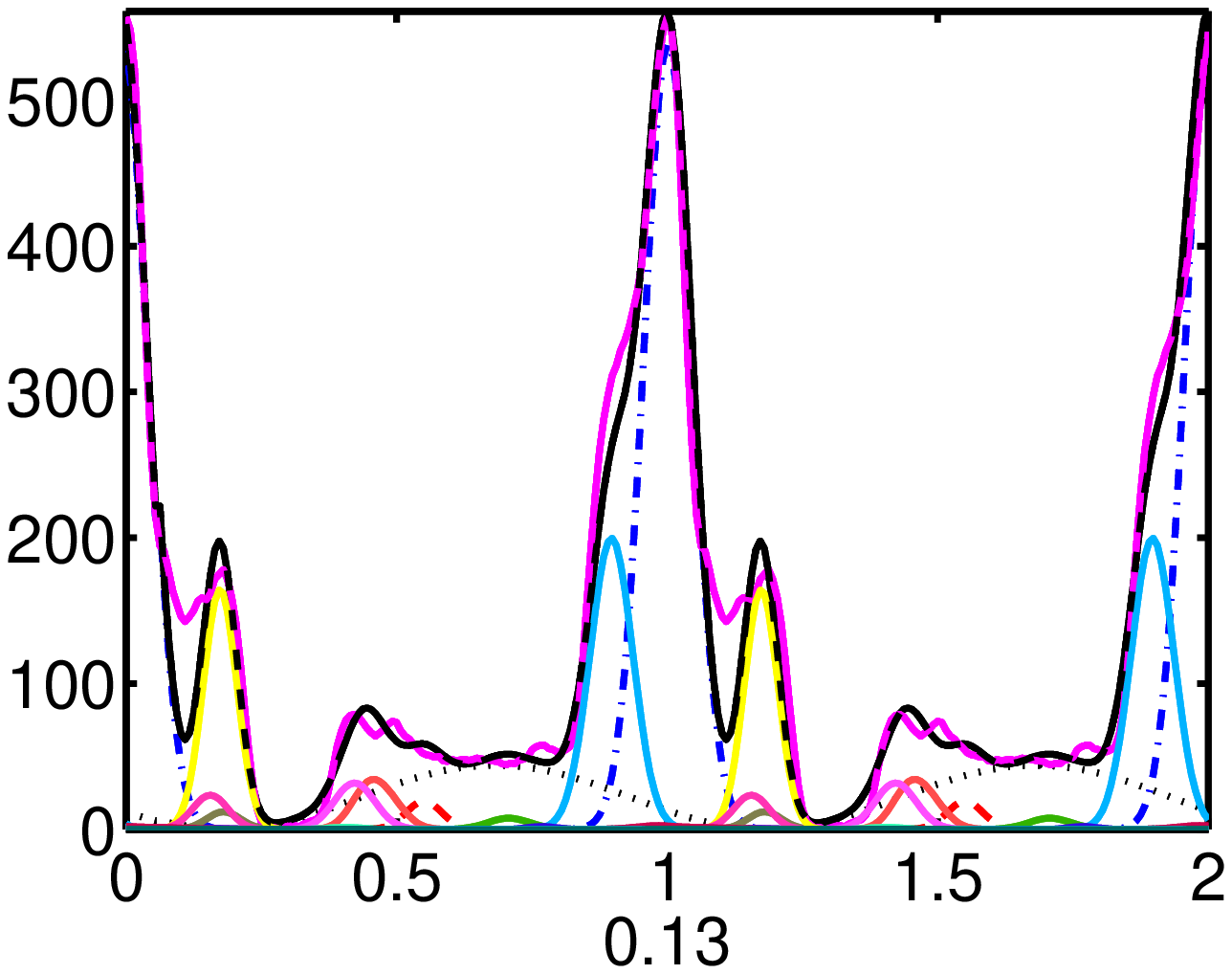}
\includegraphics[width=0.24\textwidth]{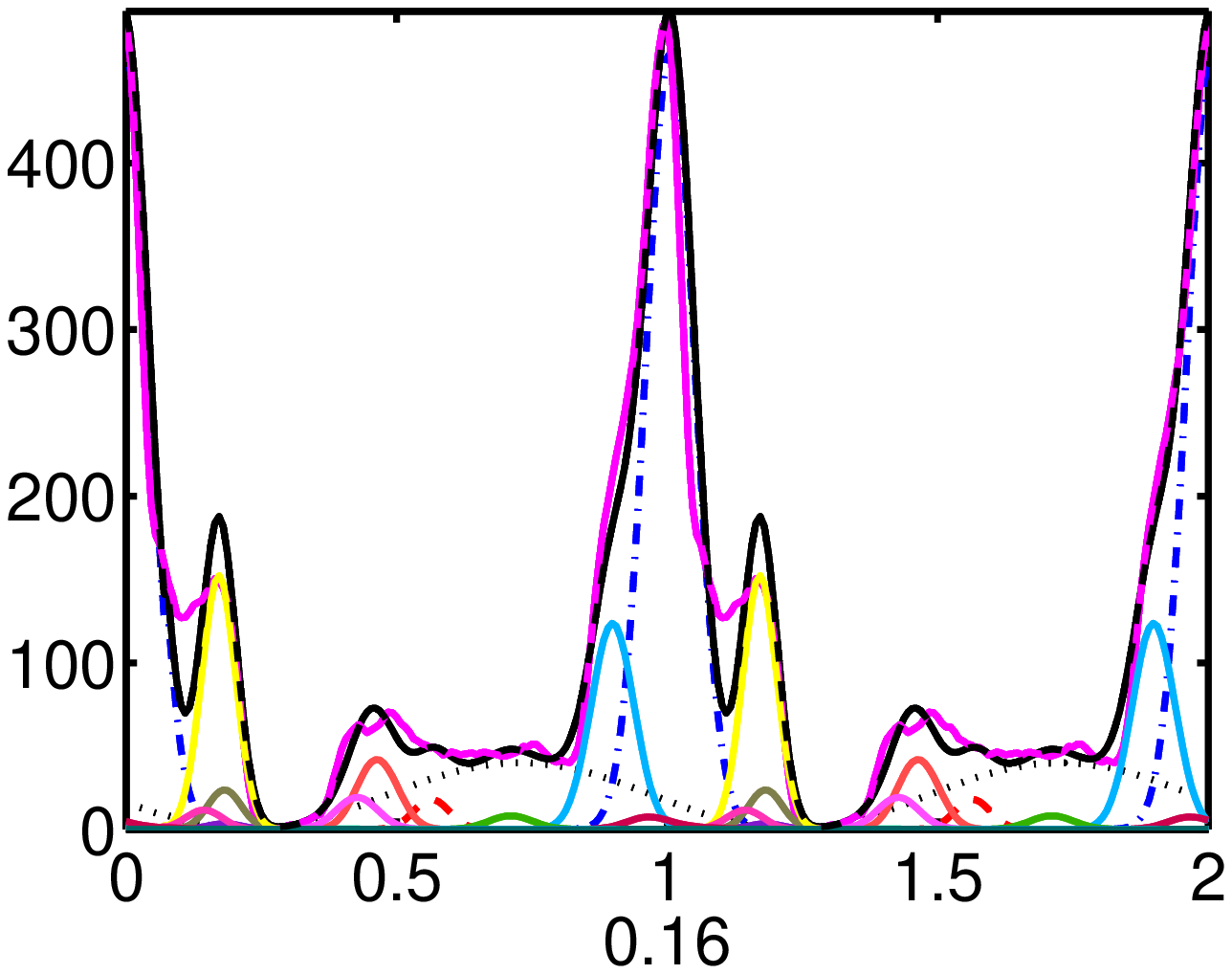}
\includegraphics[width=0.24\textwidth]{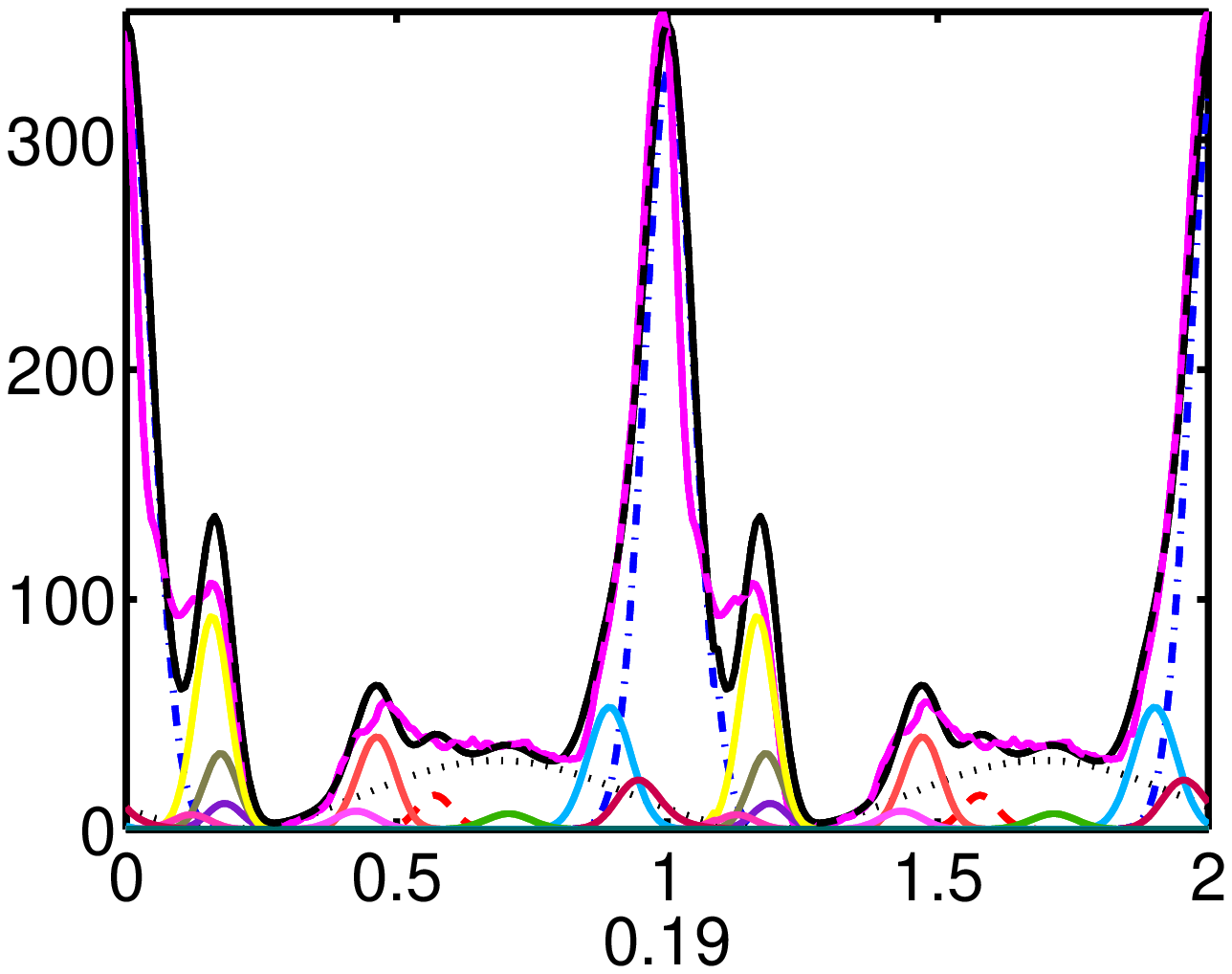}
\includegraphics[width=0.24\textwidth]{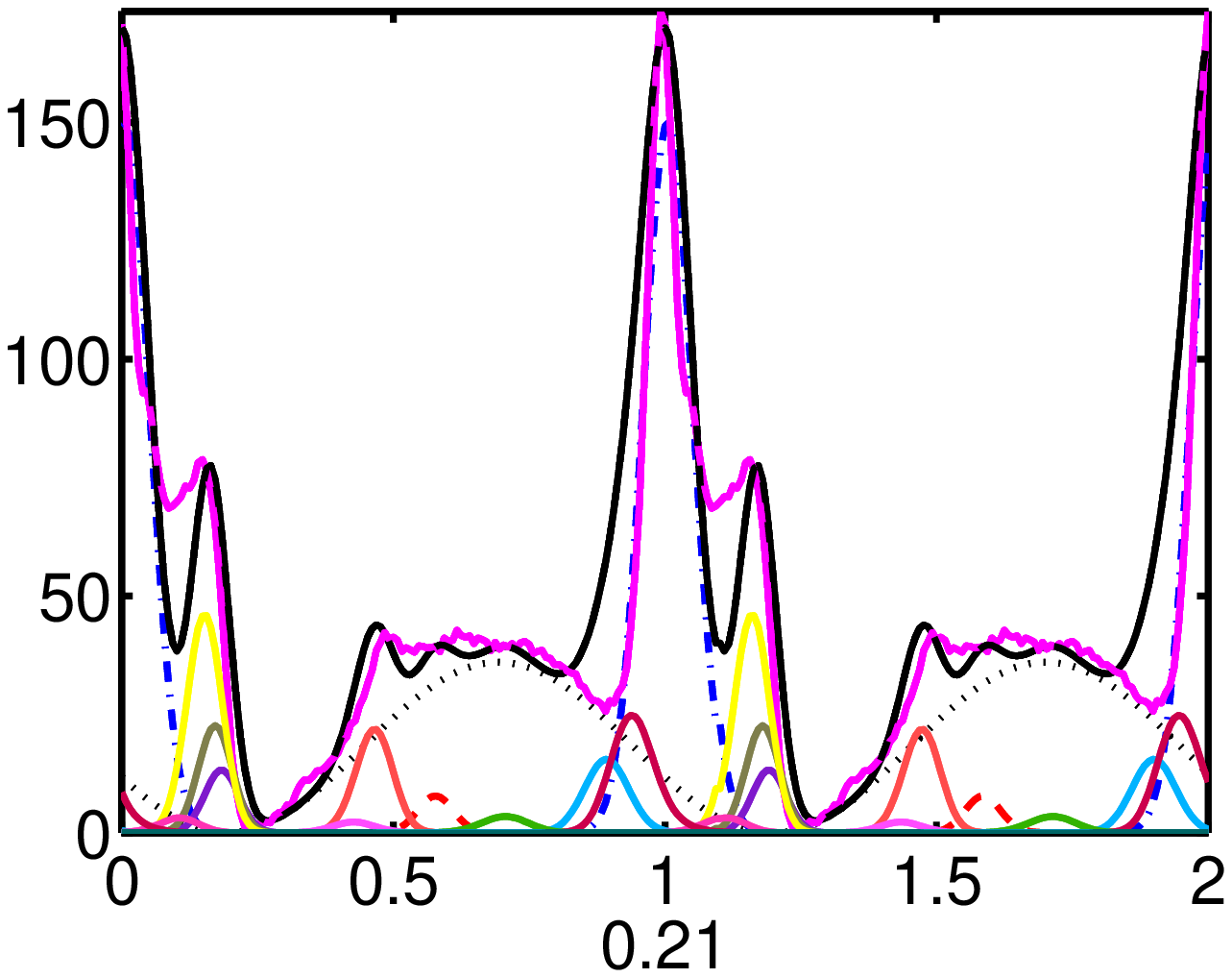}
\includegraphics[width=0.24\textwidth]{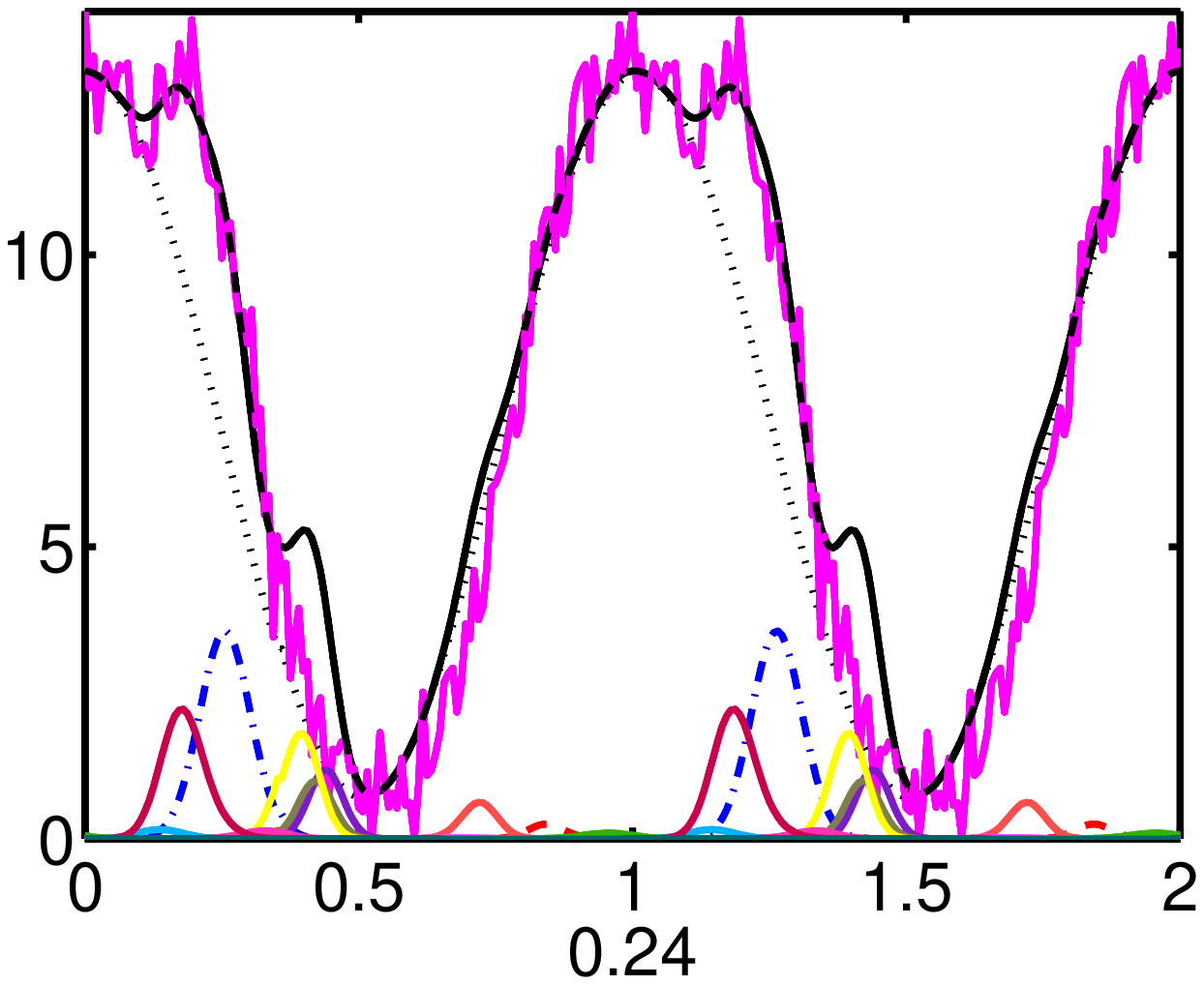}
\caption{323 RXTE}
\label{f323}
\end{figure}

\begin{figure}
%\centering
\includegraphics[width=0.24\textwidth]{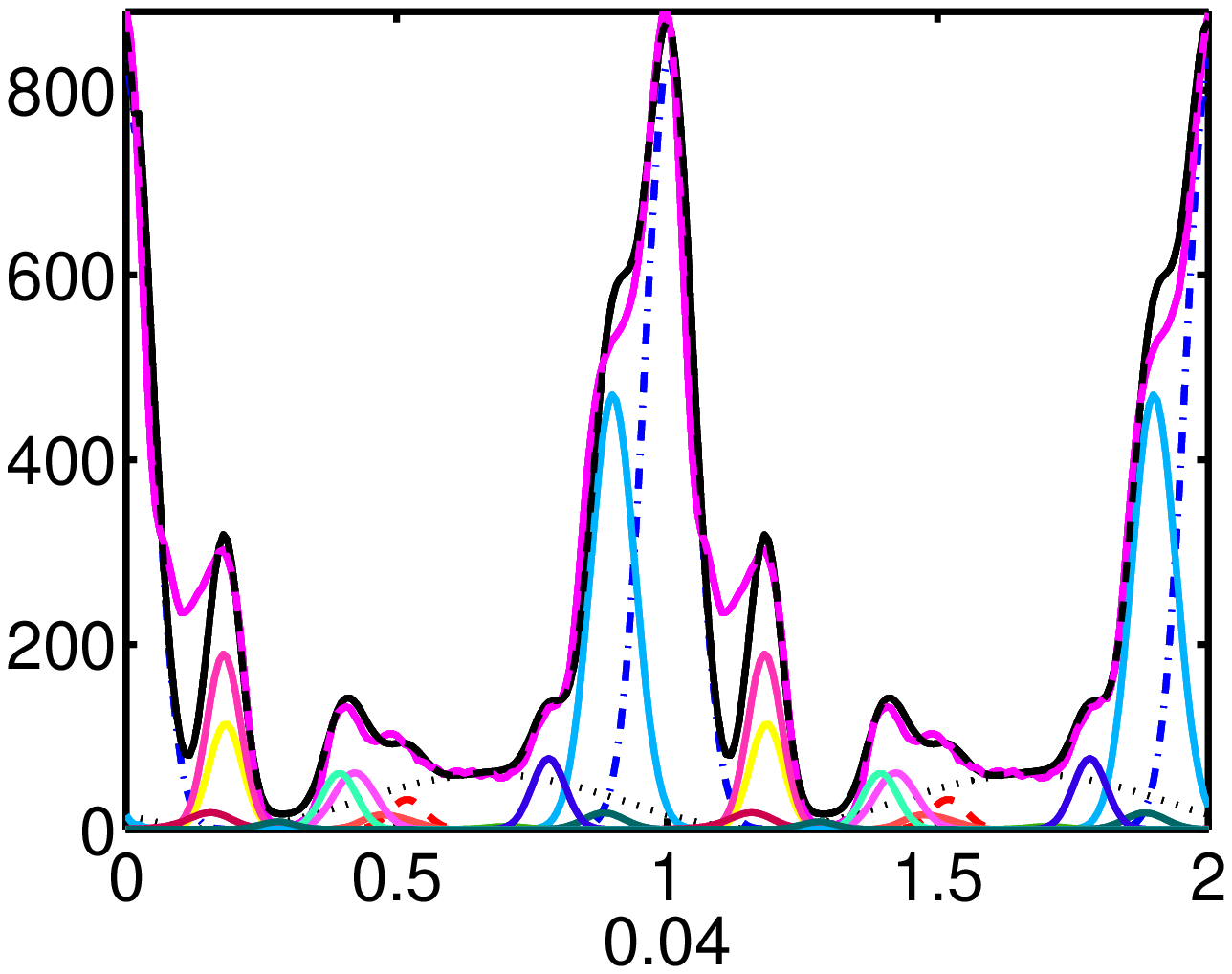}
\caption{324 RXTE}
\label{f324}
\end{figure}

\label{lastpage}

\end{document}